# Overt and covert paths for sound in the auditory system of mammals, version4

B. Auriol[1]*, B. Auriol[2], J. Béard[3], B. Bibé[4], J.-M. Broto[5]*, D.F. Descouens[6], L.J.S. Durand[7], J.-P. Florens[8], F. Garcia[9], C. Gillieaux[10], E.G. Joiner[11], B. Libes[12], P. Pergent[13], R. Ruiz[14], C. Thalamas[15].

## Abstract

Current scientific consensus holds that sound is transmitted, solely mechanically, from the tympanum to the cochlea via ossicles.

But this theory does not explain the hearing extreme quality regarding high frequencies in mammals. So, we propose a bioelectronic pathway (the covert path) that is complementary to the overt path..

We demonstrate experimentally that the tympanum produces piezoelectric potentials isochronous to acoustic vibrations thanks to its collagen fibers and that their amplitude increases along with the frequency and level of the vibrations. This finding supports the existence of an electrical pathway, specialized in transmitting high-frequency sounds, that works in unison with the mechanical pathway. A bio-organic triode, similar to a field effect transistor, is the key mechanism of our hypothesized pathway. We present evidence that any deficiency along this pathway produces hearing impairment. By augmenting the classical theory of sound transmission, our discovery offers new perspectives for research into both normal and pathological audition and may contribute to an understanding of genetic and physiological problems of hearing.

## Introduction

The scientific literature of sound transmission and perception is founded on the travelling wave (TW) principle proposed by von Bekesy (Nobel, 1961). The eardrum vibrates in response to sound. These vibrations travel via the three ossicles through the oval window and into the fluid-filled cochlea.  Inside the cochlea, acoustic signals are broken down into their component frequencies by the mechanical properties of the basilar membrane, to which the hair cells are attached.

This paper questions the adequacy of current theory to explain the transmission of high frequency sounds (above 3 kHz according to Quix[1]) along the tympano-cochlear pathway. Beginning with the tympanum, we observe that radial fibres of collagen extend from its periphery to its centre.
Thus the eardrum is a membrane attached to its periphery. At high frequencies, the Chladni model implies local resonances, which fragment its surface into vibrating zones all the more complex as the frequency increases[2].
This implies[3] that ossicular transmission is extremely weak for frequencies above 2 kHz. The magnitude and the precision of the transfer decrease as well, particularly in the 2-5 kHz range[4-5] (Cf. fig. 1a and 1b below) and this apparent flaw in current theory has not been convincingly explained[6].

**1**- MD, Toulouse, France (*auriol@free.fr). **2**- Théâtre du Capitole, Toulouse. **3**- LNCMI, CNRS, EMFL, INSA, UGA, UPS, Toulouse. **4**- Emer. Dir. of Research at INRA, Toulouse.  **5**- University Professor, UPS, Toulouse 3, LNCMI CNRS (*jean-marc.broto@univ-tlse3.fr). **6**- MD (ENT), UMR 5288, CNRS, Toulouse. **7**-  CEMES/CNRS, Toulouse. **8**- TSE, Toulouse. **9**- INRA UR875 Unité de Mathématiques et Informatique Appliquées, Toulouse. **10**- DVM - Montel Veterinary Clinic, Tournefeuille, France. **11**- Department of Languages, Literatures, and Cultures, The University of South Carolina, USA. **12**- MD, ENT, CRA, CHU Purpan, Toulouse. **13**- Veterinary Doctor, Puylaurens, France. **14**- LARA-SEPPIA, Toulouse. **15**- CIC 1436, Université Toulouse-Purpan, INSERM.

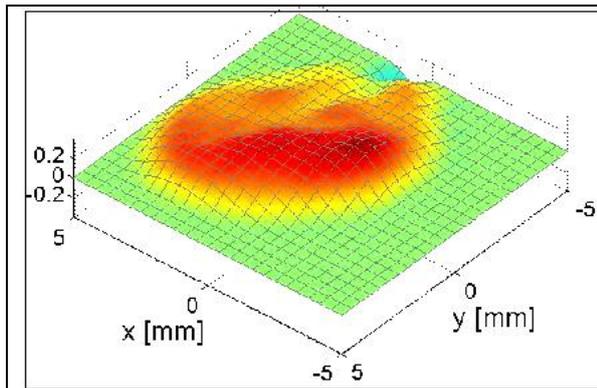 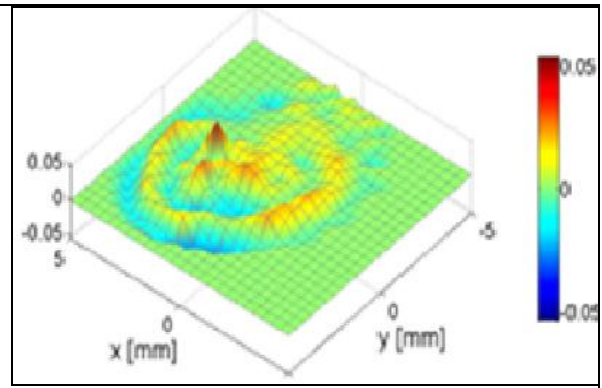

| Fig. 1a | Fig. 1b |
|---|---|
| frequency 1062 Hz | frequency 5175 Hz |
| jbio_201200186_sm_video04.avi | jbio_201200186_sm_video06.avi |

Excerpted instant photographs of OCT videos [from Burkhardt A. et al., Investigation of the human tympanic membrane oscillation ex vivo by Doppler OCT, J Biophotonics,7:434-441 (2012).].

These videos show that the tympanic surface responds to the high frequencies in a much more fragmented way (*Chladni phenomenon*) than to the frequencies under 1500 Hz.

Click here while simultaneously using the Ctrl key : This figure shows the effect of Bessel functions resulting from vibrations imposed on a roughly circular thin membrane. This phenomenon is particularly at work in the vibratory phenomena discovered experimentally by *Chladni*.

The Hindawi following figures[7], clearly illustrate the decay of the amplitude transmitted by the eardrum from 1 kHz up to highest frequencies, via the ossicles and up to the stapes.

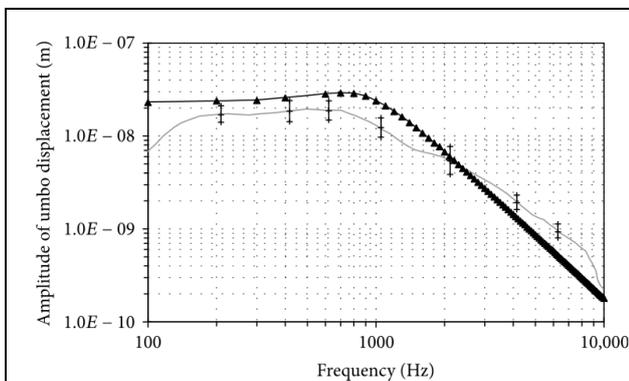 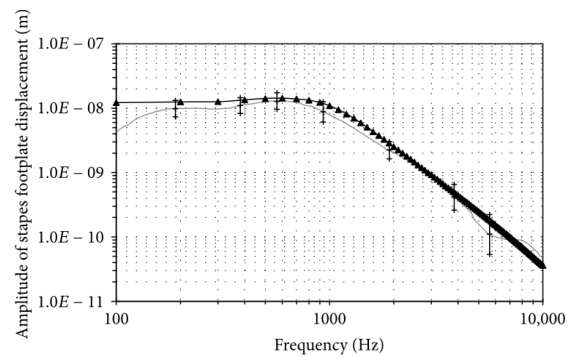

fig. 1c     fig. 1d

Amplitude of displacement versus frequency of umbo (c) and stapes footplate (d), for a range from 100 to 10,000 Hz at 80 dBSPL (sound pressure). The amplitude of the displacements, at the entrance of the vestibular canal, decreases with the frequency. It is obvious that this amplitude participate to the control of the Traveling Wave.

This work by Hindawi shows that the original TW is insufficient for the processing of high and medium frequencies.

According to Nakajima et al[8] "The significant sound pressures measured at certain frequencies (e.g. 6 kHz) after ossicular interruption suggest that sound is transmitted to both [cochlear] scalae through a path independent of the ossicular chain". Furthermore, cetaceans and other sea mammals develop only vestigial parts of the external and middle ear yet have extreme hearing capabilities[9] (Cf. Si 08). It seems, then for Röösli et al[10], that "a mechanism independent of the chain of ossicles is necessary for optimal transmission of high frequency sounds".

Our search for a mechanism that would overcome this low-pass effect has led us to hypothesize a sound signal path originating from the piezoelectricity of the tympanum and bone collagens.

Furthermore, contrarily to a common supposition, the electric response of eardrum is not uniquely due to the cochlea (Cf. Si 05).

Examining the Deiters cells and the outer hair cells (OHC) at the other end of the tympano-cochlear pathway[11], we note that stereocilia crowning the OHC have a mechano-electrical activity that transduces the acoustic TW into electrical signals[12] (Cf. Si 10). This transduction of acoustic waves, and the transmembrane transmission of resultant electrical signals, should involve the RC time constant of the plasma membrane and ion channels, resulting logically, in low-pass filtering (<1 kHz)[13]. However, some mammals hear at frequencies above 200 kHz[14]. At these high frequencies, the relaxation time constant would be $\tau \leq 1/(2\pi \times 200 \text{ kHz}) \leq 1$ μs, i.e. an order of magnitude faster than that found for ion channels[3]. Further, there is very intense debate[15] about currently accepted concepts. Several models have been advanced, but none has yet been experimentally verified, and the invoked mechanisms could not allow the transmission of frequencies higher than 12 kHz[16].

Thus, we have seen that, at the beginning of the tympano-cochlear pathway, as at its end, the high frequencies should be weakened.

## The tympanum, a piezo-electric bio-electret

The triple-helical collagen molecules are organized hierarchically into fibrils, fibers, and bundles. Sounds produce piezoelectric potentials due to the collagen fibers[17] of the tympanum. Fibers, like fibrils, are piezoelectric bioelectrets[18], having a negative pole (C-terminal) and a positive pole (N-terminal).

The voltage of the piezoelectric potentials due to isolated fibers[19] is much higher (up to ten millivolts) than the potentials we measure on bundles of millimetric fibers.

Furthermore ear canal obstruction, physical separation between eardrum and the cochlea or general anesthesia (ketamine) confirm that electrical potentials, isochronous to acoustic stimuli, do exist at the local level of collagen structures and are measurable independently of the activity of OHCs. So there is not any ambiguity at all between these two electrical activities.

Could the recorded potentials be the simple result of artifacts dependent on ambient electromagnetic phenomena such as microphonic potentials generated by coaxial conductors, or the speaker? (Cf. Si. 04).

If this were the case, the observed potentials should persist even in the absence of collagen fiber structures. In fact, during measurements concerning the patellar tendon we found that its replacement with a metal prosthesis (non collagenic) abolished the electrical response of the considered knee, while the contralateral knee, which had retained its tendon and was not equipped of prosthesis, responded electrically as the knees of all other subjects of the group.

We verified that, in the absence of an acoustical signal, the measured voltage was zero, whereas for almost all the observed series, non-zero microvoltages could be measured.

---

[3] Yet "When the OHC electrical frequency characteristics are too high or too low, the OHCs do not exert force with the correct phase to the OC mechanics so that they cannot amplify. We conclude that the components of OHC forward and reverse transduction are crucial for setting the phase relations needed for amplification" (Nam J-H, Fettiplace R (2012) Optimal Electrical Properties of Outer Hair Cells Ensure Cochlear Amplification. PLoS ONE 7(11): e50572. https://doi.org/10.1371/journal.pone.0050572).

# Material and Methods

## Aim of the study

This study was designed to test our hypothesis that the collagen fibers of the tympanum are piezoelectric. Our protocol involved stimulating the tympanum at various frequencies and using various pressures. We measured the potentials resulting from this stimulation to determine if their electric properties were dependent upon the amplitude (dB SPL) and the acoustic frequency (Hz).

## Types of Collagen

Collagens II and I are piezoelectric histological components of eardrum. Their properties are very similar and we made measurements not only on collagen II of eardrum, but also on collagen I of the patellar tendon with the purpose to know at best their piezoelectric properties (Cf. Si 01C; Si 01D). It is possible to detect an electric potential isochronous to the acoustic vibration between an indeterminate point of the tympanum and the mastoid bone[20]. It does not follow necessarily, however, that the potential measured in this type of experiment is produced by the Outer Hair Cells (OHCs). Our methodology[21] allows us to demonstrate, in vivo, and under normal physiological conditions the piezoelectricity both of collagen I in tendons and of collagen II in eardrums.

## Measurements tools for eardrum, tendons and bones fibers

For every measurements (eardrums, tendons and bone collagen) we use a lock-in amplifier to drive a loudspeaker. In this manner, we broadcast a sinusoidal sound at about one meter from the target (external auditory conduit, etc.) (Cf. Si 01E, Si 01F, Si 01G and Si 01H). We position a probe consisting of two electrodes at the center and the periphery of the tympanum, or at the ends of any targeted fibers. This probe captures the piezoelectric response of the fibers when they vibrate in response to the sound sent to the target. The lock-in amplifier makes it possible to select only those electrical responses isochronous to the acoustic stimulation. We measure electrical responses to stimulations at different acoustic frequency levels.

A Lock In Amplifier, via an electric wave of $V_{hp}$ tension (output) creates an acoustic wave at various frequencies diffused by a loudspeaker.

## Approval for the study

The official title of this research is " Non-invasive Study in vivo of the generation of microvoltages by the tendons and the tympanum when they are subjected to moderate sound stimulations".
This work was promoted by CNRS-INSB (ID RCB N°2012-A01375-38; protocole12 008), Paris (France). This study was carried out in accordance with the recommendations of Institut National des Sciences Biologiques - CNRS PARIS. This protocol was approved by the French Institutional Review Board (IRB/IEC) ad hoc : CPP SOOM 4 N° ID RCB 2012-A01375-38. It was conducted in accordance with "*Good Clinical Practices*" and with French legislation on clinical trials (Loi de santé publique n° 2004-806 du 9 août 2004, Titre V, Chapitre II (recherches biomédicales).

Approval of the study was obtained from the following:
Committee of Experts and National Institute of Physics of the CNRS (Research Laboratory: UPR 8011).
Centre d'Elaboration de Matériaux et d'Etudes Structurales (CEMES / CNRS–Toulouse).
Comité de Protection des personnes (CPP SOOM4, Jan-Apr 2013; N° CPP13006a),
    Agence Nationale de Sécurité des Médicaments (ANSM, B130246-81; March 2013),
    Agence Française de Sécurité Sanitaire des produits de Santé (ASSFAPS, June, 2013).

## The Subjects

The subjects (N=35) were healthy and their hearing was "normal", as determined by an ENT.
We did, however, include one subject (D31) who had a cicatricle eardrum and another who needed to use a hearing aid (D33). In the latter case, the subject removed the prosthesis during testing.
Other measurements were made on knees and other tendons. After being informed and signing the Inform Consent Form, all subjects gave written informed consent in accordance with the Declaration of Helsinki. Then they were tested individually in a room designated for that purpose. The piezoelectricity of collagen fibers and the impedance of the epidermis and of the dermis vary with temperature. For this reason, the testing room was heated to 22° C in colder weather.

## The measurements team

The measurements team consisted of two physicists and three physicians. Two of the physicians were ENTs.

We checked the health and medical history of each subject through a personal interview and a clinical examination (audiogram, otoacoustic emissions) performed by an ENT physician.

Then, one of the ENTs and another physician explained the protocol and installed the subject. Earlier, the two physicists had verified the correct operation of the lock-in amplifier and the connection of the probe conductors to the input of the lock-in.

The physicists set up the left and right loudspeakers (1 m from the subject's ear, and at the same height as his or her ears). They connected the lock-in output to the input of the loudspeaker to be used.

One of the two physicists was responsible for regulating the device for each of the frequencies to be studied. This included setting the frequency, sensitivity, and the amplitude of the stimulation addressed to the loudspeaker.

The other physicist supervised all these actions and noted the results, i.e., value of the voltage detected by the probe in response to each stimulation in turn.

After each testing session, the team met with the subject to obtain reactions and to determine the degree of "discomfort," if any, felt by the subject for either ear. The team also met to compare observations they had made during the testing.

## Lock-In Amplifier

To demonstrate, in vivo, and under normal physiological conditions, the piezoelectricity of tympanum fibers, we used either a digital lock-in amplifier ("Stanford SR830") or a dual-phase analog lock-in amplifier ("EG and G, M 5210") to drive a loudspeaker.

A lock-in amplifier can extract a signal with a known carrier wave from an extremely noisy environment. Signals up to $10^6$ times smaller than noise components can still be reliably detected.

## Loudspeakers

In our protocol, the lock in amplifier, via an electric wave of $V_{hp}$ tension (output), created an acoustic wave at selected frequencies broadcast by a loudspeaker. The loudspeaker was either a Harman/Kardon: DP/N 0865DV[4] or a Yamaha HS 50; For frequencies higher than 20 kHz, we used a Conrad TE300 tweeter.

## Tympanic Probe

Furthermore, the conformation of the eardrum is quite variable depending on the individual, and the experimenters must adapt to it. For example, according to some of them, measures were taken under better conditions when one of the two electrodes was resting near the handle of the hammer but not on it, while the other electrode was in place near the annulus tympanicus, but staying on the periphery of the tympanic membrane without encroaching on the annulus tympanicus (i.e. local recordings).

The advantage of that differential recording technique[22] was that it allowed us to determine the source of the potential[23].

Thus we tried two configurations (Fig. 2):
> **(a)** two electrodes equipped with a manipulable rigid elbow and flexible electrodes, easy to handle, easily affixed to appropriate parts of the eardrum, but not allowing standardization of the inter-electrodes distance... Because of the obliquity of the eardrum, other experimenters might wish to use a probe with electrodes that are easy to mold extemporaneously.
**(b) two** electrodes attached to a plastic bracket, ensuring a constant distance, but more difficult to handle.

---
[4] no longer available

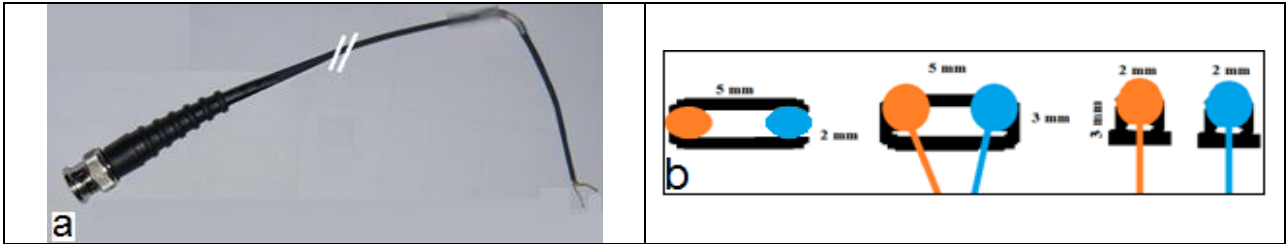

Figure 2
Probes used for measurement on the eardrums

The shielding braid is put in contact with the peripheral structure of the tympanum, away from the handle of the malleus (Fig. 3). The central copper wire is put in contact with the central structure of the tympanum: either umbo (direction "A") or handle of the malleus ("G").

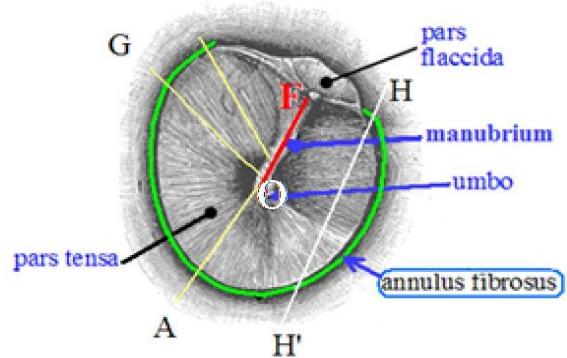

Figure 3

It is possible to detect an electric potential isochronous to the acoustic vibration between an indeterminate point of the eardrum and the mastoid bone[5]. However, it does not necessarily follow that this potential be a microphonic produced by Outer Hair Cells OHCs. On the contrary, our methodology allows us to demonstrate, in vivo, and under normal physiological conditions, that it is the result of the piezo-electricity of the collagen II of the eardrums.

The closer together the electrodes are placed, the smaller the area recorded will be (i.e. local recordings). This is one advantage of the differential recording technique[24]. This "differential technique" is necessary for determining the source of a potential[25].
In order to evaluate the electrical behavior of points belonging to the central structure (manubrium) during acoustic stimulations, electrodes can be placed at two points on the same side of the manubrium (symbolized by an F letter). This system can detect if there is electrical isochronism between these points. On the contrary, electrodes might be placed facing each other on either side of the manubrium (E). This latter system would allow us to capture the activity of a bundle of circular fibers (not completed).
The following letters were added by us: A or G: "radii" types of fibers of collagen ; HH': arbitrary cord joining two peripheral points ; F: manubrium of the malleus ; GAH'H: annulus tympanicus. In order to verify the piezoelectric activity of the tympanum, an electrode is placed on the manubrium, another on the periphery according to the straight lines either A or G.
With humans the synchronous electrical responses between two points on the same side (inner or outer) of the annulus tympanicus (HH') are generally impossible to measure.

---

[5] Gavilan C., Sanjuàn J., Microphonic Potential picked up from the human tympanic membrane. *Ann Otol Rhinol Laryngol.* **73,** 102-109 (1964).

For measurements on the tympanum (Cf. Si 01G), we used a probe consisting of a coaxial copper cable (1 mm outside diameter). The advantage of coaxial design is that electric and magnetic fields are restricted to the dielectric, with little leakage outside the shield. Electric and magnetic fields outside the cable are entirely prevented from interfering with signals inside the cable. This property makes coaxial cable a good choice for carrying weak signals without interference.

In order not to damage the epidermis of the eardrum, the ends of the two strands were either coated by a small amount of an eutectic solder or shaped to form a smooth closed loop..

The probe consists of two electrodes based on a coaxial cable:
- Central copper wire wrapped with a sheet of plastic isolator (with the end of the copper wire used for contacting the eardrum)
- Surrounding conductive braid wrapped with a sheet of plastic isolator (with its end used for contacting the eardrum)

For measurements, one electrode was apposed to the center of the eardrum (a point on the posterior side of the manubrium) and the other at its periphery (a point on the posterior limbus). The probe captures the piezoelectric response of the radial tympanic fibers when they vibrate in response to the sound sent to the tympanum. We measured electrical responses to stimulations, at different acoustic and frequency levels. We selected 16 frequencies (125 Hz to 30 kHz) and five sound pressure levels (55 to 80 dB) for this study. All measures were taken on both sides.

An electric wave of $V_{HP}$ tension is sent by the lock-in to the loudspeaker, creating an acoustic wave at the chosen frequency. Then, the electric tympanic probe receives the piezoelectric response. The voltage of the response is displayed on the digital screen of the lock-in.

## Otoscopy for optimal visual access

We used either a Zeiss OPMI 99 microscope (19X magnification) or a KAPS Som62 MAT005 halogen microscope (16X magnification).

## Units of measurement

As units of measurement, we used the International System. For acoustic frequencies, we used Hz (cycles/second), and for amplitudes the decibel sound pressure level (dB SPL).

## Possible drawbacks with respect to measurements

Our measurements on the eardrum encompass broadcasting a sinusoidal sound at one meter from the external auditory conduit. We were very cautious, yet the measurement of sound amplitude reaching the eardrum from a loudspeaker in a free field may be problematic.

Spatial fluctuations of the sound pressure might be an important factor: a frontal or lateral shifting of a few centimeters between the sound source and the ear of the subject may have significant effects on perceived and measured amplitudes. In despite of all our efforts to assure a strong stability during measurement process, the experimenters could not ensure that the clinician, or his hands, were, or were not, at times, an obstacle dampening the amplitude for the frequencies emitted by the loudspeaker.

We recall moreover, that, in addition to the implied physical parameters, in live measurements, there are psycho-physiological interactions as well. When enabled, the tensor tympani muscle pulls the malleus medially, tensing the eardrum and damping its vibrations (Cf. Si 03A ). This is also the case of the stapes muscle and of the smooth muscles that are inserted into the annulus tympanicus. Cerebral commands of OHC activity likewise follow this pattern.

In addition, the position of the electrodes should make it possible to measure the variations of potential in response to the sounds between the two ends of one and the same fiber; This is obviously very difficult to achieve, especially in vivo. It is obvious that our measures are underestimated (Cf. Si 02). Similarly, it is very difficult to standardize the pressure of the electrodes on the skin (Cf. Si 03B).

## Anesthesia during measurement on the eardrum

Measurements on the eardrum of humans were made after applying a light anesthesia to the eardrum (*cream EMLA 5%*: Lidocaine 25 ‰, Prilocaine 25 ‰) this cream was removed afterwards, using vacuum aspiration or wiping with gauze.
We did not use a conductive paste because it would be inappropriate to mix it with the anesthetic paste. Also, the quantity and surface of contact would be very difficult to standardize.

## Evaluation of the possibility of pain

Yet, an index (Likert scale, 0 to 5) of "*felt pain*" was recorded.

## Differentiation of piezo-tympanic potential and cochlear microphonics

We have considered and evaluated possible technical artifacts: (contact surface (Cf. Si 04C), microphonic of measurement cables (Cf. Si 04D)).
We have shown that our measurements correspond to a tympanic generation and that it would be a mistake to equate them with cochlear microphonics (Cf. Si 05). We performed an experiment to check whether the sound generated a piezoelectric response at the mastoid level, including by blocking the airway.
This supports our hypothesis that the electrical response of the system cannot be reduced to cochlear microphonic activity (Cf. Si 06).

## Statistics

For explaining the potential difference V, we estimated a regression model, log-quadratic in the frequencies F and linear in the sound level L, including a fixed effect component (for each individual and each side).
The model was estimated by ordinary least squares completed by the usual tests (global tests of significance and tests of significance of the parameters) with the R software[26] (*lm* command). Further analysis of individual characteristics and experimental conditions is presented in Si 14.

## Results

We tested 35 individuals, aged 17 to 87, for 16 frequencies (125 Hz to 30 kHz) and for five sound pressure levels (55 to 80 dB). All measures were taken on both sides. Due to missing data, the sample size was 453 (Cf. above Material and Methods). This sample constitutes an unbalanced panel data sample estimated by ordinary least squares with the introduction of a so-called "fixed effect" dependent on the individual and the side (left or right). For the raw data, Cf. 10.6084/m9.figshare.5671807.

The empirical evidence of a significant effect of frequency and sound pressure on the difference in potential was verified by statistical analysis using the "R" software. The explained variable is the microvoltage $V_{ijks}$ measured for individual $i$, frequency $j$, sound pressure level $k$ and side $s$ (left or right).

The relation is assumed to be log-quadratic in the frequencies $F$ and linear in the sound level $L$. The least squares estimation is:

$$\log(1+V_{ijks}) = \alpha_{is} + 0.036616\, L_{iks} - 2.142166 \log(F_{ijs}) + 0.144777\, (\log(F_{ijs}))^2 + U_{ijks}$$
$$(s.d. = 0.006437)\ (s.d. = 0.481262)\ (s.d. = 0.029839)$$

where $U_{ijks}$ is a zero mean random residual (Cf. fig 4).

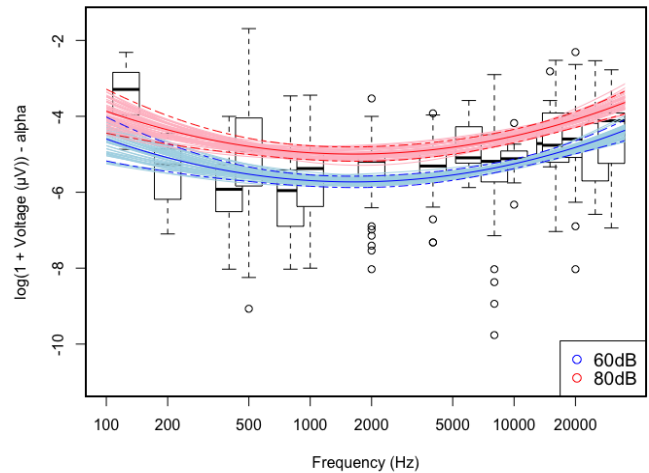

Figure 4
Relationship between piezotympanic voltage (fixed-effect corrected data) and acoustic frequency: estimated model for the two sound levels, 60 dB and 80dB (bold parabolic line) and 50 random models based on the estimated covariance matrix of coefficients (curves in between the two dotted lines and encompassing the bold line).

All the coefficients are highly significant ($p < 0.001$). The standard error of $U_{ijks}$ is 1.109 and the adjustment is measured by the $R^2 = 0.90$ (adjusted 0.89 with 376 degrees of freedom). The F-statistic is 53.57 with a p value smaller than $22.10^{-16}$. The estimated U-shaped curve is represented in fig.4.

The values of the $\alpha_{is}$ fixed effects are distributed between 5.24 and 11.69 with an average value of 8.02 dB. We established a significant ($p < 0.001$) positive correlation between $\alpha_{is}$ and two individual characteristics: "Age" and "Body Mass Index"; There is also a strong correlation between left and right sides.

Other factors could explain these fixed effects, such as pressure on the measurements probe by the ENT practitioner, position of the two electrodes of the probe relative to a unique (or not) collagen bundle (Cf. Si 02), muscular activity of the tensor tympani muscle (Cf. Si 03A) and pressure on the epidermis by the practitioner (Cf. Si 03B).

Taken together, these measurements show that the tympanum responds to acoustic stimulations by isochronous potentials, which we attribute to the tympanum's collagenous fibers. This result corresponds to the outcomes of our measurements on other collagenous fibers such as tendons: knees, Achilles, arm muscles, etc. (Cf. Si 01; Si 14B).

These isochronous tympanic potentials (dubbed "pT") are dependent upon frequency. The voltage decreases from infrasounds to middle frequencies (minimum at 1632 Hz), and then increases along with frequencies for every subset of measures.

We hypothesize that these tympanic voltages created by the piezoelectricity of the collagen fibers of the eardrum are transmitted very rapidly to the OHCs of the cochlea by an electrical path. We will now describe this hypothesized "covert path," yet up to now neglected.

## The covert path, from tympanum to trickystor
We will show that electrical responses are transmissible via a series of electrical synapses from the tympanum to the apex of the DOHC complex (Deiters Outer Hair Cell complex), where we identify a structure similar to that of a Field Effect Transistor (FET). We dub this structure Trickystor (TkS) due to its complexity.

Gap Junctions (GJs)[27] are cytoplasmic conduits possessing large pore size (10–15 Å). They allow communication between the intracellular milieus of two contiguous cells and the passage of small metabolites and signaling molecules (mass < $2.10^{-27}$ kg) between cells. GJs are composed of two hemi-channels, each made up of six connexins (Cx).

GJs are very fast conductors able to constitute an electric network. The GJs are especially useful in facilitating electrical transmissions[28]. One of the neuronal functions of GJs is thought to be synchronization between brain cells[29]. The transmission of a signal by means of these electrical synapses is not dependent upon a certain threshold. Further, such transmission is extremely rapid and takes place without diffusion (leakage) into extracellular spaces. It is noteworthy that an electric sinusoidal wave (e.g. pT voltages) can travel along an electrolytic pathway, going through the GJs with minute displacements of ions between adjacent cells. This takes place without global displacements from the first cell to the last and back. Alternating current (AC) voltages cause no net movement into the conductive medium, regardless of its length, since the charge carriers oscillate back and forth in response to an alternating electric field. Nanotube structures might be implicated in the electrical communication by GJs between ear and OHCs[30].

The cell bodies of the osteocytes act as mechanosensors of the petrosal bone. They merge to form a syncytium (based on the Cx43) capable of conveying electrical signals. Electrical transmission between osseous cells always travels in the same direction: from the interior of the bone toward its surface (periost)[31]. Electrical signals arising from the piezo-electricity of the tympanum can, thus, be transmitted to the external wall of the cochlea (the spiral ligament, which is a periosteum structure) via the syncytium of the subperiosteal cells.

Through the root cells[32], Cx43 interacts with the Cx26 of the cochlea, thus enabling the transmission of the piezotympanic signal to the cochlear Deiters Cells (DCs). A critical relationship may be established between the mutation of Cx43 proteins and non-syndromic high-frequencies deafness[33].

Yet there are two independent syncytia in the cochlea:
- The connective tissue GJ system of the lateral wall (fibrocytes): The deterioration of this system results in a progressive hypoacousis, especially with respect to high frequency sounds[34]. The Fibroblast Growth Factors (FGFs)[35], which regulate the electrical excitability of cells, appear to have a role in the maintenance of normal auditory function[36].
- The epithelial cell GJ system is composed of several types of supporting cells linked to the root cells within the spiral ligament. It is obviously the most important system for transmitting the pT signal. Root cells are present primarily in the basal part of the cochlea, the part devoted to hearing high frequencies. The epithelial cell GJ system is capable of transmitting variations of potential[37] from the root cells to the DCs, and, when it does not function, the OHCs, even if they are normal, lose their effectiveness[38].

Thus, active cochlear amplification is dependent on the GJs of supporting cells[39] - [40]: Genetic[41] or experimental alterations of either the structure of root cells or of several connexins [Cx26 (GJB2), Cx30 (GJB6), Cx31, Cx32, Cx43] have been shown to result in non-syndromic deafness[42 - 43 - 44]. Purely metabolic explanation of their usefulness seems insufficient to explain why this is so. The number of Cx26 and Cx30 declines from the cochlear apex to its base, but this finding does not weaken the hypothesis that these GJs play an essential role for all frequencies: Either mutations or a blocking[45] of Cx26 produces a reduced, or absent, distortion product of otoacoustic emission (DPOAE) and hearing loss at all frequencies.

That apparent discordance should be clarified by their role in the functioning of the Trickystor (below).

## The Trickystor

Travelling waves cause the stereocilia crowning the OHCs to move, producing a mechano-electrical transduction. The resulting electrical signals, if their frequency is under a few kHz, pass through the cuticular bilayer.

Our experiments have demonstrated the piezoelectricity of the eardrum and its adjacent bony structures. We have also presented a probable route of transmission of the tympanic electrical signals (pT) via electrical synapses (GJs) up to the DOHC complex, It has been proposed that cochlear support cells interact with hair cells in a manner similar to interneurons or astrocyte interactions with neurons in the central nervous system[46] - [6]. Each OHC is surrounded by five DCs (fig. 5). Its base is supported by the cupular body of a DC ($DC_5$) and its ciliated apex is bordered by four phalangeal apexes from four other DCis $(i=1..4)$: on the right ($DC_1$), inside ($DC_2$), on the left ($DC_3$), outside ($DC_4$); each being different from the $DC_5$.

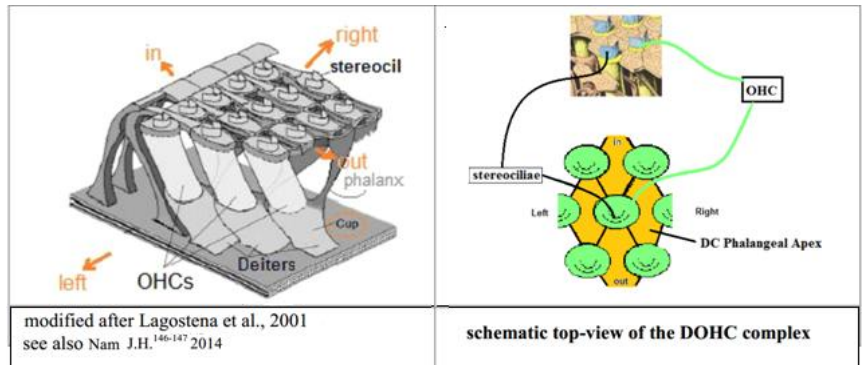

Figure 5
3D and top view of the DOHC-complex organization

Every component of an Organic Field Effect Transistor (OFET) is present in the apex of the DOHC complex (Fig. 6).

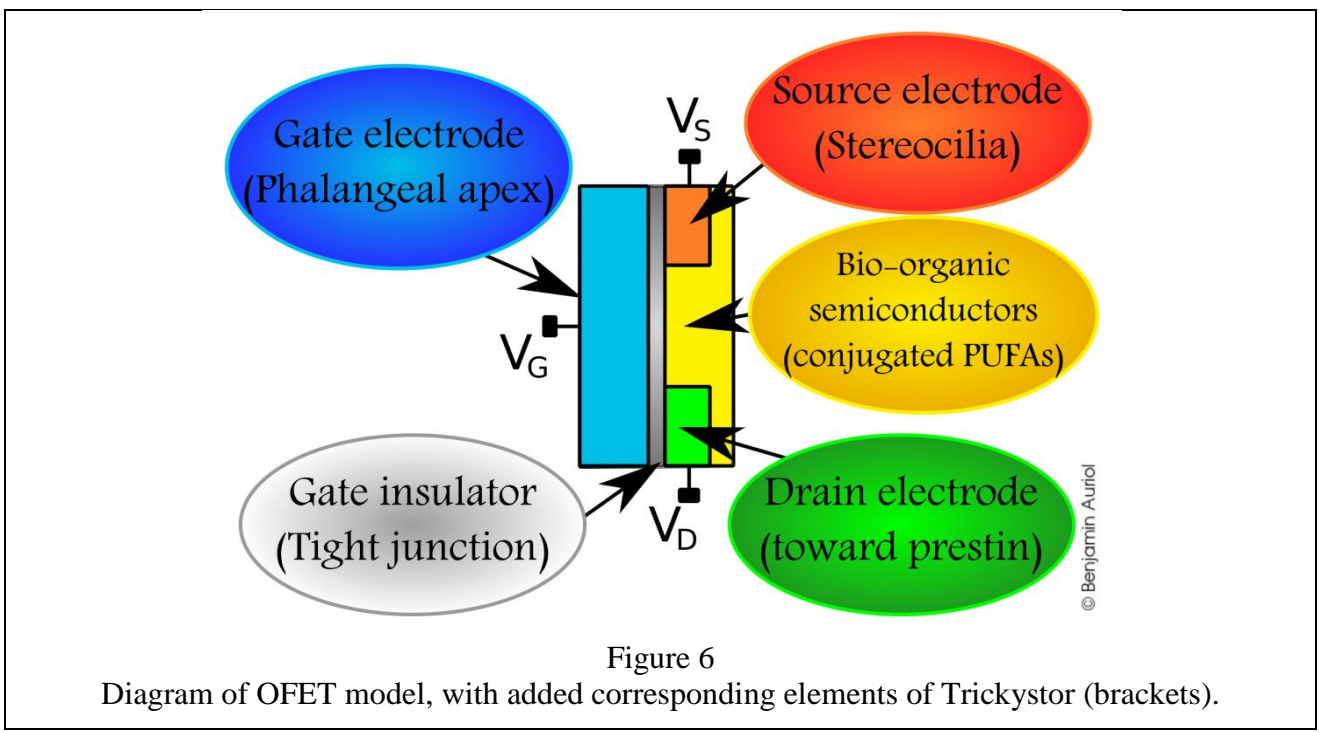

Figure 6
Diagram of OFET model, with added corresponding elements of Trickystor (brackets).

We will consider its elements in turn: the source (Cf. Si 10 A), the semiconductor channel, the gate with its electrical insulation from the overall structure, and, finally, the drain.
The source is identical to what has been shown in the classical literature: Electrical potentials result from stereociliar movement (flexo-electricity) and from the biasing potential.

---

[6] Connors and Long, Electrical synapses in the mammalian brain, "Annu Rev Neurosci" 2004;27:393-418.

A common feature of Organic FET materials is the inclusion of a conjugated π-electron system. This system serves as the active semiconducting layer and facilitates the delocalization of orbital wave functions. In its original meaning, a conjugated system is a molecular entity whose structure may be represented as a system of alternating single and multiple bonds. In such systems, conjugation is the interaction of one π-orbital with another across an intervening σ-bond. Homoconjugation is defined as "an orbital overlap of two π-systems separated by a non-conjugating group, such as CH2" [47-48]. Conjugation and homoconjugation alike give semiconductor properties to a biological molecule. When biological poly-unsaturated fatty acids (PUFAs) are either conjugated (R-CH=CH-CH=CH-R') or homoconjugated (R-CH=CH-CH2-CH=CH-R), they have the properties of a semiconductor[49].

The cuticular bilayer (fig.07) encompasses semi-conductors such as phospholipids, conjugated linoleic acid, conjugated linolenic acids, or docosahexaenoic acid ethyl ester-d5 as well as other conjugated and homoconjugated PUFAs. There is an inverse association between hearing loss and higher intakes of long-chain n−3 PUFAs and regular weekly consumption of fish[50]. Modifications of the PUFAs by genetic mutations, for example, "peroxisome biogenesis disorders " or "X-linked adrenoleukodystrophy", have deleterious consequences on the auditory processing of high frequencies (Cf. Si 10D).

Fig. 07 presents two conductive pathways capable of passing charge carriers through the cuticular membrane: The ionic channels are relevant for frequencies below 3 kHz[51] but ineffective for higher frequencies; the semiconductor channels are most likely to intervene for high frequencies, up to more than 200 kHz.

In fact, these two connecting structures might be topographically associated since conjugated PUFAs are close neighbors of ionic channels. PUFAs are incorporated into the lipid bilayer near to, but not included within, the pore domain. They affect voltage transition electrostatically. If the charge is switched, these electrostatic interactions accomplish opposite effects (Cf. Si 10B and 10C).

The leaflets of the bilayer are primarily composed of phospholipids, and anti-phospholipids can negatively affect hearing (Cf. Si 10D). Chlorpromazine, which intercalates into the inner leaflet of the phospholipid bilayers, alters OHC electro-motility without a known direct action on prestin; according to Ricci et al[52] "The conductance of the Mechano Electrical Transducer channels changes along the tonotopical position within the cochlea, suggesting differential requirements at different frequencies". Thus, the intervention of the phospholipids concerns specifically neither the action of the stereocilia, nor that of the prestin, but rather that of the cuticular bilayer.

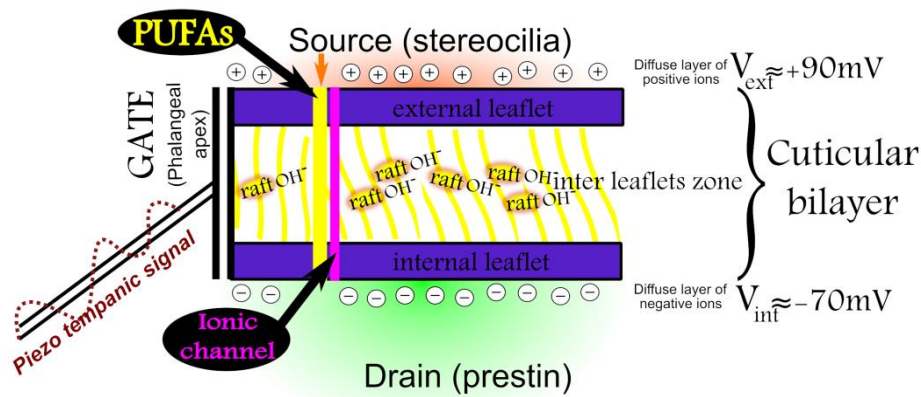

Figure 7
Simplified diagram of the cuticular bilayer

The doping of a semiconductor material consists of introducing into its matrix very small quantities of other material comprising different charge carriers. Small numbers of them can change the ability of a semiconductor to conduct electricity. In the case of conjugated PUFAs, the doping material may consist of a system of anionic cholesterol rafts nested between the two leaflets of the bilayer[53]. These cholesterol rafts modify the passage of electrons through the bilayer in certain directions and also the voltage dependence of the prestin (Cf. Si 10C).

The lipid bilayer is associated with an electronic double layer: endocochlear (+90 mV) versus intracellular (-70 mV) potential; It has a non-linear capacitance dependent upon the voltage applied. That is an essential component of a beneficial **biasing** imposed on the TkS. The ($Cx26^{+/-}$ / $Cx30^{+/-}$) digenic mutation, which decreases that bias, results in high frequencies hearing loss[54].

The gate is represented by Deiters phalanxes with their high microtubules content. These microtubules cause negative differential resistance, improve electric connectivity between their two ends, and amplify the critical frequency of the transferred signals[55].

In physiological conditions, there is a strict insulation, chemical as well as electrical, between phalangeal apexes of 4 DCs and the cuticle of an embedded OHC (fig. 5). This being the case, no electric current will flow from one to the other of their apical membranes.
However, this border, the tight junction (TJ) and other elements of the Apical Junctional Complex cannot prevent a hydrophobic intercellular electrostatic coupling unrelated to any GJ. Thus, the stereociliar signal can be amplified by the isochronous piezotympanic (pT) signal acting on semiconductors of the cuticular membrane. TJs between OHCs and DCs are, indeed, critical for normal functioning of the organ of Corti; mutations of the TJP2 gene cause autosomal dominant non-syndromic hearing loss[56] (Cf. Si 10F; Cf. also Si 10G).

As in the classical FET schema, the electric signal from the source must reach the drain (prestin) (Cf. Si 10H) within the lateral wall of the cell and stimulate it. In our view, when passing through the cuticular bilayer, the signal is driven by the TkS. Voltage variations due to the stereociliae alternatively shorten and lengthen the prestin located in the latero-basal wall of the OHCs. In the case of the highest frequencies, we make the assumption that the signal is amplified by the pT coming from the external ear (eardrum and bone collagen). The problem of the presence of prestin in the vestibular system dedicated to some infrasounds is examined in Si 10I.

## Overview of the "covert path"

The TkS is probably a decisive element for refreshing the acoustic signal (especially for high frequencies), for enhancing cochlear amplification, and for frequency analysis (tuning) of the audio signal.

We propose (Fig. 8) the design of an equivalent circuit which presents the classical pathway (in black) supplemented by the "covert path" (in red).

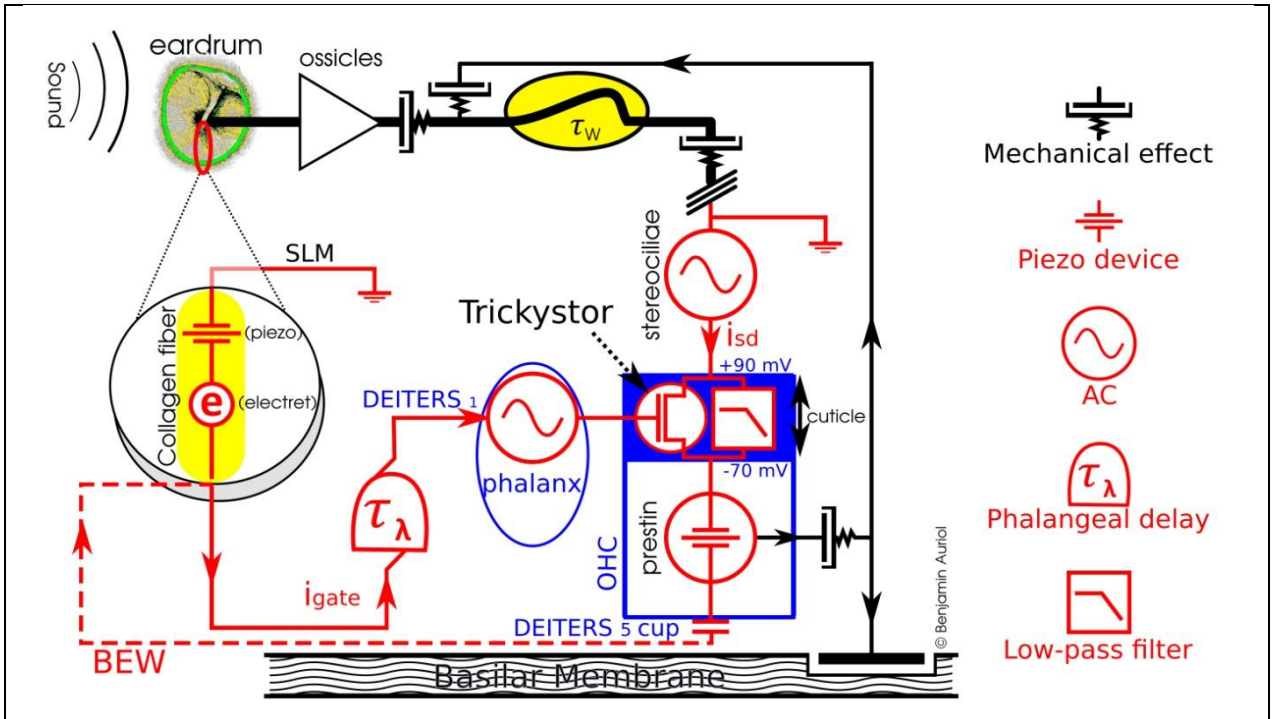

Figure 8
Equivalent circuit: classical pathway (in black); "covert path", our model (in red).
SLM (Superior Ligament of Malleus): Collagenic ligament that crosses from the head of the malleus to the roof of the tympanic cavity (tegmentum attici). BEW: hypothetic Backward Electrical Wave (Cf. Si 12 OAE vs BEW).

Every step of the covert path is indispensable for good hearing, and any deficiency (either genetic, experimental or toxic) in any of the steps of the covert path produces hearing impairment, mainly in the high frequencies.

## 'Sine Qua Non' demonstration of the covert path

In the following figure (fig.09) we number the places of the consecutive stages whose alteration has the effect of hearing loss. For each of these stages, we then listed a list of significant publications relating these hearing losses to these alterations.

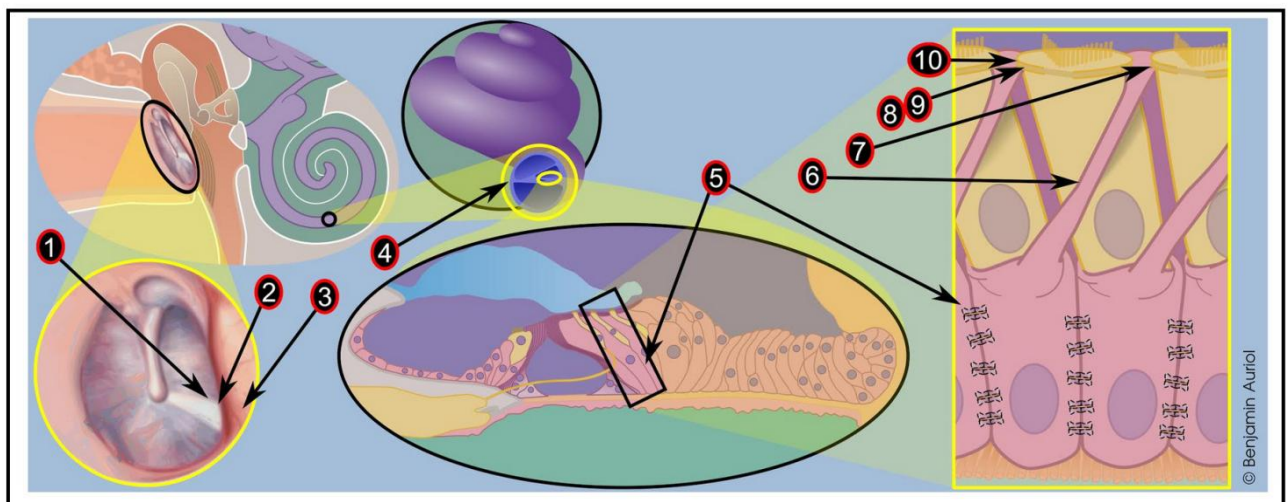

| Step 1 | Step 2 | Step 3 | Step 4 | Step 5 | Step 6 | Step 7 | Step 8 | Step 9 | Step 10 |
|---|---|---|---|---|---|---|---|---|---|
| Collagen II | Annulus T. | GJ Cx43 | Root cell | GJs Cx26,30 | Phalanx | TJs | Semi-conductors | Doping | Biasing |

Figure 9
'*Sine Qua Non*' demonstration of the covert path

For each step indexed in fig. 9, we now give a brief description, and then we cite the references establishing the negative effect of any flaw on auditory functioning:

**Step 1 Collagen II** of the eardrum and mastoid. Any flaw in the structure of type II collagen, accompanied by a defect in its piezoelectricity, causes hearing loss:
- Liberfarb RM, et al., The Stickler syndrome *Genet. Med,*.**5,** 21-27 (2003)
- *Omim120140*

**Step 2** The **annulus fibrosus tympanicus** is the thickened peripheral rim of the pars tensa of the tympanic membrane; It is connected with the bony sulcus tympanicus via radial fiber bundles, which continue directly into the tympanic bone. Osteoma of the osseous and fibrous annulus tympanicus can cause hearing impairment:
- Uno Yoshihumi, The attachment structure of the guinea pig tympanic membrane, Auris Nasus Larynx, 27,45-50(2000).
- He Z., Vibration measurements on the widely exposed gerbil eardrum, Biomedical Engineering, partial fulfilment of Master of Engineering, McGill University (2012).

**Step 3** Mutations in **GJA1 (connexin 43)** are associated with non-syndromic autosomal recessive deafness:
- Liu, X. Z., et al., Mutations in GJA1 (cx43) are associated with non-syndromic autosomal recessive deafness. Hum Molec Genet 10: 2945-2951, 2001.

**Step 4** Crossing spiral ligament via **Root-cells** may be central to pathological processes associated with various forms of hearing loss:
- Jagger DJ, Forge A. The enigmatic root cell - emerging roles contributing to fluid homeostasis within the cochlear outer sulcus. *Hear Res.* **303,** 1-11 (http:\www.sciencedirect.com\science\article\pii\S0378595512002523 2013) **.**

**Step 5 GJs** (**Cx26, Cx30)** Mutations in either Cx26 or Cx30 are the major cause of non-syndromic prelingual deafness in humans.

- Zong L, Chen J, ZhuY, Zhao HB, Progressive age-dependence and frequency difference in the effect of gap junctions on active cochlear amplification and hearing, Communications, 489, 223-227 (2017)
- Zhu, Y. et al., Active cochlear amplification is dependent on supporting cell gap junctions, *Nat Commun,***4**, 1786 (2013)**,** Omim 604418**.**
- Chang Q, Tang W, Ahmad S, Zhou B, Lin X, Gap Junction Mediated Intercellular Metabolite Transfer in the Cochlea is Compromised in Connexin30 Null Mice,. *PLoS ONE* **3 (12)**, e4088 (2008).

**Step 6 Phalanx of Deiters** (GATE). Destruction of the phalanx cytoskeleton annihilates the electric effect of the DCs on the OHCs:
- Yu N, Zhao HB, Modulation of Outer Hair Cell Electromotility by Cochlear Supporting Cells and Gap Junctions. PLoS ONE 4(11): e7923 (2009)**.**

**Step 7 Tight Junctions** between phalanx of Deiters Cell and OHC cuticle; Mutation of the *TJP2* gene causes autosomal dominant non-syndromic hearing loss (ADNSHL):
- Kim MA, et al. Genetic Analysis of Genes Related to Tight Junction Function in the Korean Population with Non-Syndromic Hearing Loss. PLoS ONE 9(4), e95646 (2014).
- Wilcox ER et al., Mutations in the Gene Encoding Tight Junction Claudin-14 Cause Autosomal Recessive Deafness DFNB29, Cell, 104: 165-172 (2001).
- Tang VW. Proteomic and bioinformatic analysis of epithelial tight junction reveals an unexpected cluster of synaptic molecules. *Biology Direct.* 2006;1:37**.**
- Itajiri S, Katsuno T. Tricellular Tight Junctions in the Inner Ear. *BioMed Research International.* 2016;2016:6137541**.**

**Step 8 Semi-conductors** (Conjugated PUFAs and phospholipids, etc.); Genetic, dietary or toxic deficiencies of conjugated PUFAs in cuticular phospholipids, induce a downward sloping audiometric pattern:
- OMIM# 253260 and 609019.
- Wolf B, Spencer R, Gleason T., Hearing loss is a common feature of symptomatic children with profound biotinidase deficiency. J Pediatr 2002, 140,2:242–246
- Tanigawa T, Adiponectin deficiency exacerbates age-related hearing impairment, Cell Death and Disease (2014) 5, e1189
- Van Veldhoven PP, Biochemistry and genetics of inherited disorders of peroxisomal fatty acid metabolism, J. of Lipid Research, Thematic Review Series: Genetics of Human Lipid Diseases, 51, 2010 , p. 2885
- Wanders RJA, Peroxisomes, lipid metabolism, and peroxisomal disorders, ASHG 2004 Meeting Toronto, Molecular Genetics and Metabolism, 83, 1–2, September–October 2004: 16–27
- Braverman NE et al., Peroxisome biogenesis disorders in the Zellweger spectrum: An overview of current diagnosis, clinical manifestations, and treatment guidelines, Mol Genet Metab 117 : 313-21 (2016)

- Zempleni J, Hassan YI, Wijeratne SS, Biotin and biotinidase deficiency. Expert Rev Endocrinol Metab **3** : 715–24 (2008)
- March J, Advanced Organic Chemistry reactions, mechanisms and structure (3rd ed.). New York: John Wiley and Sons (1985)
- Hush, N. S., An Overview of the First Half-Century of Molecular Electronics. Annals of the New York Academy of Sciences, 1006: 1–20 (2006)
- Inzelt, György "Chapter 1: Introduction". In Scholz, F. Conducting Polymers: A New Era in Electrochemistry. Monographs in Electrochemistry. Springer : 1–6. ISBN 978-3-540-75929-4. (2008)
- Bard Allen J., Inzelt György, Scholz Fritz, Electrochemical Dictionary : cuticular phospholipids, Springer Science and Business Media. (2008)
- Engelman DM. 2005. Membranes are more mosaic than fluid. Nature 438: 578–580
- Jacobson K, Mouritsen OG, Anderson RGW. 2007. Lipid rafts: At a crossroad between cell biology and physics. Nat Cell Biol 9: 7–14.
- Coskun U., Simons K. 2010. Membrane rafting: From apical sorting to phase segregation. FEBS Lett 584.

**Step 9 Doping** by cholesterol rafts. T**he modulation of Voltage-gated calcium channels (VGCCs) and Big Potassium channels (BK)** currents by cholesterol, and the associated changes in hair cell excitability may have implications for sensorineural hearing loss.

- Purcell EK, Liu L, Thomas PV, Duncan RK. Cholesterol Influences Voltage-Gated Calcium Channels and BK-Type Potassium Channels in Auditory Hair Cells. Dryer SE, ed. *PLoS ONE*; 6(10):e26289**.** doi:10.1371/journal.pone.0026289 (2011).
- https://en.wikipedia.org/wiki/Niemann%E2%80%93Pick_disease,_type_C**.**
- Oghalai JS., Pereira FA., and Brownell WE., Tuning of the Outer Hair Cell Motor by Membrane Cholesterol, J Biol Chem., 282(50): 36659–36670. (2007)[7].

**Step 10 Biasing.** The lipid bilayer is associated with an electronic double layer (endocochlear potential), which is an essential component of a beneficial biasing imposed on the TkS. The ($Cx26^{+/-}$/ $Cx30^{+/-}$) digenic mutation, which decreases that bias, results in high frequencies hearing loss:

- Mei L et al., A deafness mechanism of digenic Cx26 (GJB2) and Cx30 (GJB6) mutations: Reduction of endocochlear potential by impairment of heterogeneous gap junctional function in the cochlear lateral wall, *Neurobiol Dis.* **108**, 195-203 (2017).

# Conclusion and perspectives

The electric signal generated by tympanic collagen fibers is not conceived as the alternative origin of the mechano-sensation in the auditory system, but rather as a significative electronic contribution (Cf. Si 10E).

Our experiments have shown that the tympanum has piezoelectric properties that engender an electrical signal in response to acoustic vibrations. This signal, which is frequency-dependent, is, then, carried to the outer wall of the cochlea and from there to the DCs by means of electrical synapses (various GJs and their connexins).

The piezoelectricity of the tympanum opens up the perspective of an electrical synergistic pathway of sound transmission heretofore unknown (the covert path). This pathway from the tympanum to the cochlea is capable of contributing significantly to hearing, especially to hearing the highest frequencies, as it has a determining effect on the amplification and tuning attributed to the basal OHCs (Cf. Si 11B).

Our hypothesis of an electric pathway does not negate the established theory of sound transmission but rather expands it. For it is our idea that the mechanical and the electrical transmission of sound work together[57] - [58] to produce optimal hearing (especially for high frequencies). Thus, our findings pave the way for a better understanding of the hearing process and have important implications for both theory and practice. It is well known that age-related hearing loss primarily involves high frequency sounds, the very sounds we believe are mainly transmitted by the electrical pathway. Our work opens perspectives in the understanding of hearing and in the treatment of hearing problems.

So the discovery of this electric transmission of sound may elucidate certain as yet unexplained phenomena of auditory physiology. For example, it may lay the groundwork for a better understanding of otoacoustic emissions (OAEs). No satisfactory explanation has yet been found for the backward propagation of OAEs (elusive backward travelling wave) [59-60]. Our theory may also shed light on hyperacusis of children, which is associated with larger amplitude OAEs but with no other auditory factors[61]. Furthermore despite ossicular blockage by the mesenchyme until after birth, it has been shown (see Si 08B) that the foetus hears and memorizes the sounds of its environment several weeks before birth[8]! This makes it difficult to understand the existence - though well demonstrated - of foetal hearing from the 22$^{th}$ amenorrhea week[9-10]. So Hill concludes that "any prenatal conduction to the cochlea must be mediated through bone conduction". However, the mechanism of bone conduction is itself poorly understood, so that the explanation of foetal hearing by bone conduction would simply move the problem on. It therefore seems that another conceivable mechanism would be the "covert path".

---

[7] doi: 10.1074/jbc.M705078200; Cf. also Si 2.

In fact, it is likely that knowledge of this new pathway can shed light on how sea mammals and bats use very low or very high frequencies for echolocation. In the case of cetaceans, hearing is dependent upon a "collagenous-fatty acoustic pathway" which, according to our theory, primarily uses the piezo-electricity of collagen to amplify frequencies up to more than 250 kHz (Cf. Si 08). The eardrum of terrestrial mammalians and the collagenous-fatty acoustic bodies of sea mammalians share a double function: on the one hand, a mechanical mobilization for medium frequencies, and, on the other, a piezoelectric behavior of collagen fibers for the perception of very high and very low frequencies.

This new findings might have an important impact on the future development of hearing aids. Such a practical application would be very useful for an ageing population.

## Supplementary Informations

This article is very shortened and limited to what is most recent in our theory and results. So we propose some supplementary information (Si) to expand and discuss several aspects of our experimental and theoretical work.

| |
|---|
| The authors declare no competing financial interests. This work has not been published previously and is not under consideration for publication elsewhere. |
| All authors have approved the manuscript in its present form. Supplementary Information are available on /Supplementary Information and Source Data are available on 10.6084/m9.figshare.5671807. A figure summarizing the main results of this paper is included as Fig 04. Readers are welcome to comment online. |

## Author Contributions

BMA Proposed the idea of the "covert path" (with piezotympanic source), contributed to the physiological theoretical work;
JB, J-MB, DD and CT designed the experiments;
BMA, JB, J-MB, DD, PP and BL performed measurements;
JB, J-MB, JPF, and FG analyzed data;
BA made the artistic figures;
BMA, JB, JPF, LD, FG, EJ wrote the paper;
RR did the acoustic instrumental calibration.

---

[8] Hill, M.A. Embryology *Hearing - Middle Ear Development* (2018, October 21).
[9] Gelman SR Wood S., Spellacy WN, Abrams RM, Fetal movements in response to sound stimulation- Am J Obstet Gynecol 143: 484-485, 1982.
[10] Dunn, K., Reissland, N., Reid, V.M.,The Functional Fetal Brain: A Systematic Preview of Methodological Factors in Reporting Fetal Visual and Auditory Capacity, Developmental Cognitive Neuroscience (2015), http://dx.doi.org/10.1016/j.dcn.2015.04.002.

18
# Acknowledgments

We are very thankful to the volunteers which gave their time for participating to the experimental measurements and to the complementary tests.

We thank for scientific assistance : INP of CNRS; Andrews S., PhD, LSU Medical School; Auriol J.B., PhD, BD (New Jersey, US); Azaïs C., PhD, ex-LAMI, Toulouse University; Banquet J.P., MD, PhD, CNRS UMR, Paris; Bibé B., Emeritus Research Director, INRA, genetics and wording; Calvas F., MD, CIC, Inserm, CHU-Purpan Toulouse; Chernomordik L.V., Ph.D., Senior Biomedical Researcher, NICHD, NIH, Bethesda, USA; Csillag P., ENSEEIHT - Toulouse France; Demmou H, Senior Researcher, LAAS – CNRS, Toulouse, France; Faurie-Grepon A., MD, CIC, Inserm, CHU-Purpan Toulouse, France; Gaye-Palettes J.F., MD, EHPAD Isatis, Quint-Fonsegrives; Harnagea C., PhD, INRS, Montreal, Canada; Lagrange D., Electronic engineer, LAAS, Toulouse; Legros C., Emeritus Pr –Jean Jaurès University, Toulouse; Marlin S., INSERM, Centre de Réf. des Surdités Génétiques, CHU Necker, Paris; Marmel F., PhD, INCyL, Salamanca, Spain; Merchan M.A., PhD, INCyL, Salamanca, Spain; Millot M., PhD, LNCMI, UPS Toulouse; Mitov M., PhD, Director of Research (CNRS), leader of the liquid-crystal group (CEMES, Toulouse) ; Morell M., PhD, INM, Inserm Unit 1051; Norman B., Distinguished Pr Emeritus, University of South Carolina, Columbia, USA; Pr Petit C., for her outstanding teaching and contributions to our field; Picaud F., PhD, MCF., University of Franche-Comté, UFR des Sciences et Techniques ;  Portalès P., Electronic engineer,  Airbus Industry Toulouse, France; Szarama K.B., PhD, NIH, NIDCD, Bethesda, USA; Weeger N., ex-Pdt of Wikimedia-France.; Guijo-Pérez B. English Translator.

We thank for practical assistance:
The British Library - London; Le Neillon M., Connectic System - Toulouse; Leroy L., Veterinary Nurse – Tournefeuille, France; Auriol family for their encouragements, implication and advices (Nanou, auteure; Emmanuelle, PhD, Pr TSE).

# Supplementary Information

## Si 01 About Material and Methods

### Si 01A  Aim of the study
This study was designed to test the hypothesis that the collagen fibers of the tympanum are piezoelectric when subjected to acoustic vibrations. Our protocol involved stimulating the tympanum at various frequencies and using various acoustical pressures. We measured the potentials resulting from this stimulation to determine if their electric properties were dependent upon the amplitude (dB SPL) and the acoustic frequency (Hz).

### Si 01B Types of Collagen
Collagens II and I are piezoelectric histological components of eardrum. Their properties are very similar and we made measurements not only on collagen II of eardrum, but also on collagen I of the patellar tendon with the purpose to know at best their piezoelectric properties.

It is possible to detect an electric potential isochronous to the acoustic vibration between an indeterminate point of the tympanum and the mastoid bone[S61]. It does not follow necessarily, however, that the potential measured in this type of experiment is produced by the Outer Hair Cells (OHCs). Our methodology[S61] allows us to demonstrate, in vivo, and under normal physiological conditions the piezoelectricity both of collagen I in tendons and of collagen II in eardrums.

### Si 01C Piezoelectric proprieties of the fibrous collagen

The piezoelectric tensor of collagen I has a symmetry close to the hexagonal crystal structure[S113]. Its fibers are ferroelectric and piezoelectric[S114, S115]. Stimulating these fibers by means of high frequency sounds directly affects osteogenic cells[S116-S117]. In a similar way, the production of the collagenous fibres of the tympanic membrane (TM) is increased and modeled by acoustic stimulations: In vitro, applied mechanical forces are able to promote TM-fibroblastic differentiation, increasing the production of collagen II, that is a peculiarity of TM structure[S118].

A detailed analysis of the Piezoresponse Force Microscopy signal of collagen I "…revealed clear shear piezoelectric activity associated with piezoelectric deformation along the fibril axis." (Harnagea, 2010). Piezoelectric activity of collagen fibrils can be detected in vitro in a large range of frequencies going from a few Hz[S119] up to more than 200 kHz[S120] up to 1 MHz[S121]. This result corresponds to the outcomes of several studies with respect to collagen in vivo[S122 - S123] (Harnagea, 2010). The inverse piezoelectric effect is also demonstrable[S124]. (Harnagea, 2010). The properties of collagen I are thought to be similar to properties of collagen II[S125 - S126 -S127] (Minary-Jolandan, 2009) . To confirm this, we have measured synchronous electrical potentials not only on eardrum collagen radial fibres but also on the patellar ligaments of individuals at various ages. The amplitude of measured synchronous potentials increases with the frequency of the sound signal.

In order to verify the piezoelectric activity of the tympanum, an electrode is placed on the manubrium, another on the periphery according to the straight lines either A or G. The "standard" measurements we did, and studied statistically, were between an undetermined point of the posterior side of the manubrium and an undetermined point of the posterior limbus.

In order to evaluate the electrical behaviour of points belonging to the central structure (manubrium) during acoustic stimulations, electrodes can be placed at two points on the same side of the manubrium (symbolized by an F letter in **b**). Since the collagen fibres have their positive end linked to the malleus, the system should then measure an electrical potential close to zero. On the contrary, if the electrodes would be placed facing each other on either side of the manubrium (E), the system should allow us to capture the activity of a bundle of circular fibres, ie an potential isochronous to the acoustic waves. Isochronous electrical responses between two points on the same side (inner or outer) of the annulus tympanicus (HH') are generally impossible to measure. Regarding exceptions, it might be that electrodes had been positioned, either not on the same side of the annulus tympanicus, or on an inner circumferential collagen fibre. An analogous difficulty could be found along the manubrium of the malleus; yet it is easier to position the electrodes at one boundary of the manubrium.

### Si 01D What about the myringoplasty ?

The most successful myringoplasty provides a happy outcome for the patient who becomes able to listen to intelligible conversations: yet regarding frequencies above 2 kHz, it is limited to a poor hearing[S128]-[S129]. Tympanic Membrane Perforation (TMP) causes hearing loss which increases with the perforation size[S130]-[S131]. TMP hearing loss generally increases as the frequency increases[S132]. Audiometric results shows that the larger the TMP, the larger the air–bone gap conductive hearing loss[S133]. Patients, with a small central perforation, recover with just a conservative management, but 10–15% suffer from bone conduction loss[S134].

Myringoplasty is most successful if the graft is made up of type II collagen (cartilage) and if the perforation is central[S135]-[S136]. Whatever the case may be, a local restructured collagen II must colonize the graft before the
graft can return to its full functionality. This means that fibres should recover their previous structure and their peculiarities, including their radial disposal according to an homogenous center-periphery polarity. Very commonly, the transplant regardless of its origin and its histological nature (temporal fascia, cartilage, etc) does not perfectly restore radial or circumferential structures *ad integrum* even after several months (or years).
As a result, the hearing is restored only for low and medium frequencies in such a way that the subject permanently loses the perception of the high frequencies. This finding supports our hypothesis : A bio-electronical pathway comes from collagen fibres of the eardrum. If their geometry loses its radial appropriate design, the pT becomes ineffective, and high frequencies are poorly perceived

### Si 01E  Measurements of piezoelectric response on eardrums

For measurements on eardrums we use a lock-in amplifier to drive a loudspeaker. In this manner, we broadcast a sinusoidal sound at about one meter from the external auditory conduit. We position a probe consisting of two electrodes at the centre and the periphery of the tympanum. This probe captures the piezoelectric response of the radial tympanic fibres when they vibrate in response to the sound sent to the tympanum. The lock-in amplifier makes it possible to select only those electrical responses isochronous to the acoustic stimulation. We measure electrical responses to stimulations at different acoustic frequency levels.  See fig. below :position of the electrodes of the probe.
A Lock In Amplifier, via an electric wave of $V_{hp}$ tension (output) creates an acoustic wave at various frequencies diffused by a loudspeaker.

### Si 01G  Measurement Electrodes for the eardrums

For the measurements on the eardrums, we tried various configurations:
• electrodes attached to a plastic bracket, ensuring a constant distance
• two electrodes easy to handling, more easily affixed to appropriate parts of the eardrum, but not allowing standardization of the inter-electrodes distance.

The shielding braid is put in contact with the peripheral structure of the tympanum, away from the handle of the malleus. The central copper wire is put in contact with the central structure of the tympanum: either umbo (direction "A") or handle of the malleus ("G").

### A. Position of the electrodes of the probe

It is possible to detect an electric potential isochronous to the acoustic vibration between an indeterminate point of the eardrum and the mastoid bone[61].. However, it does not necessarily follow that this potential be a microphonic produced by Outer Hair Cells OHCs. On the contrary, our methodology allows us to demonstrate, in vivo, and under normal physiological conditions, that it is the result of the piezo-electricity of the collagen II of the eardrums.

(a)

(b)

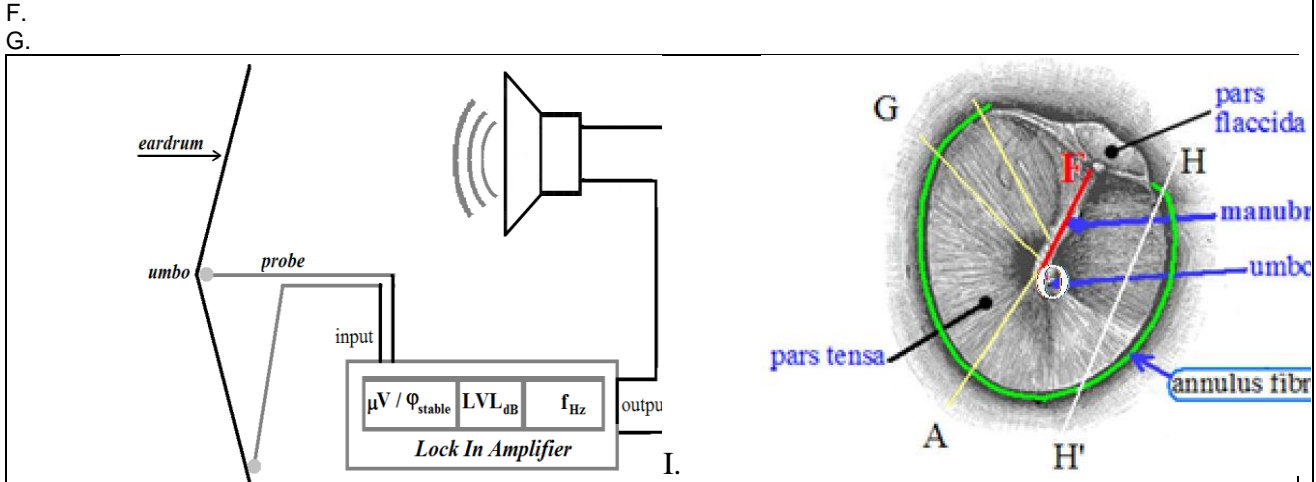

| a | b |

**Figure S01**

The closer together the electrodes are placed, the smaller the area recorded will be (i.e. local recordings). This is one advantage of the differential recording technique[61]. This "differential technique" is necessary for determining the source of a potential$s^{61}$.

### Si 01F  Measurements of piezoelectric response on knees (or other tendons)

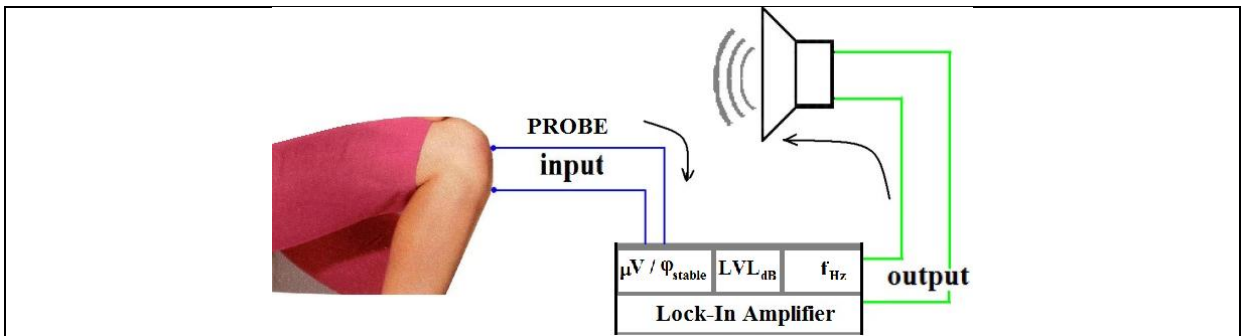

Figure S02
Diagram of the design used to measure the piezoelectric potential from the knee tendon.
□V indicates the isochronous voltage obtained when the phase j is stabilized, in response to the sound wave of frequency f (Hz) with a LVL (dBSP) amplitude.

## Measurement Electrodes for tendons

For measurements on the **mastoid** area, the patellar ligament or other tendons, we used « single use » *pregelled ECG-electrodes (Ref. "Comepa" 3.02.0200); they were connected to the* LockIn Amplifier by two *medical leadwires* AIP018 with *Touch proof connectors* DIN42802. (Table S09)

### Effect of the length of the collagen fibres

To evaluate the effect of the length of the collagen fibre bundles on their piezoelectric response, we performed some experiments on the most accessible bundles (patellar ligament). This allowed us to understand in particular why the response of these bundles was much larger than those of the eardrum.

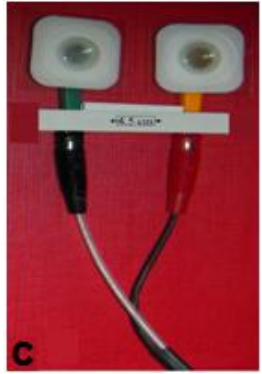

N.

O.

P. **(c)** Probe used for measurement on the **knees**

Q. As the amplitude of the voltage generated (in µV) increases with the distance (in mm) between electrodes; we, thus, ensured a fixed distance (45 mm) between the centre of the two COMEPA, 3020200 CA electrodes.

**Figure S03**

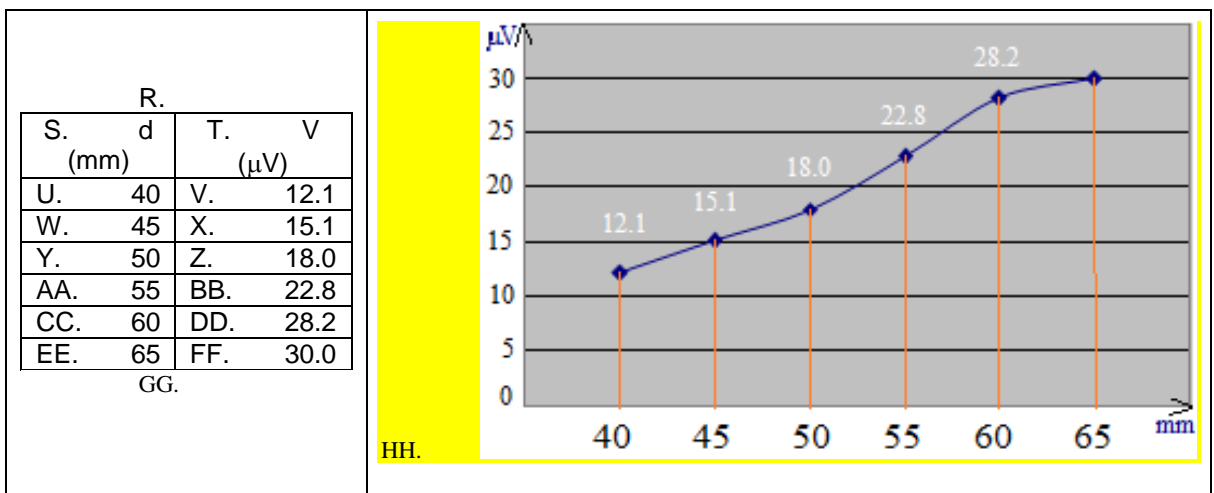

| d (mm) | V (µV) |
|---|---|
| 40 | 12.1 |
| 45 | 15.1 |
| 50 | 18.0 |
| 55 | 22.8 |
| 60 | 28.2 |
| 65 | 30.0 |

**Figure S04**
**Measurements on knee** : Acoustical frequency of the stimulus was 4000 Hz
x axis (d): distance (mm); y-axis : voltage (µV).
Result : **V (µV) = 0.76 d (mm) -19 mm ; $R^2$ = 0.98**

We have taken measures with different distances of separation between different electrodes placed in contact with the skin of a subject. The table, schema, and formula above present the effect of the distance between the electrodes. The voltage is proportional to the distance, i.e. to the length of the piezoelectric fibres.

In order to evaluate the electrical behaviour of points belonging to the central structure (manubrium) during acoustic stimulations, electrodes can be placed at two points on the same side of the manubrium (symbolized by an F letter). This system can detect whether there is an electrical isochronism between these points (synchronous potential close to zero). On the contrary, electrodes might be placed facing each other on either side of the manubrium (E). This latter system would allow us to capture the activity of a bundle of circular fibres (not completed).

The following letters were added by us : A or G: "radii" types of fibres of collagen ; HH': arbitrary cord joining two peripheral points ; F: manubrium of the malleus ; GAH'H: annulus tympanicus. In order to verify the piezoelectric activity of the tympanum, an electrode is placed on the manubrium, another on the periphery according to the straight lines either A or G.

With humans the synchronous electrical responses between two points on the same side (inner or outer) of the annulus tympanicus (HH') are generally impossible to measure (Table S10). Regarding the E48 RE (HH') exception, it might be that electrodes had been positioned, either not on the same side of the annulus tympanicus, or on an inner circumferential collagen fibre

## Si 02 The "space shift" problem

### Si 02A The "space shift" problem

According to F. Mammano[s61], and others, the tympanic potentials that we measure being low (less than one mV), they would be unable intervene in the operation of the OHCs. But, as we explained in the main part of our work, the operation of the TkS suppresses the need for strong potentials.

Yet Harnagea has pointed out that the pT potentials measured "in vitro" between the two extremities of a single collagen fibre can reach, and even exceed, 10 mV. If this is also the case "in vivo", such potentials could reach the cochlea (OHCs) with sufficient amplitude, provided that the fibres share a homogenous direction and polarity[s61]". Other authors explain that the mixed polarity of the collagen fibres in the biological structures significantly weakens their piezoelectric response[s61]. With respect to the eardrum, we took measurements by simple and soft application of the electrodes to the tympanic epidermis which is, more or less, insulating, in such a way that there was a weakening on the measured signal with respect to the genuine signal. So, we must be aware of bias caused by our measurements in vivo: We were not able to set up the two electrodes of the probe with the required precision in such a way that they would be in contact with both ends of a single fibre or bundle.
Often, there will be a mismatch between the ends of a seemingly united bundle of fibres: Let us assume that a first electrode is near the upper end of a fibres bundle and the other electrode near the lower end of the same fibres bundle; if the first electrode is located on the left of the first extremity, and if the second electrode is located on the right of the other extremity, the electrical pathway between the two electrodes does not correspond to the targeting of a single fibre: There is a space shift with implications for the conductivity and a measured voltage attenuation.

We tried to detect on the patellar ligament of knees, whether a configuration of the electrodes at the two ends of a supposedly unique collagen bundle could give higher measures than if electrodes were supposedly mismatched with the same bundle: one electrode on the external top side end of the "bundle", and the other electrode on the internal bottom side.
Our Protocol of measurement on knees was the following. We placed the subject at a fixed distance from the loudspeaker. We, then, apposed one electrode at the apex (lower point) of the patella and another electrode at the upper anterior top of the tibia. We made two trials on each knee for each of two frequencies. One trial used electrical stimulation of 70 dB SPL and the other of 80 dB SPL. As Harnagea pointed out, a maximum potential is collected if the electrodes are apposed at both ends of the same fibre, but this potential decreases or cancels if fibres overlap in all directions, as do the fibres of a "felt".
In vivo, the design of the collagen fibres is ordered by the functional necessities of the biological tissue. From a macroscopic point of view, we find very ordered structures, at the level of bones, joints, tendons, muscles or eardrum. For example, it is well known that the bone tissue is shaped according to the "law" of Wolf[s61].
The placement of the electrodes on the two ends of a single fibre or a single bundle of collagen fibres is expected to get a larger potential than if both electrodes are at the extremities of two different bundles of fibres. That is much easier to be successful if measurement is done on the patellar tendon, than on the eardrum.
We tried to measure, as locally as possible, voltage variations, synchronous to sound pressures. This is achievable without too much difficulty for collagen fibres *in vitro*. In our approach of the eardrum *in vivo*, the line that goes from a central point of contact to a peripheral one is laid out so that it corresponds as closely as possible to a bundle of radial fibres. But it is not possible for us to lay out with certainty both electrodes of the probe at the two ends of the same bundle of collagen fibres; It is likely that our measures may quite often involve two more or less neighbouring bundles; For this reason, the measurement will result in lower levels than if it were taken on one and the same bundle. This practical difficulty could help us understand why potentials measured on the eardrum are weaker than those that can be measured on the knees.

## Si 02B Effect of space-shifts on measurements of collagen piezoelectricity

The Supplementary Table 1 gives the results of our measurements for two subjects (B15, E48), on their knees (right and left) for different frequencies (f in Hz). We set the potential (Vout), sent by the LockIn to the Harman Speaker to 5V; the potential measured by the probe placed on the knee is expressed in µV.

| Sujets | B15 | B15 | B15 | B15 | B15 | B15 | E48 | E48 | E48 | E48 | E48 | E48 |
|---|---|---|---|---|---|---|---|---|---|---|---|---|
| Knee (R / L) | Right | Right | Right | Left | Left | Left | Right | Right | Right | Left | Left | Left |
| top electrode | med. | ext. s. | int. s. | med. | ext. s. | int. s. | med. | ext. s. | int. s. | med. | ext. s. | int. s. |
| bottom electrode | med. | int. s. | ext. s. | med. | int. s. | ext. s. | med. | int. s. | ext. s. | med. | int. s. | ext. s. |
| Line | {1} | {2} | {3} | {1} | {2} | {3} | {1} | {2} | {3} | {1} | {2} | {3} |
| f (Hz) | D1 | D2 | D3 | G1 | G2 | G3 | D1 | D2 | D3 | G1 | G2 | G3 |
| 500 | 30 | 25,3 | 8 | 29 | 24,1 | 16,6 | 12,5 | 33 | 20 | 28,2 | 5 | 12,3 |
| 1000 | 21,1 | 26,5 | 7,7 | 89,3 | 31,1 | 16,9 | 18,1 | 30,5 | 23,5 | 33,2 | 21,1 | 17,2 |
| 2000 | 70 | 46,5 | 9,7 | 110,2 | 41,1 | 21,3 | 29,6 | 41,5 | 28,5 | 41,1 | 28,5 | 33,1 |
| 3000 | 98,5 | 57,9 | 10 | 111,5 | 50,1 | 25 | 66 | 58,2 | 40 | 78,3 | 60,6 | 39,2 |
| 4000 | 110 | 66,6 | 18,4 | 152 | 58,2 | 26,8 | 100,5 | 73 | 41,1 | 111 | 87,7 | 44 |
| 6000 | 112,3 | 87,5 | 22,4 | 163,5 | 73 | 33 | 128,7 | 99,5 | 42,5 | 144,2 | 110 | 51 |
| 8000 | 173 | 110 | 28 | 184,3 | 88,7 | 38,5 | 180 | 105,6 | 43,7 | 201,5 | 135,8 | 53,3 |
| 9000 | 185,1 | 119 | 34,5 | 195,7 | 93,9 | 43,2 | 239,1 | 156 | 48,5 | 244,2 | 177 | 60 |
| 10000 | 215,2 | 135 | 44,3 | 185,9 | 102 | 47 | 256,7 | 200,2 | 51 | 264,1 | 189,4 | 71,1 |
| 16000 | 218,9 | 209,9 | 58,3 | 219,1 | 145 | 67,3 | 321,6 | 205 | 47,5 | 305 | 211 | 72,2 |
| 20000 | 282,6 | 270,5 | 70,8 | 381,4 | 160 | 79,9 | 450 | 233,1 | 52,7 | 412 | 258 | 79,5 |
| 25000 | 390,8 | 288,5 | 73 | 423,7 | 182 | 96,6 | 500 | 256,2 | 61 | 523 | 289,2 | 85 |
| Mean | 159 | 120 | 32 | 187 | 87 | 43 | 192 | 124 | 42 | 199 | 131 | 52 |
| Median | 143 | 99 | 25 | 174 | 81 | 36 | 154 | 103 | 43 | 173 | 123 | 52 |
| **Supplementary Table 1** *med.* means "median"; *int.* means "internal"; *ext.* means "external"; *s.* means "side" | | | | | | | | | | | | |

The Supplementary Table 2 give the results of our measurements for two subjects (B15, E48) on their two knees (right and left) for different frequencies (f in Hz), on the central line *{1}*.

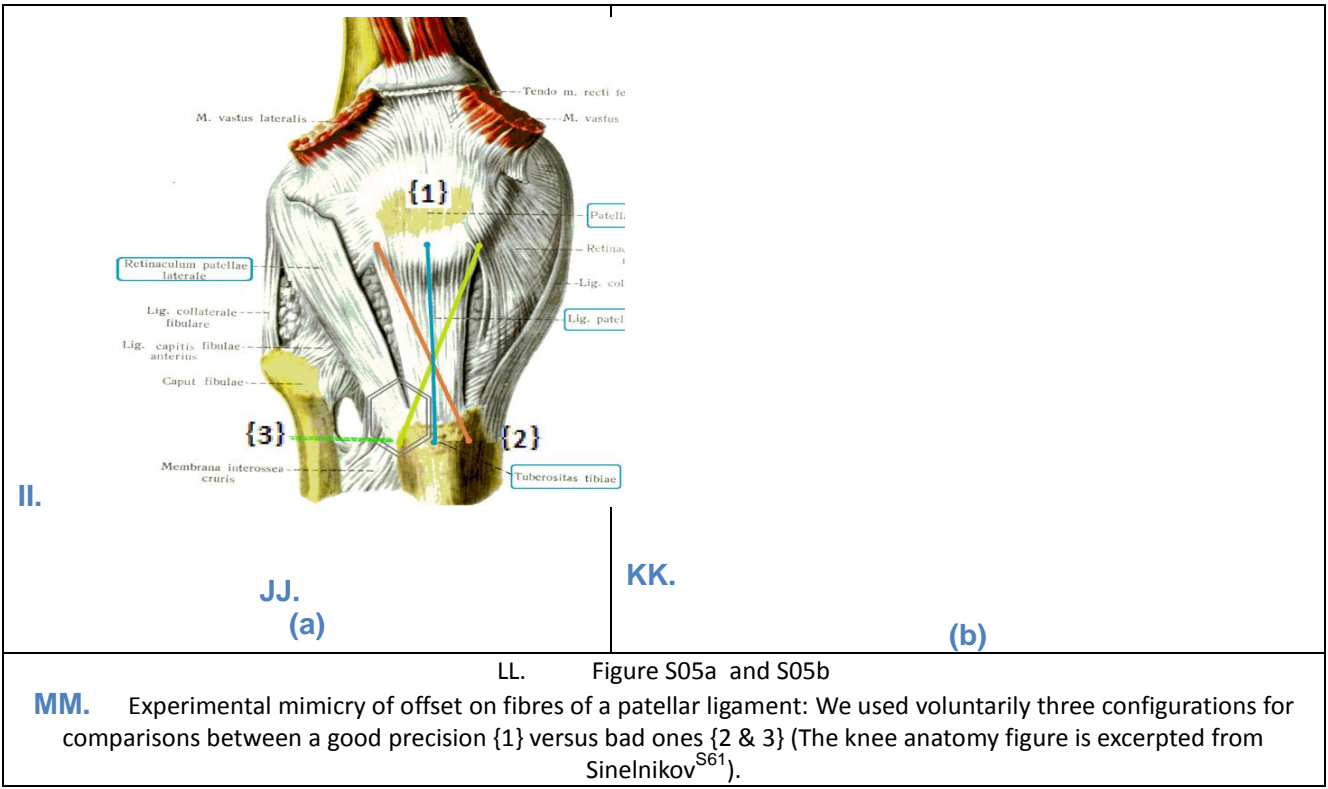

LL. Figure S05a and S05b

MM. Experimental mimicry of offset on fibres of a patellar ligament: We used voluntarily three configurations for comparisons between a good precision {1} versus bad ones {2 & 3} (The knee anatomy figure is excerpted from Sinelnikov[S61]).

| F (Hz) | {1} | {2} | {3} |
|---|---|---|---|
| 500 | 28,60 | 24,70 | 14,45 |
| 1000 | 27,15 | 28,50 | 17,05 |
| 2000 | 55.55 | 41,30 | 24,90 |
| 3000 | 88,40 | 58,05 | 32,10 |
| 4000 | 110,50 | 69,80 | 33,95 |
| 6000 | 136,45 | 93,50 | 37,75 |
| 8000 | 182,15 | 107,80 | 41,10 |
| 9000 | 217,40 | 137,50 | 45,85 |
| 10000 | 235,95 | 162,20 | 49,00 |
| 16000 | 262,05 | 207,45 | 62,80 |
| 20000 | 396,70 | 245,55 | 75,15 |
| 25000 | 461,85 | 272,35 | 79,00 |

Table 02

Ideally, measurements on collagen should concern a single fibre. But this is clinically very difficult: There may be an offset illustrated above (**a:** left panel), which we have minimized as much as possible. To check the effects of this type of configuration, we carried out measurements voluntarily strongly shifted on the tendon of a knee (**b**).

  \* The musculo-tendinous bundle concerned by the measurement line {1} includes the central collagen fibres of the tibial ligament.

  \*\* The musculo-tendinous bundle concerned by the measurement line {2} goes from the upper lateral portion of the tibial ligament down to its medial lowest part.

  \*\*\* The musculo-tendinous bundle concerned by the measurement line {3} goes from the upper retinaculum patellae down to the membrana interossea cruris (hexagonal framing at the bottom of {3}).

Harnagea showed in vitro that the response is low or non-existent if the electrodes are arranged in any two points of the culture, but this response reached tens of mV if the electrodes are placed at both ends of a **single fibre (**Denning et al. 2017)[61]. The longitudinal piezoelectric coefficient for idividual fibrils at the nanoscale was found to be roughly an order of magnitude greater than that reported for macroscopic measurements of tendon, the low response of which stems probably from the presence of poorly oriented fibrils. Thus we can assume that the pT voltages are in the mV range and may reach the DOHC complex.

Our assumption, based on the reflections of Harnagea, had made us envisage a shift between the central curve on the one hand {1}, and the two shifted curves on the other hand. We supposed that the curves {2} and {3} would be of amplitude and/or slope lower than the curve of reference {1}, which verified by our measures.

Moreover, it appears that the curve of configuration {3}(hexagonal framing at the bottom of {3}) is definitely lower than the two other curves; this is coherent with the fact that the measurement line {3} encompasses two musculo-tendinous bundles, which fact logically dampens the signal collected on this level compared to the signals collected in {1} or {2}.

| F (Hz) | D1 | G1 | D1 | G1 | | | |
|---|---|---|---|---|---|---|---|
| | Sujet B15 | Sujet E48 | *median {1}* | *mean {1}* | *sd {1}* | | |
| 500 | 30 | 29 | 13 | 28 | 29 | 25 | 8 |
| 1000 | 21 | 89 | 18 | 33 | 27 | 40 | 33 |
| 2000 | 70 | 110 | 30 | 41 | 56 | 63 | 36 |
| 3000 | 99 | 112 | 66 | 78 | 88 | 89 | 20 |
| 4000 | 110 | 152 | 101 | 111 | 111 | 118 | 23 |
| 6000 | 112 | 164 | 129 | 144 | 137 | 137 | 22 |
| 8000 | 173 | 184 | 180 | 202 | 182 | 185 | 12 |
| 9000 | 185 | 196 | 239 | 244 | 217 | 216 | 30 |
| 10000 | 215 | 186 | 257 | 264 | 236 | 231 | 37 |
| 16000 | 219 | 219 | 322 | 305 | 262 | 266 | 55 |
| 20000 | 283 | 381 | 450 | 412 | 397 | 382 | 72 |
| 25000 | 391 | 424 | 500 | 523 | 462 | 459 | 62 |
| **Supplementary Table 3** ||||||||
| Electrodes on the median line between the tip of the patella and the top of the tibia: *Configuration {1}* ||||||||

**Si 02Ba The "space shift"  (configuration 2 & 3):**

The Supplementary Table 3, give the results of our measurements for two subjects (B15, E48), on their two knees (right and left) for various frequencies F (Hz), on the line of configuration {2}.

| f (Hz) | D2 | G2 | D2 | G2 | *median {2}* | *mean {2}* | *sd {2}* |
|---|---|---|---|---|---|---|---|
| | Sujet B15 | Sujet E48 | | | | | |
| 500 | 25 | 24 | 33 | 5 | 25 | 22 | 12 |
| 1000 | 27 | 31 | 31 | 21 | 29 | 27 | 5 |
| 2000 | 47 | 41 | 42 | 29 | 41 | 39 | 8 |
| 3000 | 58 | 50 | 58 | 61 | 58 | 57 | 5 |
| 4000 | 67 | 58 | 73 | 88 | 70 | 71 | 13 |
| 6000 | 88 | 73 | 100 | 110 | 93 | 93 | 16 |
| 8000 | 110 | 89 | 106 | 136 | 108 | 110 | 20 |
| 9000 | 119 | 94 | 156 | 177 | 138 | 137 | 37 |
| 10000 | 135 | 102 | 200 | 189 | 162 | 157 | 46 |
| 16000 | 210 | 145 | 205 | 211 | 208 | 193 | 32 |
| 20000 | 271 | 160 | 233 | 258 | 246 | 230 | 49 |
| 25000 | 289 | 182 | 256 | 289 | 272 | 254 | 50 |
| **Supplementary Table 4** ||||||||
| One electrode shifted up on the outer side of the patella and the other electrode staggered downstairs on the internal side of the top of the tibia: *Configuration {2}* ||||||||

The table Supplementary Table 4 below, give the results of our measurements for two subjects (B15, E48), on their two knees (right and left) for various frequencies *f* (Hz), on the line of configuration {3} (green).

| F (Hz) | D3 | G3 | D3 | G3 | median {3} | mean {3} | sd {3} |
|---|---|---|---|---|---|---|---|
| | Sujet B15 | Sujet E48 | | | | | |
| 500 | 8 | 17 | 20 | 12 | 15 | 14 | 5 |
| 1000 | 8 | 17 | 24 | 17 | 17 | 16 | 7 |
| 2000 | 10 | 21 | 29 | 33 | 25 | 23 | 10 |
| 3000 | 10 | 25 | 40 | 39 | 32 | 29 | 14 |
| 4000 | 18 | 27 | 41 | 44 | 34 | 33 | 12 |
| 6000 | 22 | 33 | 43 | 51 | 38 | 37 | 12 |
| 8000 | 28 | 39 | 44 | 53 | 41 | 41 | 11 |
| 9000 | 35 | 43 | 49 | 60 | 46 | 47 | 11 |
| 10000 | 44 | 47 | 51 | 71 | 49 | 53 | 12 |
| 16000 | 58 | 67 | 48 | 72 | 63 | 61 | 11 |
| 20000 | 71 | 80 | 53 | 80 | 75 | 71 | 13 |
| 25000 | 73 | 97 | 61 | 85 | 79 | 79 | 15 |

**Supplementary Table 5**
One electrode shifted up on the inner side of the patella and staggered downstairs on the external side of the top of the tibia: *Configuration {3}*

**Si 02Bb The "space shift" : comparison of the three configurations**

We compare the three medians (Supplementary Table 5); "*a*" and "*b*" are the coefficients of the polynomials of the curves of linear regression of the three configurations with "*a*" parameter of slope, and "*b*" ordinate at the linear origin of the curves of regression:

| Parameters | Median {1} | Median {2} | Median {3} | Decreasing order |
|---|---|---|---|---|
| a | 0,018 | 0,010 | 0,003 | {1} > {2} > {3} |
| b | 31,12 | 27,76 | 20,05 | {1} > {2} > {3} |

Supplementary Table 6
Table of Linear regressions

To compare the effect of the three configurations, we joined together the data belonging to each one (Supplementary Table 7).

| f (Hz) | {1} | {2} | {3} |
|---|---|---|---|
| 500 | 28,60 | 24,70 | 14,45 |
| 1000 | 27,15 | 28,50 | 17,05 |
| 2000 | 55,55 | 41,30 | 24,90 |
| 3000 | 88,40 | 58,05 | 32,10 |
| 4000 | 110,50 | 69,80 | 33,95 |
| 6000 | 136,45 | 93,50 | 37,75 |
| 8000 | 182,15 | 107,80 | 41,10 |
| 9000 | 217,40 | 137,50 | 45,85 |
| 10000 | 235,95 | 162,20 | 49,00 |
| 16000 | 262,05 | 207,45 | 62,80 |
| 20000 | 396,70 | 245,55 | 75,15 |
| 25000 | 461,85 | 272,35 | 79,00 |

Supplementary Table 7
Median, depending on the frequency-test ($f_{Hz}$) according to each of the three configurations.

## Si 02C Application to fibers of the tympanum (Denning et al.)

As a preliminary remark, when the electrodes are laid out at the two ends of the handle of the hammer and on the same side, it is expected that the measured voltage be null or so. We verified it on chinchilla and human.

Otherwise, we can apply the previous observation on knees about the measurement lines {1} , {2} and {3} to our measurements on eardrum fibres: To get a maximum voltage, the optimal design would be to put the electrodes at both ends of a same radial fibre of collagen. If there is a spatial shift, there will necessarily be a lessening of the amplitude of the signal. Not being able to meet that theoretical requirement, we have to accept that constraint. In order to decreases its statistical effect, we may calculate an individual parameter for each series of measures in a given configuration (see § statistics).

Denning et al. (2017)[61] have shown that nanoscale fibrils can generate higher electrical potentials (mV) than long-tendon fibers (µV). As suggested by Harnagea they deduce that these long fibers have a polar orientation ratio of their constituent fibrils significantly different from 50/50%.

It is probable that, in the resting position, the polarity of the collagen radial fibers be oriented due to their centripetal development, resulting their positive end in the centre and their negative end in the periphery.

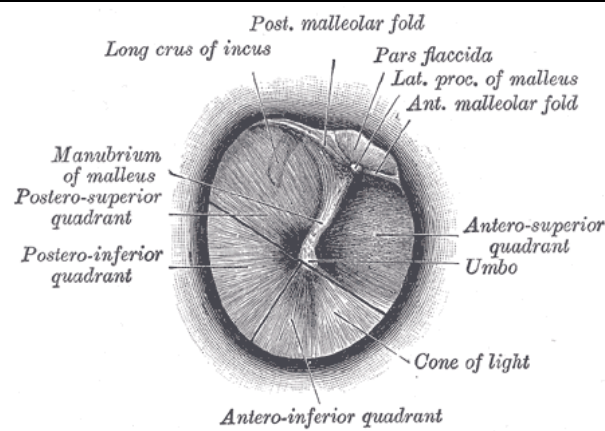

Figure S06
internal aspect of tympanic membrane

The radial fibres of collagen grow from the umbo to the periphery .
The ENT experimenter tries to put both of the probe electrodes at the ends of one single fibre. Yet such an utmost precision is unreachable because every fibres are extremely thin and not separately identifiable. So it happens an unwished offset on fibres of the tympanum.

## Si 03 Variations

### Si 03A Variation of eardrum tension

The subject on which we take the measurements is equipped with several types of very efficient biological control mechanisms, which are usually managed by reflex mechanisms (unconscious and involuntary activity), in a non-controllable way. For this reason, the acoustical strain of the eardrum (and by implication, its piezoelectric response) may vary in a non-controllable manner according to the expectations of the subject, his psycho-physiological state, the potential effect of local anaesthesia on the subject's biological reactions.

Eardrum tension depends mainly on two types of muscle. The "tensor tympani" is a striated muscle, that some subjects may contract, voluntarily, on demand, for a fairly short period ([S61]-[S61]). Furthermore, for everybody, it contracts by reflex, involuntarily and unconsciously, when it is appropriate to attenuate or target the perception of low frequency sounds.

In several species of mammals, and interestingly in those which have the best sense of hearing (bats), a lot of radial myofibrocytes are included within the annulus tympanicus (which encircles the eardrum). These myofibrocytes are closely and individually connected to the radial fibres of collagen[S61] attached to the bony circumference of the eardrum. So their activity may help to regulate the tension of the eardrum by means of that ring comprising radially oriented smooth muscles[S61]. This arrangement suggests a role in creating and maintaining tension on the tympanic membrane much like a drum tuning key which is used to adjust the "*tension rods*" and pitch of a drum.

Cf. below : Si 08

### Si 03B Variation of contact pressure of the electrodes on the skin

A stronger pressure of an electrode on the skin might increase admittance of the pathway between the electrode and the collagen under the skin (owing to better contact and/or capacitive modification). It is therefore likely that the strength of pressure from the electrodes on the skin of knees should have some slight consequences on the results. However, we did not accurately study it.
Regarding the measurement on the eardrum, we face two antagonist phenomena: more pressure, better contact and better conductivity; But also, more pressure, less freedom of movements, so damping of vibrations and decrease of the associated piezoelectricity.
That is the conundrum that we face without being able to solve it at the moment.

## Si 04 Possible Artefacts ?

### Si 04A Artefact resulting from links to ground
If two pieces of electronic equipment are plugged into different power outlets, there will often be a difference in their respective ground potentials. The two ground connections may cause ground loops when the components are interconnected by signal cables (Fig. S02 below). There will be a parasitic standing wave created at the AC mains base frequency (50 or 60Hz) and the harmonics thereof (120 Hz, 240 Hz, and so on). So it is preferable to have "single-point grounding", with the system connected to the building ground wire at only one point. To avoid this phenomenon as much as possible, we have measured the validity of ground link.

Furthermore we have connected the different outlets of the system (lock-in, speaker) on the same power strip and the same ground socket.

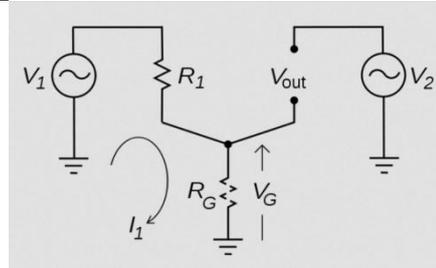

Figure S07
"ground artefact" excerpted from en-WP[S61] (R resistance, G ground, V Voltage, I intensity; in this schema there are three links to the ground).

### Si 04B Artefact resulting from a "hand effect"

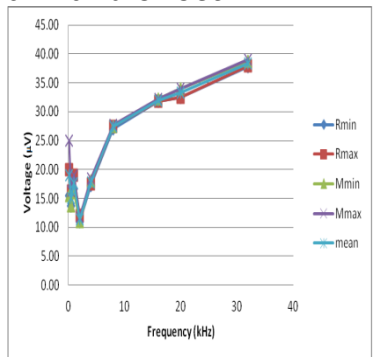

Figure S08
Mean Voltages ($\mu$V) measured for several frequencies
(R means Results without Hand Effect; M means results with Hand effect). In all configurations, when there is a change, the "hand effect" causes an increase of the voltage <1.5 $\mu$V for voltages around 25 $\mu$V, ie 6% (Fig. S08 above).
Additional connection of a conducive bracelet around the wrist of the experimenter engenders a higher "hand effect" unless the metallic case is also connected to the ground. The most practical solution, in order to avoid any hand effect, is to remove the hand of more than 20 cm from the casing.

### Si 04C Artefact depending on the contact between electrodes and the skin-target

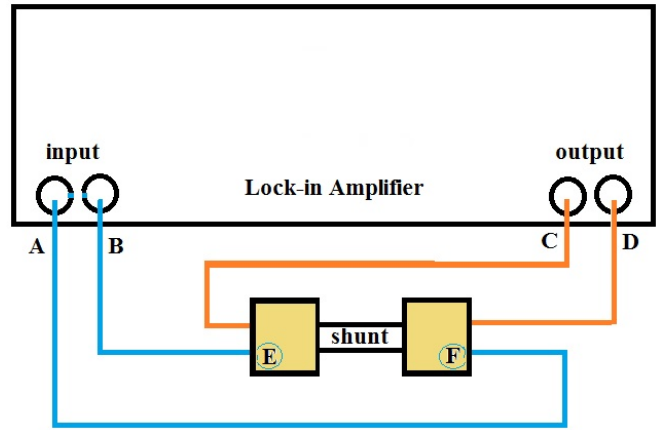

Figure S09

In fig. S04 (above) we replaced the skin of the subject by a simple copper shunt, with a resistance Rs = 0.2 Ω (AOIP 02.2351); This shunt is purely resistive up to a few tens of kHz. And the current from the output of the Lock-in (C and D) is introduced (by red circuit: coaxial cable and alligator clips) into the shunt.
The electrodes (E and F), normally applied to the skin, here are connected by the blue wires to the Lock-in Amplifier; they have different forms and surfaces (round or square COMEPA, or Alligator Clips).
Since our measures on tendons present values in the range from 10 to 200 µV, we adjust the settings so that the potential difference between E and F is close to 200 µV. As the resistance of the shunt is 0.2 Ω, we must send a current of 200 µV / 0.2 Ω, ie 1mA flowing through this shunt. To arrive at this value of 1 mA, if we take the device as a whole, including the two resistors in series, the output of the Lock-in (Ro = 50 Ω), and the resistance of the shunt (Rs = 0.2 Ω), we have to adjust the output voltage of the Lock-in to about 50 mV since {1 mA * (Ro + Rs)} = 1 * 50.2 = 50.2 mV. The resulting voltage between E and F (in µV) to the terminals of the shunt is measured at the input of the Lock-in (be A and B; fig. S04 above) with the blue cables. By varying the frequency (kHz), we obtain the results presented in Table S01 and Fig. S05.

| | two square electrodes | one round and one square electrode | two round electrodes | two alligator clips |
|---|---|---|---|---|
| f (kHz) | (µV) | (µV) | (µV) | (µV) |
| 0.2 | 0.135 | 0.142 | 0.164 | 0.196 |
| 0.8 | 0.136 | 0.138 | 0.166 | 0.198 |
| 2 | 0.138 | 0.145 | 0.165 | 0.196 |
| 4 | 0.142 | 0.146 | 0.160 | 0.196 |
| 8 | 0.134 | 0.135 | 0.143 | 0.202 |
| 20 | 0.185 | 0.188 | 0.198 | 0.219 |

Ro=50 Ω et Rs=0.2 Ω
**Table S08**

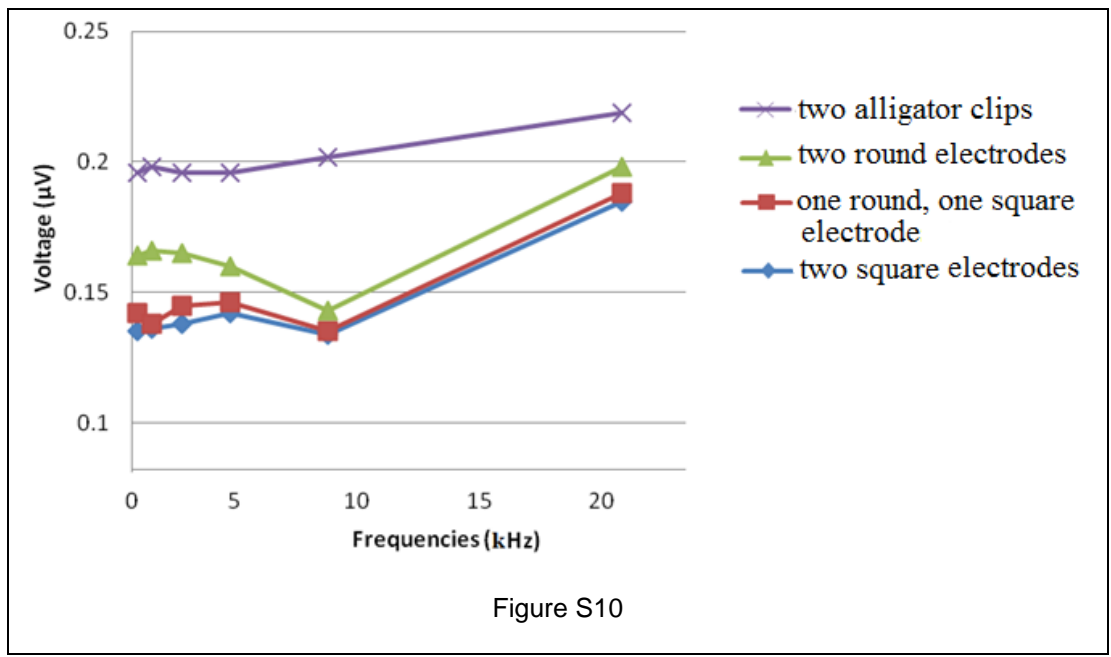

Figure S10

## Si 04D Is there artefact depending on microphonic of measurement cables ?

Mechanical noise can be translated into electrical noise by microphonic effects. Physical changes in the cables (due to vibrations for example) can result in electrical noise over the entire frequency range of the lock-in. For example, consider a coaxial cable connecting a detector to a lock-in. The capacitance of a coaxial cable is a function of its geometry. Mechanical vibrations in the cable translate into a capacitance that varies in time typically at the vibration frequency. Since the cable is governed by CV= Q[61] taking the derivative yields: C(dV/dt) + V(dC/dt) = dQ/dt = i. Mechanical vibrations in the cable which cause a dC/dt will give rise to a current in the cable. This current may affect the measured signal [S61]. We conducted an evaluation of this effect in the context of our measurements: We set the power of the loudspeaker by the lock-in with an increased voltage (5V against 2.5V for previous experiments) and with the speaker very close to the coaxial cable that we use. Variations in capacitance of the coaxial cables that we used, are very low given the broadcast sound power. We observe no voltage cable, short circuit, or open on the range 0.1 - 40 kHz.

In these measurements on metal, no sound is emitted; it is a classical voltage measurement. The increase here is due to the capacitive effect of the measurement contacts. For 20kHz, the increase is less than 30% against a factor 3 to 10 in the case of the tendons or the eardrum; it cannot, therefore, be an artefact in the case of the tendon or the tympanum. This increase is observable also in the measurement using alligator clips. This parasitic capacitance comes not only from the bad contact between the electrodes and the skin, but also from the contact between clips and electrodes.

In the case of electrophysiological recordings, it is usual to optimize skin impedance by cleaning the skin (Poch-Broto et al., 2009) and drying it, when necessary. The impedance of the path of contact also depends on the thickness of the subcutaneous fat layer, which might be roughly assessed by the Body Mass Index (BMI). There were not a significant correlation between that measure and the voltages results; So we think influence of this parameter is weak (see §statistics).

## Si 05 About a supposedly cochlear origin of the isochronous response which we measured between the tympanic and mastoidian fibers ends

It is possible to detect an electric potential synchronous to the acoustic vibration between an indeterminate point of the tympanum and the mastoid bone[S61]. It does not follow necessarily, however, that the potential measured in this type of experiment is produced by the Outer Hair Cells (OHCs).

### Si 05A Value of the microphonic potentials

Microphonic potentials (119), measured near the cochlea, are on the order of µV at the threshold. They increase from 100 to 400 µV for average sounds and up to a maximum of 800 µV for the most intense sounds$^{S61}$.

Proto-microphonic potentials (about 10% of the microphonic that persists despite the functional elimination of the OHCs), would be around 1 to 80 µV. These values correspond to 1/1000 of the experimental values measured in vitro on collagen I (149). That residual potential persists in cases where the OHCs or IHCs are destroyed or no longer function for whatever reason; This persistence obviously cannot be attributed either to the OHCs or to the IHCs. In addition, the destruction of the IHCs (chinchilla) does not alter the cochlear microphonic and does not alleviate the Electrically Evoked Oto Acoustic Emissions (EEOAE); moreover, these responses tend to increase at high frequencies $^{S61}$. That being the case, we suggest that the residual potential is due to the piezoelectricity of the tympanum (the piezo-tympanic signal or "pT").

We have been able to record, from the region of the mastoid bone, synchronous potentials which are not weakened when the auditory canal is occluded (Cf. § hereafter). Similarly, if a sound is sent to the eardrum and not to the superficial mastoid, recorded synchronous potential is lower than if the sound is sent to the superficial mastoid and not to the eardrum. Therefore, we demonstrate that if a sound is sent in the direction of the mastoid area, the synchronous evoked potential is not from the cochlea, but from local generators. This means that neither the ossicular chain nor the Bekesy Traveling Wave (TW) is involved. Rather, the synchronous potential is attributable to the collagen present in the mastoid region.

## Si 05B Could tympanic or bone fibers isochronous potentials be from cochlear origin ?

It has been sometimes accepted in the literature that synchronous potentials recorded at the level of the mastoid are purely of cochlear origin. If this were the case, obstructing the external ear canal should indeed lead to a reduction of these synchronous potentials.

The cochlear microphonic (CM) is an alternating current (AC) isochronous to the acoustical stimulating waves$^{S61}$. Its threshold is very low. Above 60 dB SPL its amplitude decreases almost at once. On the contrary, for moderate incident sounds (< 40 dB SPL), an amplification, proportional to the sound level occurs after a certain delay. This amplification is, for the most part (80-90%), attributable to OHCs, but it resists the destruction of these OHCs. The origin of this poorly understood residual 10-20 %, has been mistakenly attributed to the IHCs alone$^{S61}$ or to the piezoelectricity of the membrana tectoria (Offut, 1984). Our hypothesis is that this residual microphonic is generated, at least in very large part, by the piezoelectricity of the eardrum (and/or petrosal collagen). The so-called mastoid microphonic is not reducible to the diffusion of the cochlear microphonic. In Fig.S06 we describe this commonly accepted theory. The cochlear microphonic would be detectible near the eardrum$^{S61}$ and it would also be detectible near the mastoid$^{S61}$. This mastoid potential would be the emergence of the cochlear microphonic alone$^{S61}$ and would allow for audiometric testing$^{S61}$. Mammano$^{S61}$ thinks that the synchronous potentials that we highlight could be interpreted as the result of the activity of the OHCs reaching as far as the eardrum (or mastoid) ....

Furthermore, in order to build a tympanum restricted to its mechanical effects, the evolutionary process should have led to the generation of collagen fibres in a convenient geometrical arrangement, but with random polarity.

We will show below that :
        - The cochlear microphonic is not dependent either on the IHCs, or on cochlea
        - The so-called mastoid microphonic is not reducible to the diffusion of the cochlear microphonic.
        - Attempts to define a mastoid audiogram has appeared problematic.
        - Bone conduction allows to hear ultrasounds.

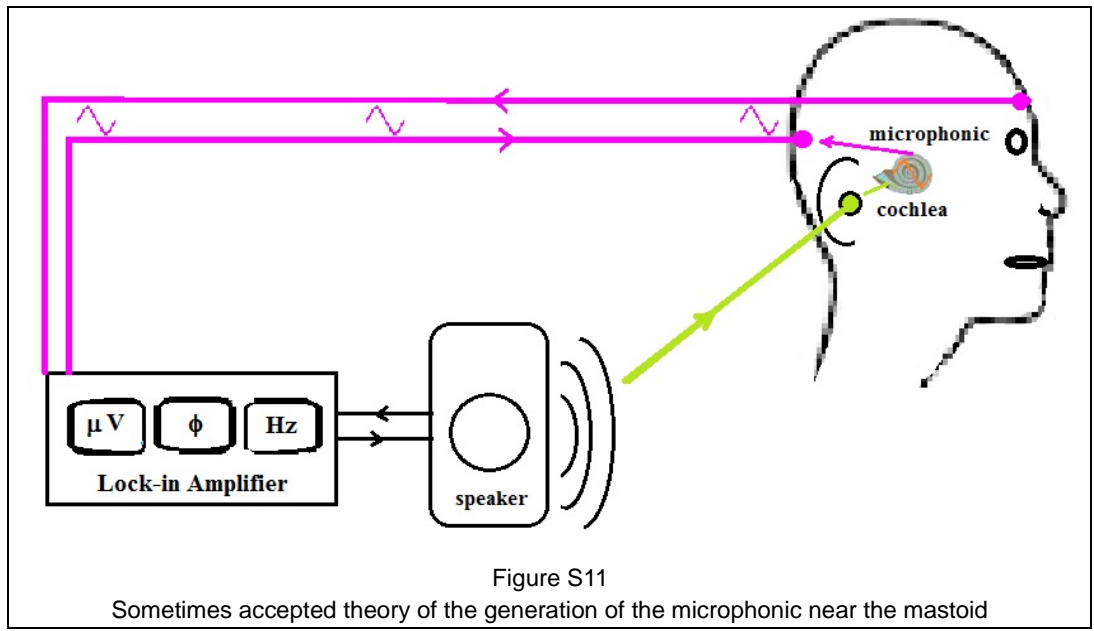

Figure S11
Sometimes accepted theory of the generation of the microphonic near the mastoid

The acoustic vibrations of the eardrum are transduced, by the cochlear OHCs, into a synchronous electric potential, known under the name of microphonics[S61]. Furthermore, such a synchronous electric potential can be recorded at the level of the skin behind the ear (with the other electrode being placed, for example, on the forehead). It is postulated that the cochlear generator alone would be responsible for that recorded potential. In other words, the recorded potential would result only from the scattering of the "cochlear microphonic" and for no other reason.

## Si 05C The synchronous potentials, tympanic or mastoidian, are not uniquely dependent on the cochlea

In order to test whether pure-frequency sound stimulation produced a piezoelectric response, we measured the tympanum of a pig a few hours after its death, and after removal of the cochlea performed without damaging the anatomy of the eardrum.

**Piezoelectricity measurements on the pig tympanum at the Puylaurens abattoirs (1 07 2018).**
We carried out measurements on pig eardrums shortly after their slaughter, in the Puylaurens Archeology Club premises (room temperature 24 ° C).

**Acoustic measurements**
Sound source: Yamaha DBR12 loudspeaker, powered by the LockIn Stanford SR 830 (frequency and amplitude adjustment).
The sound intensity measurement was carried out using a Bruel & Kjaer sound level meter, Type 2231 (measurements were made by Bernard Bibé, INRA). For the tests, we opted for a relatively moderate acoustic amplitude of about 75 dB SPL
archeology of Puylaurens (room temperature 24 ° C).

**Measurements of electrical responses:**
Lock in Standford SR 830 (measuring frequencies 150 Hz-30kHz, 300ms time constant).
The probes consist of a coaxial cable that is connected to the input of the LockIn by a BNC plug.
This cable has two components: a fiber core consisting of a single fiber and a peripheral shield made of very fine braided metal fibers. The end of these two components is completed by a small metal ball.

**Measured samples:**
1. **Tympan 1** taken from a head of the day, not scalded.
2. **Tympan 2** taken from a scalded head
3. **Tympan 3** taken from a head of the day, not scalded but damaged during preparation.

4. **Bone conduction**: Bone taken from the external auditory canal, this bone is hemicylindrical (semicircular section). It consists of a wall of X mm thick, and a "channel" of Y mm width; this wall and this channel are oriented like generatrices of the bone cylinder.

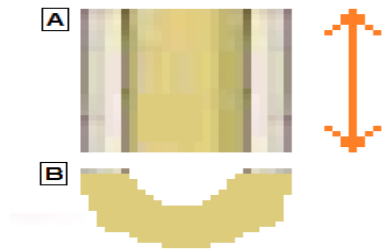

A ) Superior view of the osseous canal.
On the right: double arrow symbolizing the potential difference between the ends of the bone channel, when it is stimulated by a stable frequency acoustic wave.
B ) Sectional view of the 1/2 bony channel

6 measurements were made:

1) Tympanum 1: loudness 70dB
2) Tympanum 2: loudness 74dB
3) Tympanum 2 after removal of the cochlea: loudness 74dB
4) Tympanum 3: loudness 74dB

Hemisection of the bone canal:
5) loudness 70dB
6) loudness 90dB

We have calculated (Excel) the polynomial-type trend curve for each of the measurements made

**Measurement 1**: Tympanum 1 taken from a head of the day, not scalded: stimulus 70dB SPL

| Frequency (Hz) | Stimulus (dB SPL) | Electrical response ($\mu$V) |
|---|---|---|
| 200 | 70 | 34 |
| 500 | 70 | 9 |
| 1000 | 70 | 10 |
| 2000 | 70 | 4 |
| 4000 | 70 | 0 |
| 8000 | 70 | 3 |
| 16000 | 70 | 1,5 |

Tympanum 1; y($\mu$V)= 2E-07 $x^2$ - 0,005 x + 18,59; $R^2$ = 0,47
ordinates: voltages ($\mu$V) ; abscissa: frequencies (Hz)

**Measurements 2**: Tympan 1  (Stimulus amplitude 70dB SPL)

| Frequency (Hz) | Piezoelectric signal ($\mu$V) |
|---|---|
| 500 | 0,2 |
| 1000 | 0,5 |
| 2000 | 0,2 |
| 4000 | 0,06 |
| 8000 | 0,08 |
| 16000 | 0,3 |
| 20000 | 0,27 |
| Although the signal was very weak, the measurements were stable during the measurement and the repetition of the measurement gave an identical result. ||

y = 2E-09 $x^2$ - 5E-05 x + 0,3214; $R^2$ = 0,30
ordinates: voltages ($\mu$V) ; abscissa: frequencies (Hz)

**Measurements 3**: Tympan 2 taken from a head of the day, not scalded, after ablation of the cochlea, sound intensity 74dB SPL.

| Frequency (Hz) | Piezoelectric signal (μV) |
| --- | --- |
| 250 | 4 |
| 500 | 0,25 |
| 1000 | 0,15 |
| 2000 | 0,18 |
| 4000 | 0,07 |
| 8000 | 0,11 |
| 16000 | 0,2 |

$$y = 2E\text{-}08\ x^2 + 1,62\ ;\ R^2 = 0,26$$
ordinates: voltages (μV) ; abscissa: frequencies (Hz)

**Measurements 4**: Tympanum 3 (stimulus amplitude 74dB SPL)

| Frequency (Hz) | Signal piezo (μV) |
| --- | --- |
| 250 | 20 |
| 500 | 6 |
| 1000 | 3 |
| 2000 | 2 |
| 4000 | 0,4 |
| 8000 | 0,8 |
| 16000 | 1,4 |

$$y = 1E\text{-}07 x^2 - 0,003\ x + 10,51;\ R^2 = 0,44$$
ordinates: voltages (μV) ; abscissa: frequencies (Hz)

**Measurements 5**: Piece of External Bone Canal (70dB)

| Frequency (Hz) | piezoelectric signal (μV) |
| --- | --- |
| 150 | 0,87 |
| 500 | 0,5 |
| 1000 | 0,75 |
| 2000 | 1 |
| 4000 | 0,3 |
| 8000 | 1,1 |
| 16000 | 2,7 |

$$y = 1E\text{-}08 x^2 - 8E\text{-}05 x + 0,7737;\ R^2 = 0,91$$
ordinates: voltages (μV) ; abscissa: frequencies (Hz)

**Measurements 6**: Piece of External Bone Canal (90dB)

| Frequency (Hz) | piezoelectric signal (μV) |
| --- | --- |
| 150 | 25 |
| 500 | 25.74 |
| 1000 | 4 |
| 2000 | 6,5 |
| 4000 | 1,7 |
| 8000 | 6 |
| 16000 | 15,6 |

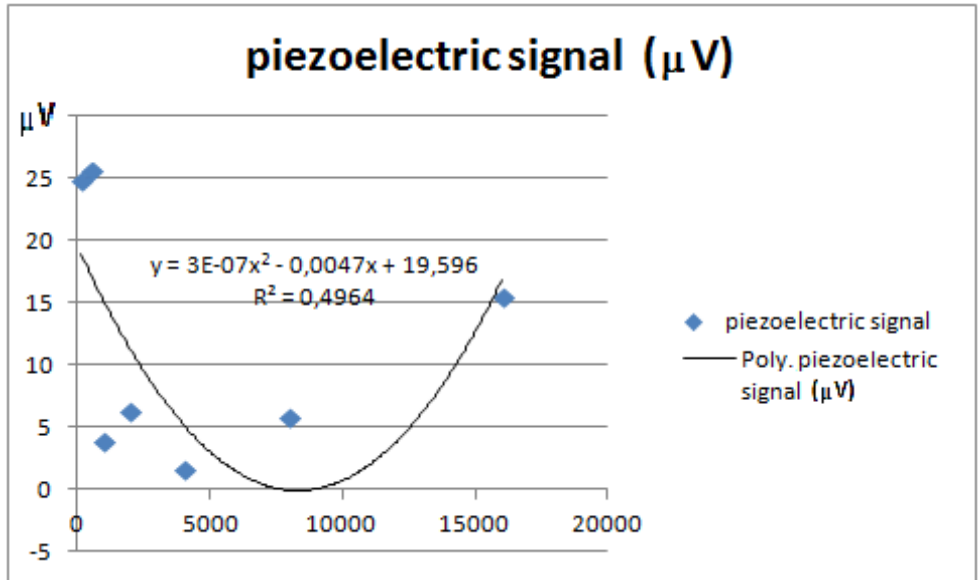

$y = 1E-06x^2 - 0,025x + 96,433; R^2 = 0,20$
ordinates: voltages ($\mu$V) ; abscissa: frequencies (Hz)

## Si 05D Bone Conduction

**Si 05Da Attempts to define a mastoid audiogram have been problematic**

These findings and their restrictive interpretation led Poch-Broto and al. (2009) to try to create an audiometric test based on the recording of the electric potential of the mastoid in response to auditory stimulations. In this theory, the schema of transmission without earplug complies with fig. S07.

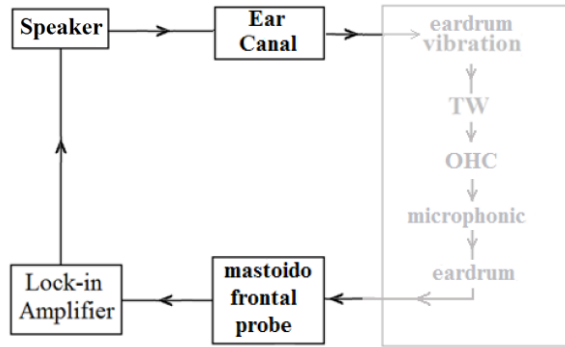

Figure S12
Capture of the mastoid "microphonic". Dimmed: Interpretation of Poch-Broto, attributing all the microphonic amplification to the TWs alone, via the OHC[S53].

But, if we compare the classic audiogram and this " mastoid " audiogram, we can see that very often (>10%) they differ by more than 10 dB HL mainly in the area of high frequencies, which is significant (fig. S08).

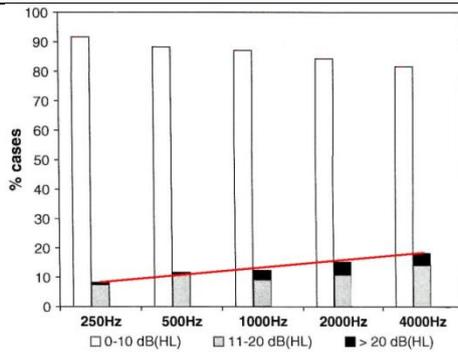

Figure S13
Modified chart based upon the data of Poch-Broto et al., 2009[S61]:
We grouped the black rectangles and the gray rectangles to represent all errors (> 10dB) combined. This clearly shows (red lines) that the error rate increases with frequency. The X axis displays the tested frequencies; the Y axis displays the % of cases according to whether the divergence between subjective audiometry and "microphonic" audiometry is less than (clear bars), or superior to (gray and black bars) 10 dB HL.

Since for all frequencies, in more than 10% of cases, the difference between subjective audiometry and measurement of the microphonic is greater than 10 dB, it can be concluded that the cochlear generator is not the sole origin of collected potentials (cf the line in red); this is particularly true for the high frequencies. We contacted Pablo Gil Loyzaga (corresponding author of the Group of Poch Broto and al.) and asked for their raw data. Yet, he answered that the raw data was no more available (Pablo Gil Loyzaga, com. pers., 07/18/2012).

**Si 05Db Bone conduction permits the hearing of ultrasounds**

Bone collagen is particularly adapted for the transduction of ultra-sounds[S61]. Ultrasounds between 20 kHz and 33 kHz are heard when they are being administered by bone conduction[S61 - S61]. They act directly on the hair cells (ie their hearing is induced by the ultrasounds themselves). And furthermore, in the case of extremely high pitched sounds, toxic damage of the auditory system damages the air-conduction hearing and improves bone-conduction hearing.
It is well demonstrated that the deterioration of collagen II is causative of auto-immune deafness[S61].

## Si 05E Experiments involving closing off of the external auditory canal (EAC)

The schemas, tables and curves below show the synchronous potentials, depending on whether a plug is not or is inserted into the auditory external conduit (earplug interposed between the sound source and the eardrum, but not between the sound source and the mastoid) : fig. S09, S10, S11, S12.

**Si 05Ea Mastoid 'microphonic' with EAC closed off :**

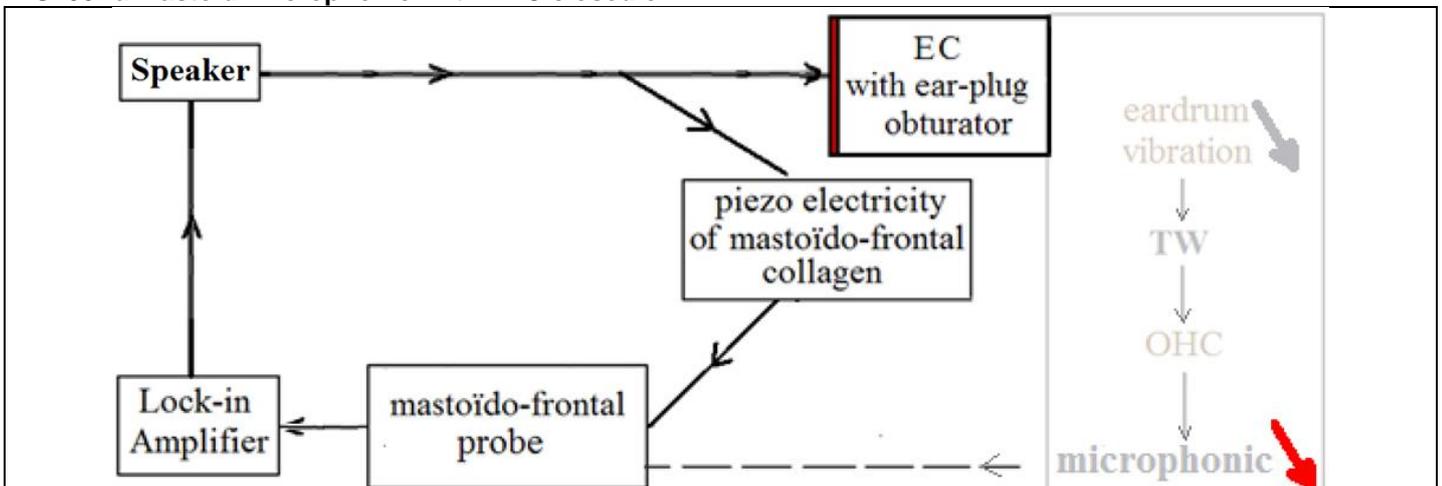

Figure S14
Capture of the mastoid 'microphonic' with the external auditory conduit closed off. (Dimmed and red arrows on the right side of the figure indicate the decrease of the mechanical vibration and of the electric potential as well).

**Si 05Eb Mastoid potential with EAC closed off by a standard shutter of trade:**

We can insert a plug to seal off the external auditory canal (EAC). The theory of Poch-Broto et al. (2009) implies that the earplug decreases the amplitude of the vibrations of the eardrum. This necessarily results in a significant decrease of the cochlear "microphonic"; And of the mastoid synchronous potential as well if the latter potential results solely from the cochlear "microphonic". If the mastoid synchronous potential is composite, and if it is generated, at least in part, by the piezoelectric activity of the mastoid collagen, the mastoid synchronous potential should be only slightly diminished by the closing off of the external auditory conduit. But, if for some subjects, the synchronous potential remains the same, whether the ears are plugged or not, we have to consider another phenomenon.

We, therefore, conducted the following experiments. Three subjects were tested (D37, D39, D44). We display the electrical response depending on whether the EAC is open or obstructed (F is the frequency of the auditory stimulus, in kHz; Vs is the electrical potential of the response, in $\mu V$).
The schemas, tables and curves below show the synchronous potentials, depending on whether an earplug is (fig. S09 above) or is not (fig. S07 above) inserted into the external auditory conduit (earplug interposed between the sound source and the eardrum, but not between the sound source and the mastoid). The earplugs used were "*quies*", made from natural wax, with an assumed protection value reported by the manufacturer of 27 dBs[61] (Noise Reduction Rate) for the subject D37 (Table S02**a**: Fig.S10), subject D39 (Table S02**b**): Fig.S11), subject D44 (Table S02**c**; Fig.S12).

| | Mastoïdo-frontal potential [Vs (µV)] without earplug ("open") or with earplug (obstructed) | | | | | | | | |
|---|---|---|---|---|---|---|---|---|---|
| Table N° | **a** (D37, 51 years old) | | | **b** (D39, 17 years old) | | | **c** (D44, 25 years old) | | |
| Figure N° | S10 | | | S11 | | | S12 | | |
| f (kHz) | **Open** | **closed** | Δ | **Open** | **closed** | Δ | **Open** | **closed** | Δ |
| 0,2 | 0,2 | 1,7 | -1,5 | 1,7 | 2,2 | -0,5 | 12,5 | 6,2 | -6,3 |
| 0,5 | 0,5 | 1,7 | -1,2 | 1,7 | 2,2 | -0,5 | 10,3 | 6,6 | -3,7 |
| 1 | 2,2 | 2,8 | -0,6 | 10 | 12 | -2 | 12 | 9,5 | -2,5 |
| 2 | 2,5 | 3,7 | -1,2 | 11 | 19 | -8 | 22 | 14 | -8 |
| 4 | 3,3 | 5,2 | -1,9 | 17 | 29 | -12 | 37 | 24 | -13 |
| 8 | 5,6 | 7,2 | -1,6 | 50 | 50 | 0 | 55 | 46 | -9 |
| 10 | 6,5 | 8,5 | -2 | 55,2 | 64 | -8,8 | 73 | 56 | -17 |
| 15 | 8,3 | 10,4 | -2,1 | 95,2 | 90 | 5,2 | 67 | 70 | 3 |
| 20 | 10,4 | 11,3 | -0,9 | 139 | 120 | 19 | 80 | / | na |
| **Mean (µV)** | 4,7 | 5,9 | **-1,4** | 53,9 | 54,9 | **-1** | 29 | 36,1 | **-7,06** |
| Table S09 (**a**, **b**, **c**) Synchronous mastoid voltage (µV), broadcasted by a loudspeaker Harman/Kardon, stimulated by a voltage of 2 Volts issuing from the lock-in amplifier; Δ = Vs *with (closed)* minus Vs *without (open)* | | | | | | | | | |

| | |
|---|---|
| NN. | OO. |
| PP. Figure S15<br>**a** (D37, 51 years old).. | QQ. Figure S16 (Ord. Volts)<br>RR. **b** (D39, 17 years old) |

Figure S17
**c** (D44, 25 years old)

The three figures (S10, S11 and S12) show that the mastoid potential with the ear plugged was very close to the mastoid potentials of the same ear when not plugged. Furthermore, regarding the relation between the with earplug and without earplug representative lines, there are three configurations. In the case of D37 (fig. S10), the magnitude of the "mastoid potential" with airplug (earplug "quies", pure wax 27 dB) is above that of the without airplug situation. In the case of D39 (fig. S11), the magnitude of the "mastoid potential" "with" airplug is intertwined with that of the "without" situation. In the case of D44 (fig. S12), the magnitude of the mastoid potential "with airplug" is below that of the "without" situation.

However, when the ear is clogged, the signal that reaches the cochlea is necessarily reduced (standard theory) and therefore the cochlear potential should be weakened as well, in all cases! Thus, as logical consequence, the mastoid potential we recorded cannot depend solely upon the cochlear microphonic. Moreover, our measures are consistent with the hypothesis that the mastoid potential depends not only on the impact of sound on the eardrum, but also on the mastoid reaction. So we think that the 'mastoid potential' results, at least in part, from the electrical signal generated by the piezoelectricity of collagenous fibres, located between the points of application of the electrodes (mastoid and forehead). According to our theory linked to the standard interpretation, this piezoelectric response is added to the electric signal from the cochlea.

**Si 05Ec Experiment using 'active noise cancelling headphones' plus 'custom moulded earplugs' (formed in place)':**

However, on the three cases above, the occlusion of the external auditory canal, by means of a standard over-the-counter earplug ("quies", pure wax 27 dB) was, as such, limited to a maximum of 27 dB HL. So, the effects obtained with "earplugs" should be even more marked with "custom-made earplugs (fig. S13) + active noise cancelling headphones (fig. S14 & fig. S15)". We, therefore, repeated the previous experiments, using active noise cancelling headphones in the context of the situation 'occluded ear'. We filled the external auditory canals of one subject with custom-made earplugs (Audial, Toulouse, Fig. S13), completed by active noise cancelling headphones, (Panasonic RP - HC 500; Fig. S14); Together : fig. S15.

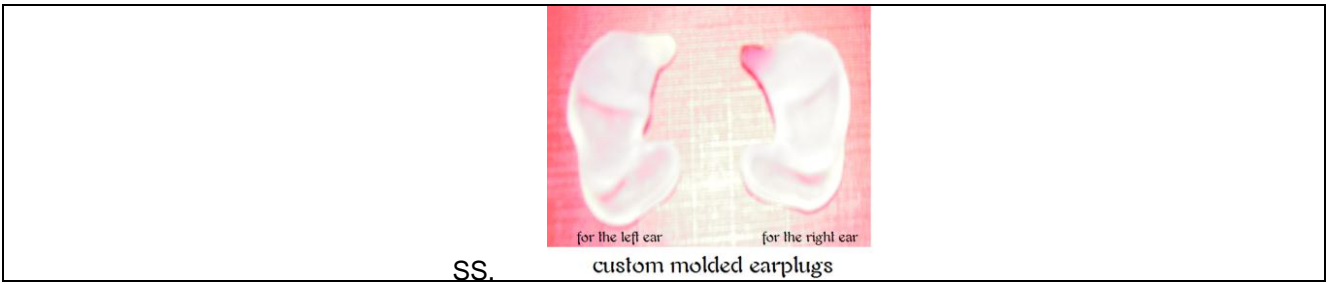

Figure S18
custom-made earplugs

We measured the audiometric thresholds using only custom-made earplugs, then using the active noise cancelling headphones alone (fig. S14), then using both (fig. S15). We used the same placement of the electrodes as in the previous experiments. The previously positioned electrodes, held in place by surgical tape, are pressed by the subject himself, the potential measured depending on the pressure on the skin. The Yamaha HS5 speaker is installed at 1 m from the ear of the subject, its centre facing the subject at the height of the ear to be tested. We performed a calibration of acoustic amplitude with a *Bruel and Kjaer 2231* sound level meter, the rod bearing the microphone being perpendicular to the face of the speaker. The voltage of the output of the Lock-in, plugged into the speaker is set so that the sound level meter displays about 80 dB.

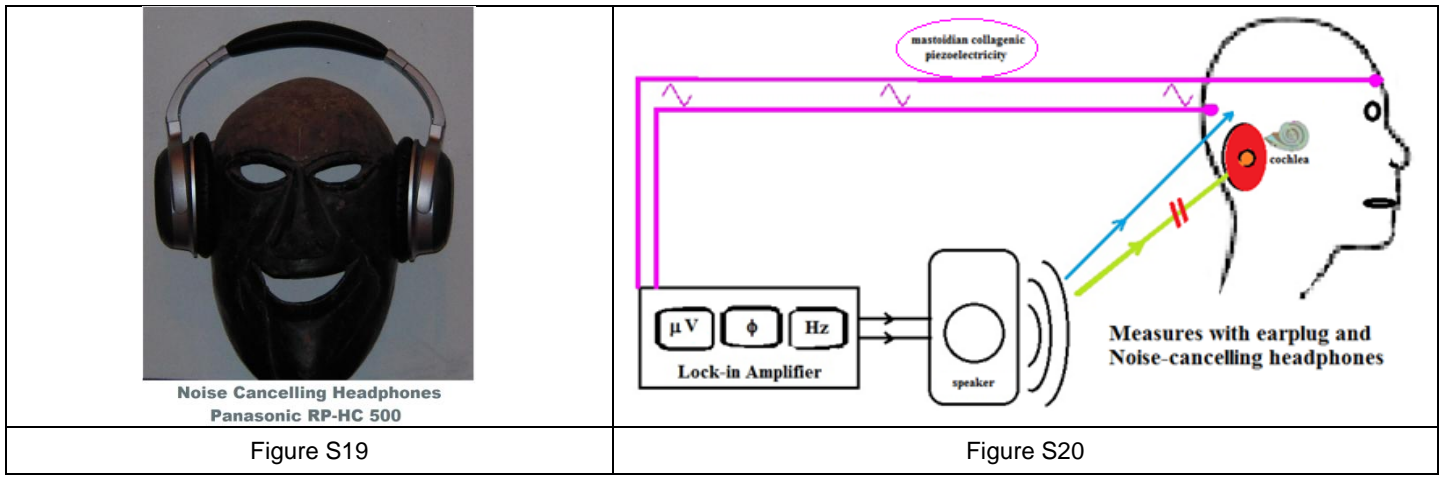

| Figure S19 | Figure S20 |

We present below (subject D38, Table S03, Fig. S16), the voltage output of the Lock-in amplifier and measurements of the synchronous electrical potentials, with (Plugged: P) and without (Unplugged: U) occlusive equipment. The data are displayed according to the sound frequency (f kHz) emitted by the speaker, either toward the right auditory canal (REAC) or toward the left one (LEAC).

| f (kHz) | Output voltage (to the speaker) | Vs (μV) U-REAC | Vs (μV) P-REAC | Vs (μV) U-LEAC | Vs (μV) P-LEAC |
|---|---|---|---|---|---|
| 0.5 | 0.4 V | 5 | 5 | 6 | 6 |

| 1 | 0.3 V | 6 | 7 | 5 | 7 |
|---|---|---|---|---|---|
| 2 | 0.5 V | 6 | 5 | 5 | 6 |
| 4 | 0.2 V | 9 | 7 | 8 | 7 |
| 8 | 0.2 V | 18 | 15 | 13 | 14 |
| 16 | 0.4 V | 44 | 42 | 45 | 43 |
| 20 | 1 V | 123 | 96 | 110 | 105 |
| 32 | 1 V | 144 | 143 | 170 | 145 |

Table S10
External Auditory Canal (EAC); subject D38,
Right (R), Left (L), Unplugged (U), Plugged (P)

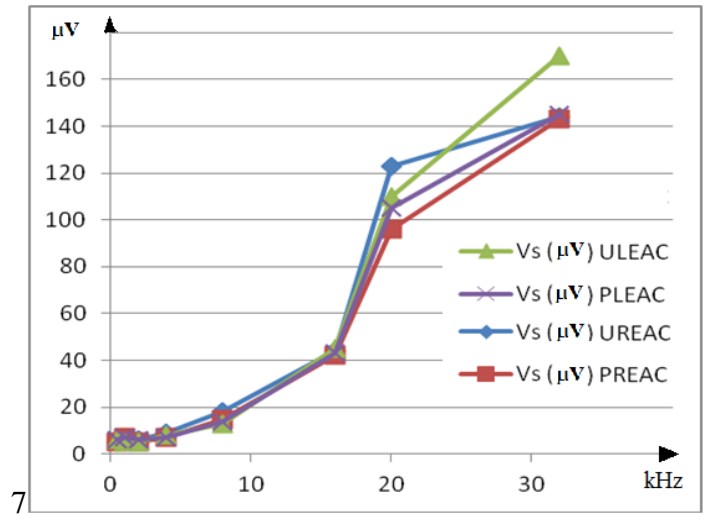

Figure S21

We found that the mastoid potentials with the ear plugged was very close to the mastoid potentials of the same ears when not plugged. As the mastoid potentials are not greatly weakened by plugging the EAC, the classical theory seems to be insufficient and we think that another phenomenon coexists. We hypothesize that this phenomenon is the piezoelectricity of the mastoid collagen. We, therefore, propose the two diagrams below (Fig. S17) in which we have added the effect of the piezoelectricity of the mastoid collagen (blue line) to the classical schema (green line) for purposes of comparison.

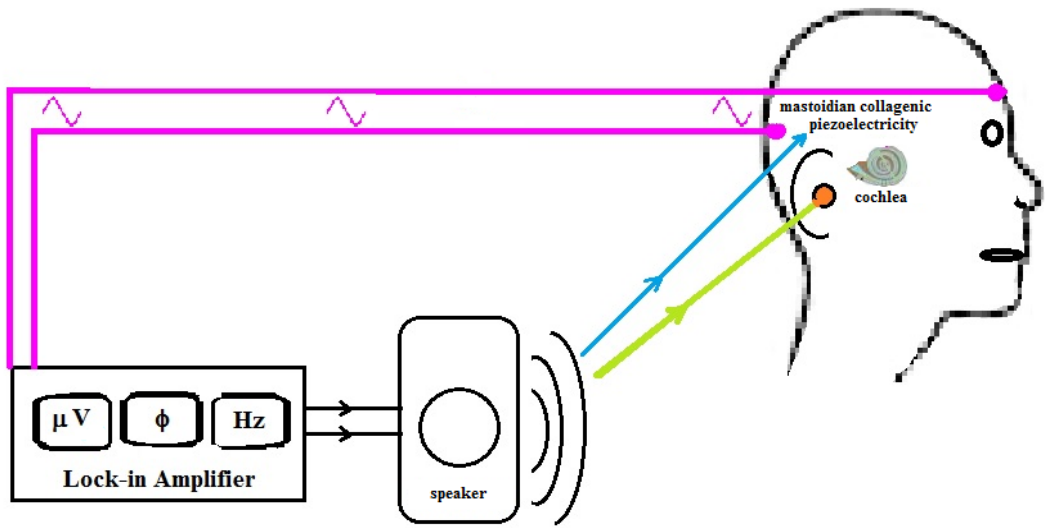

TT. Figure S22
Stimulation restricted to the external auditory conduit (stimulation by headphone only) compared with stimulation targeting the mastoid (stimulation by loudspeaker only)

**Si 05F Conclusion on Si 05 :**

If one puts an *earplug* (in red) or/and *standard headphones* and/or "*active noise cancelling headphones*" on the ears of the subject, the acoustic vibrations (in green) encounter this obstacle which necessarily weakens the vibrations of the eardrum; Thus, in the accepted theory, the cochlear generator will produce a synchronous potential of a lower amplitude. The result would be that the synchronous electric potential measured at the level of the skin behind the ear (µV) should logically be reduced. Since it does not, we conclude that the cochlea is not the only generator of this potential.

Collagenous fibers of the mastoid region have roughly the same structure as the fibers belonging to the patellar ligament. It follows that the collagenous fibers of the mastoid are able to generate a piezoelectric potential (synchronous to the acoustical vibrations) just like the potential we measured at the level of the patellar ligament. Acoustic vibrations produced by the speaker (blue arrow) stimulate an electric non-cochlear generator, i.e. the piezoelectric collagen fibers of the mastoid.

Plugging the ear canal should not only decrease very sharply the cochlear microphonic, but it should also decrease the petrous resonances caused by the vibrations of the eardrum ! It is possible that the slight difference in favor of the non-blocked ear comes, in part or entirely, from the decrease in the phenomenon of resonance when the ear is plugged.

We may conclude that our measures are consistent with the hypothesis that the mastoid potential depends not only on the impact of sound on the eardrum, but also on the mastoid structures. We think that the 'mastoid potential' results mainly from the electrical signal generated by the piezoelectricity of collagenous fibres located between the points of application of the electrodes (mastoid and forehead).

## Si 06 Source of the mastoid synchronous potential

A diagram of the reception of the external synchronous potential, with the mastoid piezoelectricity eliminated, is given on the Figure S18: the sound is not emitted by a loudspeaker but only through a headphone.

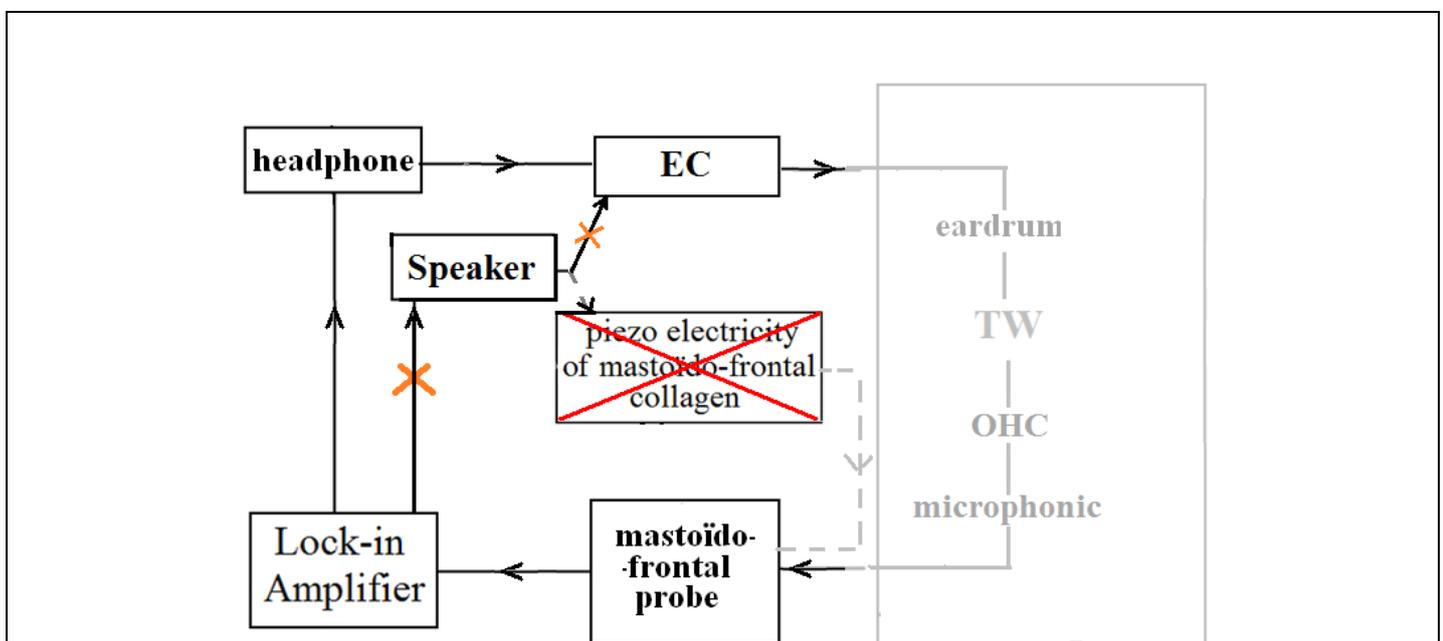

**Figure S23**
Stimulation restricted to the external auditory conduit (stimulation by headphone only) compared with stimulation targeting the mastoid (stimulation by loudspeaker only)

If the mastoid synchronous potential originated from the cochlea alone, then it would be about the same no matter whether the sound was broadcast through a loudspeaker or through the headphone.

If, on the other hand, the mastoid synchronous potential originates locally (at the mastoid), then the sound emitted by the headphone would produce a very low synchronous potential compared to the same sound broadcast through a loudspeaker.

## Si 06A Testing hypotheses: purely tympanic way vs. mixed path

The standard theory states that, if the sound reaches the eardrum without reaching the surface of the mastoid, the synchronous potentials will not weaken (Hypothesis H0). The alternative hypothesis (H1) states that the synchronous potentials will weaken. We checked H0 versus H1.

To allow comparison of stimulation through the auditory canal alone and stimulation by the loudspeaker (free field), we replaced the loudspeaker by a headphone and we fed it in such a way that the subjective magnitude would be roughly the same as with the loudspeaker (prior calibration).

For predetermined frequencies (200 Hz, 1 kHz, 8 kHz, 15 kHz), we provided an operatic singer (D40) with a reference acoustical amplitude by means of a loudspeaker placed at approximately 1 m from his ear. Then, we sent the same frequency by headphone (audio InterSound HD75, without volume attenuation) fed from the output of the lock-in amplifier. We asked the singer to modify the electrical voltage generated at the output of the lock-in amplifier, connected to the headphone, until he got the same subjective impression as with the loudspeaker (see results Table S04).

| Frequency (kHz) sent by the lock-in amplifier | Voltage (mV) sent by the lock-in amplifier to the loudspeaker |
|---|---|
| 0.2 | $5.10^3$ |
| 1 | $3.10^3$ |
| 8 | 104 |
| 15 | 200 |

Table S11
Calibration: D40, a professional operatic singer, is stimulated by a pure frequency tone, in turn via the headphone and via the loudspeaker; owing to the voltage from the lock-in amplifier, successive settings amplify these two transducers, in such a way that a subjectively identical amplitude of sound is fixed for different frequencies

But for his subjective convenience, we did not impose to D40, an identical standard amplitude for every frequency. For each of the tested frequencies, we allowed him to fit the respective amplitudes of the headphone and the loudspeaker, each in turn, by trial and error, so that the heard amplitude would be as much as possible identical for both, without aiming exactly the level proposed at the beginning.

Using the results of this calibration, the subject received an equivalent stimulation for each preselected frequency in turn, via either the headphone alone (situation A) or the loudspeaker alone (situation B). According to Hypothesis $H_0$, the mastoid synchronous potential comes from the cochlear microphonic. Consequently, the measured synchronous potential should be lower in situation B (loudspeaker only) than in situation A (headphone only).
According to Hypothesis $H_1$, the mastoid synchronous potential comes mainly from the mastoid collagen piezoelectricity. In this case, the measured synchronous potential would be expected to be lower in situation A (headphone only) than in situation B (loudspeaker only).

We present the results in Table S05 for D40 and for two other subjects (D41 and D42) in Table S06 for D41, and in Table S07 for D42. In the tables, a dash indicates that no synchronous potential could be measured: absence or extremely low amplitude.

| D40 | Situation A Headphone (µV) | Situation B Loudspeaker (µV) | $H_1$ vs $H_0$ |
|---|---|---|---|
| 0.02 kHz | - | 4.6 | $H_1$ supported |
| 1 kHz | - | 3.6 | $H_1$ supported |
| 8 kHz | - | 0.27 | $H_1$ supported |
| 15 kHz | 3.7 | 1.2 | $H_0$ supported |

Table S12 (D40)

| D41 | Headphone (µV) | Speaker (µV) | $H_1$ vs $H_0$ |
|---|---|---|---|
| 0.02 kHz | - | 20 | $H_1$ supported |
| 1 kHz | 3.7 | 16 | $H_1$ supported |
| 8 kHz | 6.5 | 2.5 | $H_0$ supported |
| 15 kHz | 3.3 | 6.1 | $H_1$ supported |

Table S13 (D41)

| D42 | Headphone (µV) | Loudspeaker (µV) | $H_1$ vs $H_0$ |
|---|---|---|---|
| 0.02 kHz | - | 180 | $H_1$ supported |
| 1 kHz | 1.8 | 440 | $H_1$ supported |
| 8 kHz | 1.8 | 36 | $H_1$ supported |
| 15 kHz | 11.5 | 90 | $H_1$ supported |

Table S14 (D42)

## Si 06B Conclusion about the source of the mastoid synchronous potential :

Hypothesis H1 is borne out ten times on twelve measures. In more detail:

- Hypothesis H1 is clearly supported with regard to frequencies 0.02 and 1 kHz. For these frequencies, the synchronous mastoid potential seems, for the most part, to be produced by the mastoid region collagen fibres and, therefore, does not come from the cochlear microphonic...
- Only D40 at 15 kHz, and D41 at 8 kHz, do not support our theory.
Furthermore, the specific results observed for frequencies 8 kHz and 15 kHz can be explained by the fact that vibrations imposed on the eardrum should be transmitted to the mastoid by resonance..
Finally, it should be noted that D40, involved in the calibration, had characterized his subjective assessment regarding 15 kHz as a very rough approximation.
- But note also that hypothesis H1 is fully validated for all measurements in the case of D42, whose potentials are all very high, and who has, in addition, high-performance hearing thresholds and a cut-off frequency in the high pitched sounds above 22 kHz (which is largely above the audiometric standards).

Even if error bars are higher than the current estimate of half a unit. it remains that the standard theory (H0) is largely invalidated by our experiments.
This experiment suggests that the mastoid synchronous potential disappears or weakens if vibrations mobilize the eardrum rather than the surface of the mastoid. This being the case, we must add to the effect of the piezoelectricity of the tympanic fibres, the effect, though somewhat lesser, of the piezoelectricity of the mastoid fibres.

Moreover, we have demonstrated that the synchronous potentials, measured at the level of the mastoid, are not reducible to the cochlear microphonic in its classical conception, according to which they are solely the result of the mechanical TW, amplified by the OHCs of the cochlea. An important part of the mastoid synchronous potential has a local origin: it derives from the piezoelectricity of the collagen (type I) of the mastoid.

Of course, our measurements are different from the audiometric classic ones, but an attentive examination of both for comparison would be interesting to do$^{S_{61} - S_{61}}$.

## Si 07 The pT voltages are they able to reach the DOHC complex ?

Our measures of pT may seem of a too low tension to act on the cochlea; but we must take into account the fact that the result of our measures is diminished by the interposition between the probe and the source of the electric potential of an more or less insulating layer, the epidermis of the eardrum. When we take into account the weakening of conduction by the epidermal tympanic layer, these values seem consistent with our measurements on the eardrum in vivo (hundreds of microvolts for an acoustical stimulation of approximately 70 or 80 dB SPL).

Furthermore, Harnagea showed in vitro that the response is low or non-existent if the electrodes are arranged in any two points of the biological culture, but this response reached tens of mVolts if the electrodes are placed at both ends of a single fiber... That was confirmed by Denning et al. 2017[s61-s61].

Thus we can assume that the pT voltages are in the mV range and may reach the DOHC complex.

# Si 08

## Si 08A The mystery of the cetacean auditory mechanism

The minimum intensity detectable by mammals is a function of the frequency. It is noteworthy that they show a weaker sensitivity to very low and very high frequencies according to an U-shaped pattern. Cetaceans, most notably the small odontocetes, possess extraordinary auditory faculties, (...). They make extensive use of sound in echolocation as well as communication behaviours. They have a U-audiometric curve, typical of mammals[s61]. In all species tested, the maximum sensitivity is at frequencies greater than 15 kHz.

The greatest sensitivity of cetaceans lies between 30 and 80 kHz. Their sensitivity decreases between 40 kHz and 1 kHz, and also between 80 kHz and 150 kHz (cf. Table S08).

| Species | Best sensitivity | Min frequency | Max frequency |
|---|---|---|---|
| Beluga whale | 30 kHz | 20 kHz | 120 kHz |
| Killer whale | 40 kHz | 0.5 kHz | 100 kHz |
| Tursiops spp | 50 kHz | 2 kHz | 135 kHz |
| Inia geoffrensis | 80 kHz | 1 kHz | 105 kHz |
| Table S15 | | | |

With respect to the frequency selectivity capabilities of cetaceans (for ex. T. truncatus) they are comparable to humans at lower frequencies, and the best reported for any mammal above 20 kHz. Moreover, the smaller toothed whales, in particular, appear to combine high temporal resolution with extremely sharp frequency tuning. (...)

"There are very short response latencies and an extremely rapid conduction of the information through the auditory pathway. The results describe an auditory system that is highly specialized for the extremely rapid conduction of auditory information from the periphery to more central structures"[S61].

The external auditory canal of cetaceans is, , vestigial, although there may be an opening (very small)[S61]. Their eardrum is an extended, thickened membrane, the function of which is unclear[S61]. Ossicles are non-existent or hindered by additional or reinforced collagen ligaments, very stiff and heavier in structure (more massive, denser, ... etc.). The attachment of the stapes at the oval window of the cochlea is very tight and seems rigid in postmortem specimens[S61]. Although the remains of the outer ear and the middle ear are detectable in odontocetes, it is not possible to attribute to them the recognized role in terrestrial mammals.

The attachment of the stapes at the oval window is very tight and seems rigid[61]. Although remains of the outer ear and the middle ear are detectable in odontocetes, it is not possible to attribute to them the same role as which recognized in terrestrial mammals.

______________________

The transmission of acoustic vibration from the surrounding sea water toward the cochlea is thought to result from a "fatty transmission" via perimandibular "acoustic fat bodies". These acoustic fat bodies are in direct contact with the malleus ossicle of the ears, especially in odontocetes but also in some mysticetes.

In addition, fibroblasts by inserting themselves into the network of collagen fibres, are also suitable for turning into adipocytes[61], the main constituents of the lipid acoustical pathway.

The role of the eardrum and these fatty acoustic bodies are very similar ; They provide double transduction. Acoustic vibrations mechanically mobilize ossicles or their remains; This phenomenon is useful for the perception of low acoustic frequencies. Ear fat attaches to the tympanoperiotic complex, inserting itself into the "triangular opening", at the entry to the middle ear, where the ear fat contacts the hammer. Thus, the "acoustic signal" transmitted by the acoustic fat bodies, like the tympanic vibrations of terrestrial mammals reaches the hammer.

- Since the fibrous collagens are piezoelectric electrets, we may deduce that this structure is organized enough, like the eardrum of terrestrial mammals, to produce electrical potentials isomorphic to the surrounding acoustical waves; In a similar way, these potentials would be transmitted by GJs to the DOHC complex, and, thus, reach the phalanx of Deiters cells (i.e. the gate of the TkS). This hypothesis explains the mysterious functioning of the whales' hearing as well as the extremely high frequencies heard by them.
- Yet there seems to be a drawback: how the electric potential could to be born, and then reach the DOHC complex, if every parts of the circuit would be bathed into a conductive medium (the sea salted water) ? Maybe, the solution is into the question : the collagen fibres are not in contact with the conductive salted water, but with the so called "acoustic fat bodies" which are electrical insulators. As such, their fatty structure would be needed by the sea medium, not essentially as a result of their acoustical properties, but instead due to their electrical and thermic insulator quality.

It is admissible that the electrical potentials resulting from the mobilisation of collagen content of the same fat bodies would be transmitted to the surrounding bone, and be conducted through GJs to the stereociliae of the OHCs as well as to the cup and phalanx of Deiters Cells. The large mysticetes, apart from their ultrasonic abilities and their use of echolocation, are able to hear bass sounds and even infra-sounds. The transmission of acoustic vibration from the surrounding sea water toward the cochlea is thought to result from a "fatty transmission".

This paradoxical data set may find the beginning of a solution, if we add an electrical "covert path" to that purely mechanical transmission. This fatty pathway is made of a very complex histological structure of adipocytes and a dense network of collagen fibres[S61]. This covert path would be similar to the one we describe between the collagen fibres of the eardrum and the DOHC complex, including the TkS. This system does not encompasses any operative tympanic structure. Yet, the tympanic ligament (homologous to the ancient tympanic membrane) attaches to the sigmoid process, a portion of the tympanic ring[S61].

It is accepted that there is a particular type of transmission, via perimandibular "acoustic fat bodies". These acoustic fat bodies are in direct contact with the malleus ossicle of ears, especially in odontocetes[S61], but also in some mysticetes [S61]. It is interesting to note that these acoustic fat bodies have an acoustic conductivity very similar to that of the sea water; Even so they are given the role of an 'acoustic lens' in the sound emission melon of delphinidae.
The structure of these "acoustic fat bodies" is very complex. They consist of layers of which each one has a specific acoustic conduction. These layers are, of course, generated by adipocytes and fibroblasts whose role must be coherent with measurement of specific "acoustic fat bodies" conductivity, which we have just evoked. Fibroblasts are the obvious source of the "dense network of collagen fibres" which are interlaced with adipocytes and fatty bodies, specific to each layer. These collagen fibres are piezoelectric like the collagen fibres in general, regardless of their type (T I, II, ...). These piezoelectric variations of the collagen network might be transmitted to the OHCs by the GJs of the cochlea. It is very probable that this last stage of the "covert path" be very fast and very complex (the ganglion cells are much more numerous in the cochlea of marine mammals).

It is interesting to note as well that the eardrum membrane of terrestrial mammals is essentially the result of the action of fibroblasts, which are generators of collagen fibres. Otherwise, fibroblasts by inserting themselves into the network of collagen fibres, are also suitable for turning into adipocytes[61], the main constituents of the lipid acoustical pathway.

Ear fat attaches to the tympano-periotic complex, inserting itself into the "triangular opening", at the entry to the middle ear, where the ear fat contacts the hammer. Thus, the "acoustic signal" transmitted by the acoustic fat bodies reaches the hammer, like the tympanic vibrations of terrestrial mammals.
The roles of the eardrum and these fatty acoustic bodies are very similar: they provide double transduction:
- Acoustic vibrations mechanically mobilize ossicles or their remains; This phenomenon is useful for the perception of low acoustic frequencies.
- Since the fibrous collagens are piezoelectric electrets, we may deduce that structure is organized enough, like the eardrum of terrestrial mammals, to produce electrical potentials isomorphic to the surrounding acoustical waves; In a similar way, these potentials would be transmitted by GJs to the DOHC complex, and, thus, reach the phalanx of Deiters cells (i.e. the gate of the TkS). This hypothesis might explain the mysterious functioning of the whales' hearing as well as the extremely high frequencies heard by them. Furthermore, cetaceans and bats use both the protein which fulfils the role of drain in the TkS (prestine). Our description of the covert path should also be applied analogously to the hearing of mammals endowed with significant ultrasonic abilities (bats and sea mammals).

## Si 08B The human fetus is in a similar situation to marine mammals.

The mesenchymal environment of the ossicles present in the middle ear of the foetus is an obvious obstacle to their correct functioning.

"The early stages of auditory ossicles development all occur within the solid mesenchyme of the pharyngeal arches until the eighth month of development[S61], then within a fluid-filled space for the final month, and finally only postnatal in the air-filled tympanic cavity of the neonate". This transition in auditory ossicle environment means that the middle ear does not function correctly until after birth"[S61]. This makes it difficult to understand the existence - though well demonstrated - of fetal hearing from the 22$^{th}$ amenorrhea week[S61] - [S61]. So Hill concludes that "any prenatal conduction to the cochlea must be mediated through bone conduction".

However, the mechanism of bone conduction is itself poorly understood, so that the explanation of foetal hearing by bone conduction would simply move the problem on. It therefore seems that another conceivable mechanism would be the "covert path" as could be the case for sea mammals as well ....

Furthermore the frog "Sechellophryne Gardineri" is able to hear without eardrum, even vestigial, and its hearing organ is just its mouth[S61].

## Si 09 Electrical pathway

### Si 09A Electrical links of the eardrum to the cochlea

The superior ligament of malleus (K), made of collagen I, attaches the head to the tegmen tympani. The SLM may electrically connect the central structure of the eardrum to bone conductive fibres and to the ground (the electric common return). Peripheral ends of the radial collagen fibres of the eardrum are connected to the osteocytic GJs network.

The eardrum consists of four layers (fig. S19): the epidermis (B), two fibrous collagenic layers (AA'), and an internal mucous membrane (C). The three ossicles are the stapes (F) with its muscle (J), the uncus (E), and the malleus which consists of three parts: the head (D), the lateral process (H) and the manubrium (I). The malleus can be mobilized by its muscle, the tensor tympani (G) which, when enabled, pulls the malleus medially, tensing the eardrum and damping its vibration.

| 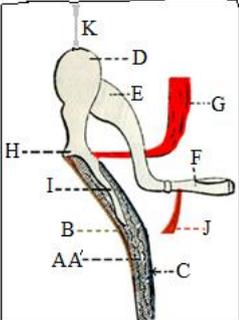 | 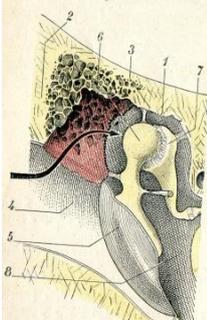 |
|---|---|
| Figure S24 | Figure S25 |
| Section of the Eardrum through the handle of the malleus[S61]: Schema of the Superior Ligament of the Malleus (SLM) ensuring the connection of the umbo to the electric common return (K). | Ligaments of the Malleus are inserted into the attic; The Superior Ligament of Malleus (SL M(3)), is inserted in the tegmen tympani[S61]. |

The superior ligament of malleus (K), made of collagen I, attaches the head to the tegmen tympani. The SLM may electrically connect the central structure of the eardrum to bone conductive fibres and to the ground (the electric common return). Peripheral ends of the radial collagenic fibres of the eardrum are connected to the osteocytic GJs network (fig. S19 and S20).

## Si 09B The DOHC complex

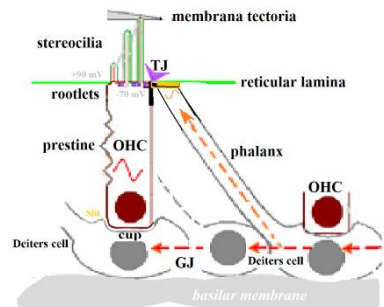

Figure S26
Schema of the DOHC complex
Stereocilia are implanted on the apex membrane of OHCs (cuticular plate / reticular lamina).
They are bathed in endolymph and mobilized by its movements (TW).

# Si 10 Trickystor

### Si 10A **Source : Ionic channels and time constant**

The question of the reality of the contractions at high frequencies of the walls of the OHC in physiological context is poorly understood;the cycle polarization - depolarization created by the movement of the ciliary tuft, induced by the sound, in fact involves the electric plasma membrane time constant; the value of this time constant is not compatible with frequencies of contraction up to 20 kHz in humans and 150 kHz in some bats[S61]. A persistent difficulty in understanding the role of electromotility in amplification is the membrane time constant; the resistance and capacitance of the membrane of an OHC cuticle would suggest that a voltage-driven process should diminish in sensitivity above a corner frequency of only 1 kHz[S61 - S61].
But, the stereociliar signal can trigger prestin electromotility at higher frequencies, provided that it is combined with a second stimulation, whose effects would not be subject to the membrane time constant[S61 - S61].
In reality, such a potential is recordable in the gap (slit) separating the DC cup from the base of the OHC[S61- S61 - S61 - S61 - S61 - S61 - S61 - S61] (Cf. Fig. S21, DOHC5: « cup »). For high frequencies (> 8 kHz), variations of potential of the cup are rapidly transmitted to the gap by means of hemi-channels made of Cx26 and Cx30[S61 - S61]; This being the case, the prestin could contract at the same frequencies[S61 - S61]. It is also possible that variations in voltage of the Deiters cup would produce a capacitive effect, through the Deiters membrane, the cupular gap and the membrane of the base of the OHC.

This extracellular cupular signal has been attributed to electrical stereociliar phenomena, and this attribution has caused it to be confused with the "cochlear microphonic potential ". But if we refer to the prevailing opinion, according to which the microphonic potential is due, for the most part, to the mechano-electrical activity of the prestin, we are faced with a vicious circle: The contractions/extensions of the prestin would produce the microphonic and this microphonic, then, is invoked to explain the phenomena of contraction / extension of the prestin!

However, this explanation might be relevant if the initial origin of the extracellular potential were attributed to the non-OHC part of the microphonic, i.e. the pT. A fact supporting this hypothesis is the following: The DC cups themselves are the seat of intracellular[S61] voltage variations (AC) with the same frequency and amplitude as the extracellular voltage[S61]. Variations in the intracellular voltage of the cupule of Deiters could be the result of the pT and, as such, the source of that extracellular potential, instead of its consequence[S61].

The hypothesis of the trickystor, with its FET features, might indeed overcome this problem. The transconductance of an FET is not submitted to a low pass limitation; it is expressed by the ratio $g = I_{DS}/U_{GS}$ ($U_{GS}$ is the Gate-Source voltage. $I_{DS}$ is the Drain-Source current). The higher g is, the greater the gain of the transistor is. This gain allows it to cancel the intensity decay which would otherwise occur because of the time constant. Its high input resistance and its low input capacitance give to a FET characteristics similar to those of a triode. The interest of such an FET lies in its excellent selectivity and its low noise factor (narrow bandwidth). which enables it to play a role of preamplifier and/or oscillator

## Si 10B The cuticular plate bilayer

Like all cellular membranes, the cuticular membrane is a lipid bilayer (fig. 03) which consists mainly of amphiphilic lipids (generally phospholipids). They have one head group that is hydrophilic ('polar') and two hydrocarbon tails that are lipophylic ('non polar'). One tail typically is saturated, while the other tail is unsaturated with one or more cis-double bonds (conjugated or homoconjugated). Each cis-double bond creates a small kink in the tail. By forming a double layer, with the polar ends pointing outwards and the non-polar ends pointing inwards, membrane lipids form a 'lipid bilayer' which keeps the watery interior of the cell separate from the watery exterior.

The "lipid bilayer" plays a dual role in the life of the cell: both as insulator and filter.
a - Its insulating lipid molecules, arranged in a 5 to 10 nm thick bilayer, form an impermeable barrier to the passage of most water soluble molecules. They block the passage of inorganic ions (K +, Na +, Cl-, Ca2+, ...) and hinder the diffusion of polar organic solutes such as amino acids.
b - It is a filter as well: protein molecules regulate transmembrane exchanges;within itself, the membrane is permeable only to small hydrophobic molecules (O2, N2, glycerol, ...).

## Si 10C The structure of the phospholipid bilayer is a liquid crystal (of smectic type).

The findings from the neutron diffraction work suggest that the cholesterol molecule stays at the center of the bilayer and is constrained to a maximum displacement inferior to 6 Å relative to the cellular body as a whole[S61]. The sterols show a strong preference for embedding themselves between bilayer leaflets, which is an unusual placement for them [S61].

Yet cholesterol has a poor affinity for phospholipids containing PUFAs and that enhances its mobility. This may affect the apical, as well as the lateral or basal, distribution of cholesterol. OHCs have an inhomogeneous distribution of free cholesterol with higher concentrations in the basal and apical membranes and lower levels in the lateral wall ([S61] - [S61]).

There are also quantitative differences in lipid lateral mobility of cholesterol, and rafts, among the apical, lateral, and basal regions of the OHC (Fig. S27).

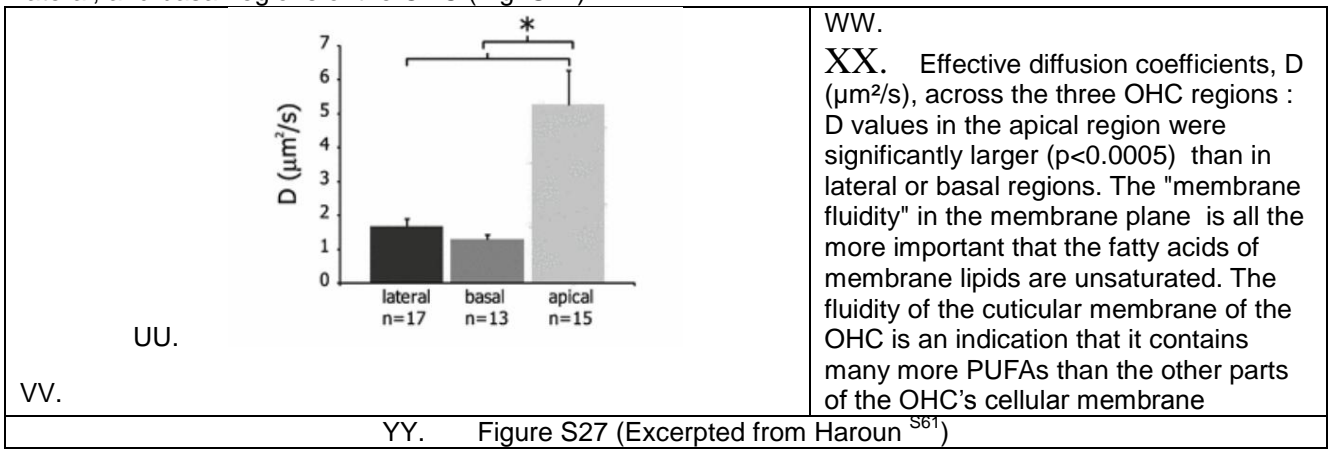

UU.
VV.

WW.
XX. Effective diffusion coefficients, D (µm²/s), across the three OHC regions : D values in the apical region were significantly larger (p<0.0005) than in lateral or basal regions. The "membrane fluidity" in the membrane plane is all the more important that the fatty acids of membrane lipids are unsaturated. The fluidity of the cuticular membrane of the OHC is an indication that it contains many more PUFAs than the other parts of the OHC's cellular membrane

YY. Figure S27 (Excerpted from Haroun [S61])

.

## Si 10D Semi-Conductors of the TkS

The hearing is impaired by genetic disorders of the conjugated and homoconjugated (fig. S24 below) PUFAs semiconductors (Role of peroxisome)[S61]. Peroxisomes are organelles found in all eukaryotic cells. There are at least 32 known peroxisomal proteins that participate in the process of peroxisome assembly. Their deficiency gives impairment hearing of high frequencies and they are involved in catabolism of very long chain fatty acids, branched chain fatty acids, D-amino acids, and polyamines, reduction of reactive oxygen species – specifically hydrogen peroxide – and biosynthesis of plasmalogens, i.e. ether phospholipids.
Peroxin (or peroxisomal/peroxisome biogenesis factor) is a protein found in peroxisomes.

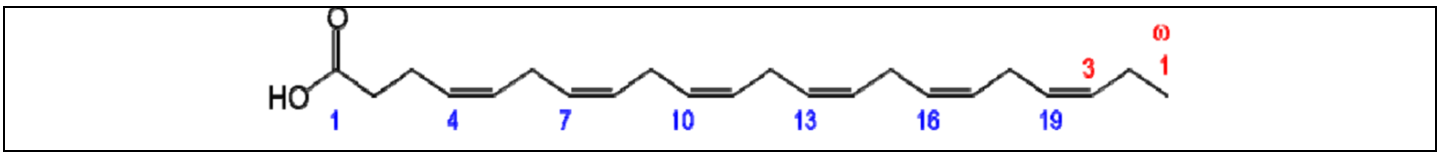

| |
|---|
| Figure S28 |
| DHA is an example of long chain homoconjugated fatty acid. |

## Si 10E **The electrical as well as mechanical state of DCs drives the electromotility of the OHCs**

The separation of a DC phalanx from a related OHC cuticle, or of a cupule from the OHC that it supports (or the elimination of GJs between the bodies of Deiters cells) results in deterioration of the electrical behavior of the concerned OHCs (Yu & Zhao, 2009). There being no electrical synapse (GJ) between DCs and OHCs, it has been proposed that this electric effect on the OHC is due not to an electrical interruption between DCs and OHCs, but rather to an interruption of a purely mechanical nature.
Stimulation of the DCs, either electric or mechanical, can modulate the electromotility of the OHCs (Yu&Zhao, 2009) and in vivo active cochlear amplification depends on GJs between the DCs. Cx26 expression in the cochlear supporting cells plays a critical role in active cochlear amplification and its targeted deletion can eliminate active cochlear amplification.
 It is coincident with a large reduction in distortion product otoacoustic emission (DPOAE) and severe hearing loss at high frequencies (changes are greater in the shortest OHCs).

## Si 10F **Gate ; Field effect of Deiters Cells on the cuticles of OHCs**

The **tunnel field-effect** transistor (TFET) is an experimental type of transistor. Even though its structure is very similar to a metal-oxide-semiconductor field-effect (MOSFET), the fundamental switching mechanism differs, making this device a promising candidate for low power electronics. TFETs switch by modulating quantum tunneling through a barrier instead of modulating thermionic emission over a barrier as in traditional MOSFETs. Because of this, TFETs are not limited by the thermal Maxwell–Boltzmann tail of carriers, which limits MOSFET drain current subthreshold swing to about 60 mV/decade of current at room temperature (exactly 63 mV/decade at 300 K[1]). The concept was proposed by Chang et al while working at IBM [2]. Joerg Appenzeller and his colleagues at IBM were the first to demonstrate that current swings below the MOSFET's 60-mV-per-decade limit were possible.

According to Yu and Zhao (2009), the electric effect of DCs on an OHC must be assigned to an intermediate phenomenon of a mechanical nature. Yet there is an electrostatic link (a field effect) between the cytoskeleton apex of the DCs phalanxes and the OHC cuticle. This link exists without a significant current, and so without electrical conduction by gap-junctions. That *ephaptic effect*[61] is the field effect of a biological electric conductor, for example the field effect of an axon on its neighbourhood. It is distinguished from the phenomena of simple diffusion in the extra-cellular environment; It is also distinguished from communications between neurons by means of synapses either chemical or electrical [S61, S61, S61].
It is clear that this field effect must be detected mainly by the effects it produces within the cuticular bilayer, where it likely acts as a gate on the communication between the set source {stereocilia / endolymph} and the set drain {OHC cytoplasm, prestin, baso-cellular membrane, cupular slot}.
 Indeed, the field effects in the DOHC suggest that it could operate in a way similar to a triode[S61]. A Field Effect Transistor (FET) uses an electric field applied by a gate, to control the conductivity of a "channel" in semiconductor materials. Applied voltage on the gate electrode controls the amount of charge carriers flowing through the system.
In physiological conditions, lipids do not cross TJs: There is a strict insulation, chemical as well as electrical, between the two bound cells, and no current will flow from one to the other of their apical membranes[S61]. For this reason, no electric current can cross the barrier between the phalangeal apexes of the DCs and the cuticle of the embedded OHC. However, this border cannot prevent a hydrophobic intercellular electrostatic coupling unrelated to any GJ[S61 -S61].

## Si 10G **The destruction of the cytoskeleton annihilates the electric effect of the DCs on the OHCs**

The apex of the phalanxes of the DCs is connected to the cuticular plate of OHCs by Tight-Adherens Junctions (TAJs). These TAJs bring into contact the cytoskeleton of adjacent cells. It would be interesting to design "*experiments that could separate changes in intracellular ion concentration, from mechanical resistance by the cytoskeleton* [s61] *as a step towards understanding the signal transduction pathways that might involve microtubules*" (Szarama, pers.com., 2012). Furthermore, the destruction of the cytoskeleton of the DCs negates the effect of electric stimulations of the DCs on OHC electromotility (Yu and Zhao, 2009).

This implies that the cytoskeleton of DCs plays a critical role in the modulation of OHC electromotility. According these authors, this is evidence that, if electrical stimulations of DCs influence OHC electromotility, it must be through the DC-OHC mechanical coupling rather than by extracellular field effect. However, the electric voltage of the phalangeal apex necessarily causes variations of the electric field in the OHC cuticle (via the intercellular border at the level of the TJ). Indeed, the destruction of the phalangeal cytoskeleton removes the electrical activity of the phalangeal cytoskeleton$^{S61}$ and, as a consequence, its field effects. The measurements used by Yu and Zhao (2009) were taken using the whole-cell patch clamp method, i.e. with electrodes implanted into the cytoplasm of the DC and the OHC, but not within their respective bilayers; Thus, weak capacitive interactions, consistent with the action of an FET gate, escaped measurement. Cupular phenomena represent a more sensitive issue. The {DC Cup /OHC base} junction is very unusual (see below S24 - S25): at the level of the cupule it includes hemi-channels, whose activity is confined in the cupular slot; It is an extremely small dedicated space, without obvious communication with what is generally regarded as the extracellular medium or the Nuel space.

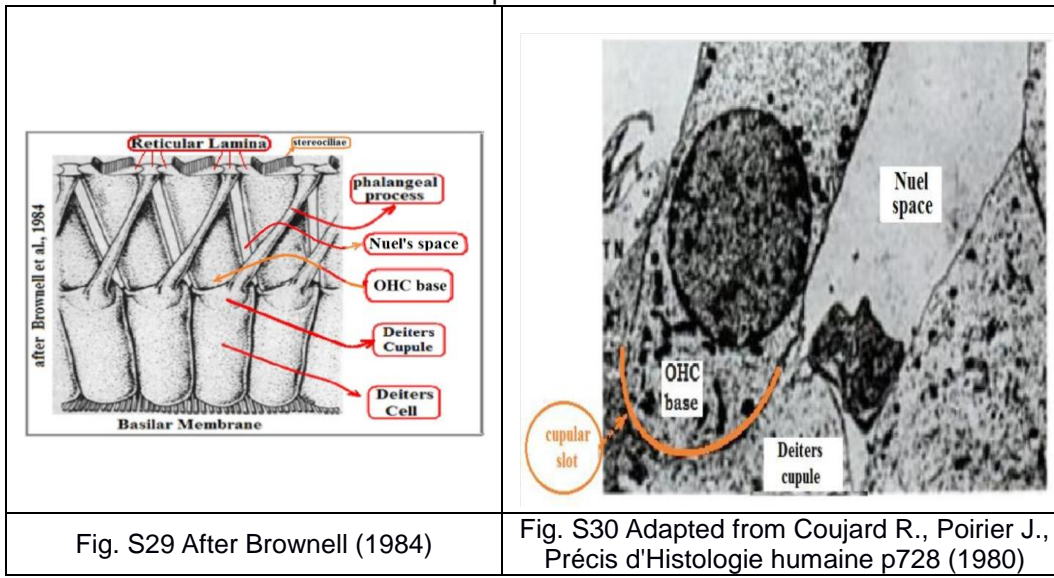

| Fig. S29 After Brownell (1984) | Fig. S30 Adapted from Coujard R., Poirier J., Précis d'Histologie humaine p728 (1980) |

Yu and Zhao (23) have put the extracellular milieu into communication with the earth, probably allowing effective control of the extra cellular environment after the breakdown of the cupule-OHC junction. This control is irrelevant, however, as regards variations affecting the slot between the cupule and the OHC as a specific space. The same is true for the variations within the intra-membranous TAJ cuticle-phalanx space$^{S61}$.

In another experiment, these authors blocked the ionic channels of the OHC basal membrane in such a way that the ionic changes in the cupular slot could not elicit charge-carrier exchanges between OHC and cupule. But, of course, when the physical joining between the DC cup and the OHC base is preserved, this precaution does not prevent the intervention of a capacitive electrical signal between cupular and OHC membranes. So, the blocking of ion channels is not sufficient to eliminate all of the effect of the Deiters cupule in the OHC membrane.

The OHC and the cupule are joined at the membrane level. The two membranes act as capacitors, and mechanically severing the coupling between them changes the capacitance of the set radically. In the case of the separation between cupule and OHC, the cupular slot enclosure is totally open and allows a fast diffusion of ions from the cupular hemi-channels: it follows that ionic changes from the cupule get "lost" in the extracellular medium and cannot reach a level sufficient to act upon the OHC.

In each phalanx, the micro-tubules set, might manage an electrical resonance which would optimize the tuning.

When a potential (eg pT amplified and filtered by phalanxes) is applied to the bio-organic semi-conductors of the cuticular bilayer, a current can flow through this cuticular bilayer, between the suitably biased source and drain "electrodes".

## Si 10H The Drain:  Prestin

There is good evidence that prestin has undergone adaptive evolution in mammals [6]. This is associated with acquisition of high frequency hearing in mammals.[7] The prestin protein shows several parallel amino acid replacements in bats and dolphins that have independently evolved ultrasonic hearing and echolocation, and it represents a rare case of convergent evolution at the sequence level. Unlike the classical, enzymatically driven motors, this new type of motor is based on direct voltage-to-displacement conversion and acts several orders of magnitude faster than other cellular motor proteins. A targeted gene disruption strategy of prestin showed a >100-fold (or 40 dB) loss of auditory sensitivity[S61].

The piezo tympanic signal reaches a $DC_5$ cup, which supports the OHC in question. $DC_5$ is distinct from the four other DCi (1..4), the phalanxes of which are in contact with the cuticle of the OHC. At the same time, the pT signal is carried by the microtubules of the phalanxes of the four $DC_{i(1..4)}$ toward the TJs. There is no electrical conduction between the phalanxes and OHCs, but there is a capacitive effect (Scut) between phalanxes and the intra-cuticular membrane space. The cuticular bilayer comprises a great many homo-conjugated PUFAs which are semiconductors. This property makes the phalangeo-cuticular structure similar to a FET (*Trickystor*). The characteristic frequency of the DOHC complex is inversely related to its distance from the oval window (Greenwood relation), the length of the phalanxes of the DCis, the length of the OHC and the length of stereociliae.

For the classical pathway, the duration of the mechanical signal path increases with the distance between the eardrum and the Greenwood area corresponding to a given frequency. For the hidden pathway, the duration of the electrical signal path increases with the length of the phalanx, and depends on characteristics of the cytoskeleton[S61]. If the Scil (ciliary signal) and the $S_{cut}$ are approximately synchronous (□d ≈ □w), this results in an amplified mechanical signal ($S_{amp}$). If there is a precession either of the $S_{cut}$ signal on the $S_{cil}$ signal, or of the $S_{cil}$ signal on the $S_{cut}$ signal, the *trickystor* does not immediately acquire an maximal effectiveness since this effectiveness requires synchronous interaction of both signals.
But, the acoustic signal (TW) is transduced into an electrical signal by the stereocilia; this electrical signal is amplified by the *trickystor*; then it is once more transduced into a new overamplified mechanical signal by prestinic movements. These movements act on the basilar membrane and the supra-reticular region in such a way that the local amplitude and accuracy of the TW are enhanced.

The wavelength depends on the velocity of propagation of the wave in the medium it crosses. When the wave passes from one medium to another, in which its velocity is different, its frequency remains unchanged, but its wavelength varies (Dic. Phys.). As a result, this feedback process could set up a phase lag[S61] (τfb), increasing with each cycle. The increasing phase lag would tend, by progressive desynchronization, to self-limitation, damping and extinction of this type of amplification (fig. S31).

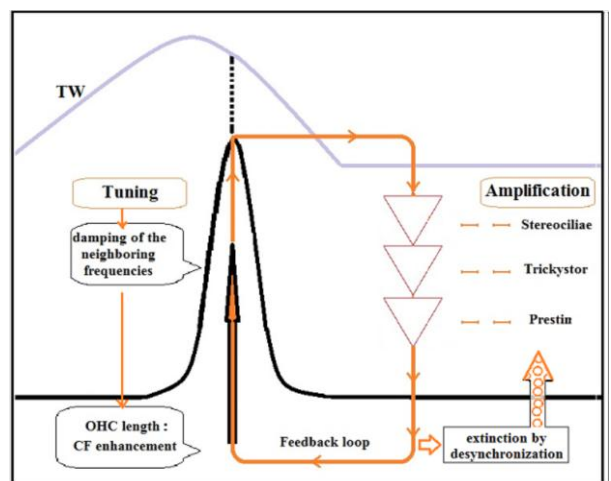

Figure S31
Schematic view of some putative processes regarding
amplification, tuning, damping by DOHC and also creation of OAE.

## Si 10l About prestin and low frequencies

The conductance of the Mechano Electrical Transducer channels changes are depending upon the tonotopical position within the cochlea, and that suggests differential requirements at different frequencies[S61].

Oddly enough, some profoundly deaf people are able to hear some so-called infra-sound frequencies[S61 - S61] through the labyrinth (sacculus). The cochlea of normally hearing people has mechanisms that weaken or suppress these sound elements before they are transmitted to the brain. Ashmore, points out that "prestin is expressed […] surprisingly, in the cells of the vestibular system[S61]".

Thus, into its vestibular locus, prestin (otherwise the motor protein of OHCs), might increase or decrease the sound amplitude according to whether the frequency is above or below 50Hz[S61].

It is possible that this "suppressive" effect would be produced by a phase inversion during the amplification process of these infra-sound frequencies, thus producing an attenuation of these stimuli either when they are not useful, or would become disruptive.

## Si 11 **Phase and Times**

### Si 11A Phase shift of the TW and pT for f > 3 kHz (Quix)

Discussion about the phase lag of TWm relatively to the without delay signal of the pT:
The princeps (originator) TW of Von Bekesy, derived from ossicles, retains the frequency of the acoustic stimulus but with a delay resulting from the speed of the sound waves in water (1.500 m/s)[S61], whereas the shorter delay of the pT results from the speed of the electric signal through the GJs (ie, approximately, the speed of an electric signal in salt water, i.e. a speed of 226 000 m/s) [S61]. The pressure variation on the eardrum, sine-type (TW), is transduced into electrical potential by the stereocils ("source" of the TkS).

Another sine of the same origin and frequency produces a pT potential and reaches the "gate" of the TkS. Since the movement of TW is slower than the movement of the pT electrical signal, there is a phase shift between these two waves that may tend to lessen the effect of the TkS.
To evaluate the effect of this phase shift, we can take two examples. Let us use for this evaluation 4 khz and 5 khz (Table S09 below).

| **Effects of phase shift on 4 kHz.** Calculation of the time delay for f = 4 kHz | **Effects of phase shift on 5 kHz.** Calculation of the time delay for f = 5 kHz |
|---|---|
| Be the delay of TW: $D1_4$. If a 4 kHz acoustic stimulation is used, the TW must travel 8.7 mm, ie 0.0087 m. To traverse this distance at the speed of a sound in water, i.e. 1,500 m/s, the delay will be → $D1_4$ = 0.0087/1500 = 0.0000058 = $5.8 * 10^{-6}$ sec. $D_{pT}$ being the $D2_4$ time of the pT potential: Delay of pT to reach the same zone 0.0044 m, at the speed of the electric wave in salt water (roughly identical to the speed via the Gap junctions) = > 0.0087/226 000 = $3.85 * 10^{-8}$ sec. For 4 kHz, the difference $dD_4 = D1_4 - D2_4 = 5.8 * 10^{-6}$ sec - $3.85 * 10^{-8}$ sec; $D2_4$ represents 6 thousandths of $D1_4$ and can be considered negligible because the value of this delay is included in measurement errors or calculation rounds and can be accepted.<br><br>dD = D1-D2 ± ε = $5.8 * 10^{-6}$ ± ε sec. The phase shift at 4000 Hz : The period of a wave of 4000 Hz is T = 1/4000 = 0.00025 sec. Δ (φ) = $5.8 * 10^{-6}$ * 360/0.00025 = 8° 35 = 0.15 rad In percent, 0.15 radian/6.28 Radian = 0.02 (i.e. 2% of a cycle).<br>**Generalization** : if f2>f1 the travel of f2 will be shorter than that of f1(Greenwood relation). The shift will be thus all the more small as F will be large. In fact the calculation shows that the phase shift of TW and pT for f > 3 kHz is < 3% and therefore should not interfere with the operation of the TkS. | Be the delay of TW: $D1_5$. If a 5 kHz acoustic stimulation is used, the TW must travel 6.3 mm, ie 0.0063 m. To traverse this distance at the speed of a sound in water, i.e. 1,500 m/s, the delay will be → $D1_5$=0.0063/1500 = 0.0000042 = $4.2 * 10^{-6}$ sec. $D_{pT}$ being the $D2_5$ time of the pT potential: Delay of PT to reach the same zone 0.0063 m, at the speed of the electric wave in salt water (roughly identical to the speed via the Gap junctions) = > 0.0063/226 000 = $2.79 * 10^{-8}$ sec. For 5 kHz, the difference $dD_5 = D1_5-D2_5 = 4.2 * 10^{-6}$ sec - $2.79 * 10^{-8}$ sec; $D2_5$ is smaller than 7 thousandths of $D1_5$ and can be considered negligible because the value of this delay is included in measurement errors or calculation rounds and can be accepted.<br><br>dD = D1 ± 0.007 = $4.2 * 10^{-6}$ ± 0.007 sec. The phase shift at 5000 Hz : The period of a wave of 5000 Hz is T = 1/5000 = 0.0002 sec. Δ (φ) = dD * 360/T = $4.2 * 10^{-6}$ * 360/0.0002 = 7°56 = 0.13 rad As a percentage, 0.13 radian/6.28 radian = 0.02 (ie 2% of a cycle). |
| Table S16 Effects of phase shift according frequency | |

### Si 11B **Precession of the electric pathway on acoustical pathway ?**

A more specific work on the phase shift between the vibratory movements of the Tectoria Lamina (TL) and the basilar membrane (BM) has determined that the vibration of the BM is leading that of the TL[S61]. Yet, the movements between the top and the bottom are in phase opposition.
Note that for the most part the movements have reverse phase between the top and bottom except for the Best Frequency. There is evolution so that it synchronizes with the best frequency.
It is difficult to draw clear lessons from this.
We believe that the electrical signal is faster than the mechanical signal and that (despite a delay in the phalanx) this phase shift suggests a causal series of the type:

---
ZZ.    pT → GJs →Deiters → Phalanx (+delay)  → Gate of TkS → signal between Reticular Lamina and Prestin
→ Prestin contractions → BM and LR in phase opposition
→ evolution (by resonance) towards synchronization at the best frequency.

---

The difference in delay between TW and pT implies a precession of the onset of the pT wave in relation to the TW and the stereociliar potential it produces. Fidberger et al.[S61] had discovered that: "Electric potentials inside Corti's organ preceded BM velocity" [...] And they went on "If such an obvious phase lead were present also at the start of the stimulus, it would mean that electric potentials were produced before BM motion". But unsatisfied by that result incompatible with the accepted design, they oddly enough modified their experimental protocol and, "using a 3 kHz stimulus at 100 dB SPL", they saw that the TW was moving before that "_detectable_" potentials could be generated. Indeed, the 3 kHz frequency they chose is included within the limits under the Quix frequency and does not conflict in any way against our own view.

### Si 11C About the Quix frontier between low and high frequencies

Is there any possibility of defining a boundary between these two sets ?
The frequency of Quix (3 kHz), concerns the minimum of the pT curve, the minimum of the iso-sonic curves, and many other phenomena of the physiology of human hearing.
With regard to humans, it is logical to use the frequency of Quix which, for greater rigor should be provided with a ± Δf (1 Khz?).

$$Quix = 3 \text{ khz} \pm \Delta f = 3 \pm 1 \text{ khz}$$

The study of other mammals invites one to consider that Quix should be applied to the animal species under consideration (cetaceans; bat, etc.) whose "Quix area" is much higher.
The greatest sensitivity of cetaceans lies between 40 and 80 kHz. Their sensitivity decreases between 40 kHz and 1 kHz, and also between 80 kHz and 150 kHz :

$$cetaceans\ f_{QUIX} = f_{QUIX} \pm \Delta f = (60 \pm 20) \text{ kHz}$$

# Si 12 OAE versus BEW

The existence of OAEs (Oto Acoustic Emissions) has led to the postulation of a "Backward Travelling Wave" (BTW). Until now, however, no clear evidence has been found for it [S61].
The process of the Electrically Evoked Otoacoustic Emissions (EEOAE), includes electrical stimulation of the OHCs. The effect of this cochlear stimulation is extremely rapid and its upper cutoff frequency is > 80 kHz[S61, S61]. This permits the recording of an acoustic response of the same frequency at the level of the external auditory conduit[S61].

Under the postmortem condition, the electrically evoked basilar membrane (BM) vibration almost disappears while **the EEOAE shows no significant change**. These results indicate that the BM vibration is not involved in the backward propagation of the EEOAE. To explain this, it has been proposed to accept the existence of fast "backward compressional sound waves" instead of the elusive "slow backward traveling waves" [S61]. Yet, that hypothesis is puzzling because it is difficult to figure out how this acoustical compressive phenomenon might be attributable to the motility of the OHCs given that, after death, the OHCs would cease to function actively.

We have shown that electrical potentials that we measured near either the eardrum or the mastoid cannot be explained solely by the diffusion of the cochlear microphonic. But if a partial diffusion of this electrical signal does exist, this 'backward electric wave' (BEW) may cause slight acoustic vibrations of the eardrum by inverse piezoelectric effect. Thus, this BEW could be responsible for the EEOAEs and, what is more, of all OAEs !

## Si 13  The global system using together the overt and covert paths

The whole system (overt path and covert path together) constitute a complex system which is a biologic Micro-Electro-Mechanical-System, similar to an industrial MEMS. This acronym MEMS is interesting since these few letters include: the mechanical point of view (or mechanical aspect), very present in our model, the piezo-electric aspect of collagen and the electronic aspect of the trickystor (with its biological semi-conductors); the term "micro" is perfectly justified since the size of several of its components are of the micrometric order.

## Si 14 Notes Concerning Statistical Analysis

The table of raw data is available as an excel file: raw_data_covert_path.xlsx (https://figshare.com/s/3338464a25212da726cf)

Si 14A Description of the data

We did 221 experimental series of microvoltage measurements in function of the sound frequencies (Hz) and the sound level (dB). These experimental series were measured on 62 individuals (A01-A10, B11-B20, C21-C30, D31-D35, E43-E44, E48-E50, F1-F12). Of these recordings, 110 correspond to a male and 111 to a female. The ages of individuals were < 93 years$^{s61}$. From these 221 series of measurements, 79 were obtained at the level of an eardrum. To check the properties of collagen fibers different from those of the eardrum, we also made these measurements on the forearm, the Achilles tendon, the nasal cartilage and especially, in a very systematic way, the ligaments patellar right and left: 123 series of measurements were obtained at the level of a knee. For eardrums 40 were the left one, and 39 the right one. For the knees, 60 were a left knee and 63 a right one.

16 specific sound frequencies were globally tested  (125, 200, 400, 500, 800, 1000, 2000, 4000, 6000, 8000, 10000, 15000, 16000, 20000, 25000, 30000 Hz), but 9 main frequencies were more particularly considered (500, 1000, 2000, 4000, 8000, 10000, 16000, 20000 Hz). We also tested 5 sound pressure levels from 55 to 80 dB (55, 60, 65, 70 and 80 dB). The data table is not complete, and there are missing data for some frequencies.

Si 14B Body Sites of measurements regarding piezoelectricity

Si 14Ba Forearm
Piezoelectric response of collagen at forearm level (μV / kHz)
For exploratory purposes, we measured synchronous potentials responding to sounds of different frequencies on four subjects, between two points of the left forearm, according to the axis of the limb, (Table S10; Fig. S27).

Si 14Bb Achilles tendon

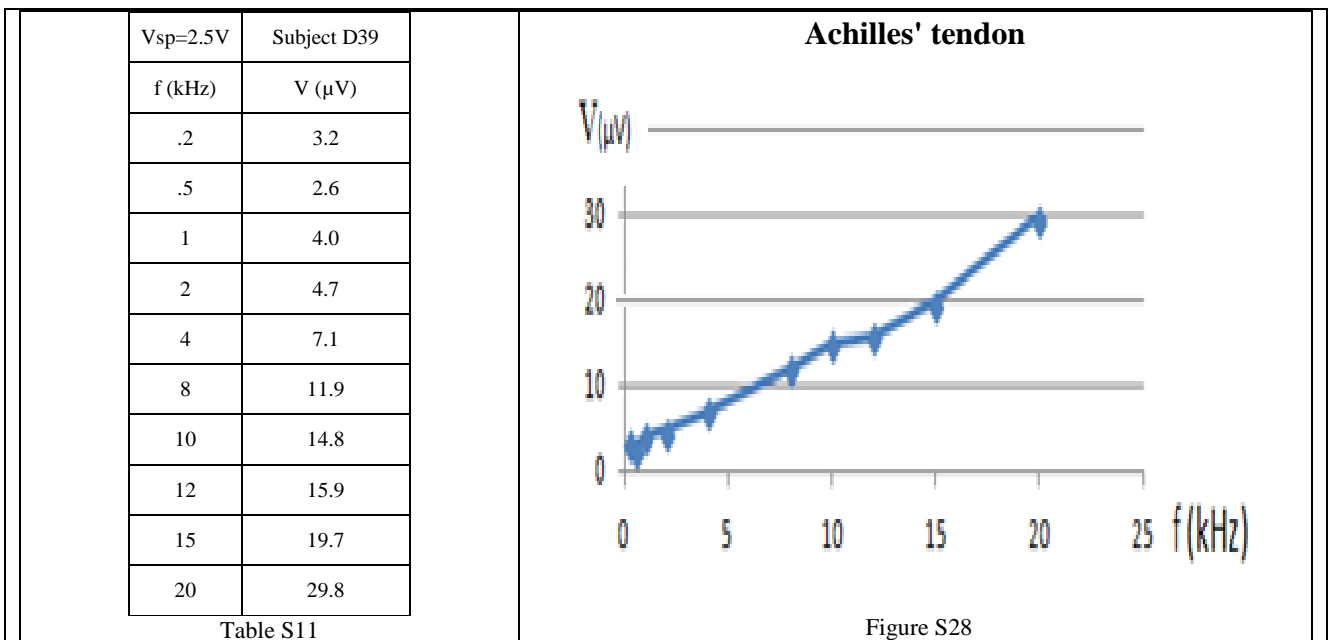

| Vsp=2.5V | Subject D39 |
|---|---|
| f (kHz) | V (μV) |
| .2 | 3.2 |
| .5 | 2.6 |
| 1 | 4.0 |
| 2 | 4.7 |
| 4 | 7.1 |
| 8 | 11.9 |
| 10 | 14.8 |
| 12 | 15.9 |
| 15 | 19.7 |
| 20 | 29.8 |

Table S11

Figure S28

# AAA.

We used the lock-in amplifier to measure synchronous potentials V (μV) of the Achilles tendon in response to the acoustic stimulation of frequency f (kHz), emitted by a Harman/Kardon Loudspeaker: DP/N 0865DV. It was stimulated by a potential Vsp of 2.5 V, according to an installation modelled on that which we have described for the fibres of the eardrum (Table S11).

**Si 14Bc Knees**

As was the case for the Achilles tendon, we used a lock-in amplifier to measure synchronous potentials V (μV) of the patellar tendon in response to an acoustic stimulation of frequency f (kHz), emitted by a Harman/Kardon LoudSpeaker: DP/N 0865DV stimulated by the VLS (Volts) potential. The magnitudes of the synchronous potentials (μV) recorded on the patellar tendon, in function of the acoustic frequency, are reported below for different subjects (Table S12 and Fig. S29).

| f (Hz) | D36 | D37 | D38R | D38L | D39 | Total |
|---|---|---|---|---|---|---|
| 100 | 14 | 5 | 8 | 8 | 8,5 | 43.5 |
| 500 | 12,5 | 4,2 | 6 | 6 | 6,3 | 35.0 |
| 1000 | 15,5 | 5,8 | 7 | 6,1 | 8,5 | 42,9 |
| 2000 | 21 | 9,1 | 8,4 | 7,6 | 10,4 | 56,5 |
| 4000 | 27 | 13,4 | 10,9 | 9,8 | 16,5 | 77,6 |
| 8000 | 42 | 23,7 | 17,1 | 14 | 26,7 | 123,5 |
| 10000 | 54 | 28,4 | 20,5 | 17 | 32,5 | 152,4 |
| 15000 | 73 | 42 | 27,2 | 22 | 47 | 211,2 |
| 20000 | 92 | 61,4 | 35 | 29 | 62 | 279,4 |
| Total | 351 | 193 | 140,1 | 119,5 | 218,4 | 1022 |
| age | 30 | 51 | 73 | 73 | 17 | |

Table S12
Speaker LVL 2.5 V ; F (Hz) ; responses (µV)

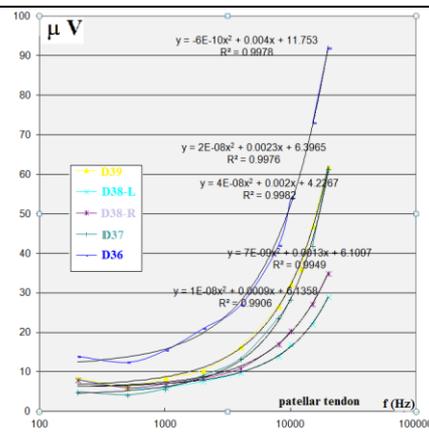

Figure S29
VLS = 2.5 V. Synchronous voltage vs acoustic frequencies for 5 patellar tendons. Every R² > 0.99

Every musculo-tendinous formation that we tested (mastoid, arm, Achilles tendon, patellar ligament) produced electric potentials synchronous to the acoustic stimuli. The piezoelectric effect sheds some light on this phenomenon.

# Si 15 Effect of pathological history on measures

Regarding eardrums, we noted some phenomena related to pathological histories;
For example, weakness of piezoelectric responses in the case of allergy inside the ear canal (B16, B17).
Anomaly of measurements carried out on a scar zone (D31 on the left eardrum).

D33 uses hearing aids "PS60 Resound" on both ears. Measures were made without any of the two prostheses present. However, the electrical response is particularly high on the eardrum G of this subject. In this case, would the histological structure of the eardrum had evolved by eliciting a more important piezoelectric response, so that the deafness would be more or less compensated ?

## Si 16 Pain felt during the measurement taken on the eardrum

Regarding the 79 eardrum series, an index of "*felt pain*" (0 to 5) was recorded. 51 recordings describe no pain (0) and 12 a gene or major pain (between 3 and 5). Sometimes the placement of the probe was too painful to result in a measure concerning the ear {A07 / D34}.

## Si 17 Explaining fixed effects by characteristic variables of the individuals

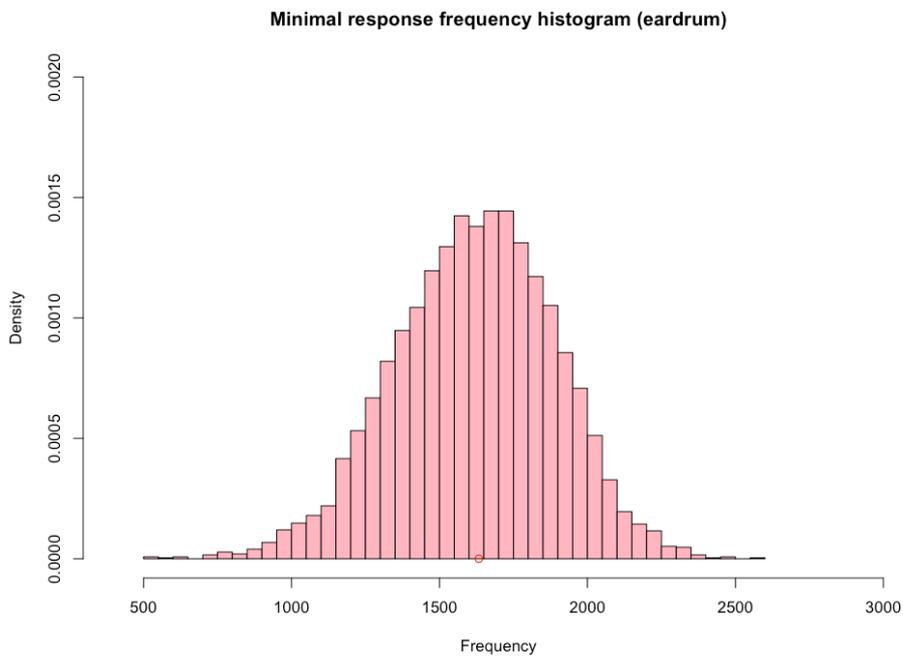

Figure S30
Distribution of the minimal frequency value for eardrum data.

For eardrums the minimal frequency (Quix frequency small red circle in Fig.S30) is lower than the knees one : It appears that separate estimates of the log-quadratic model in the frequencies F for the two sound levels 70 and 80 dB lead to two distinct minimal values 2791Hz and 3375Hz (Fig. S31). Contrary to what we observe for eardrums, it seems that there is for knees a significant interaction effect between the sound frequency and the sound level.

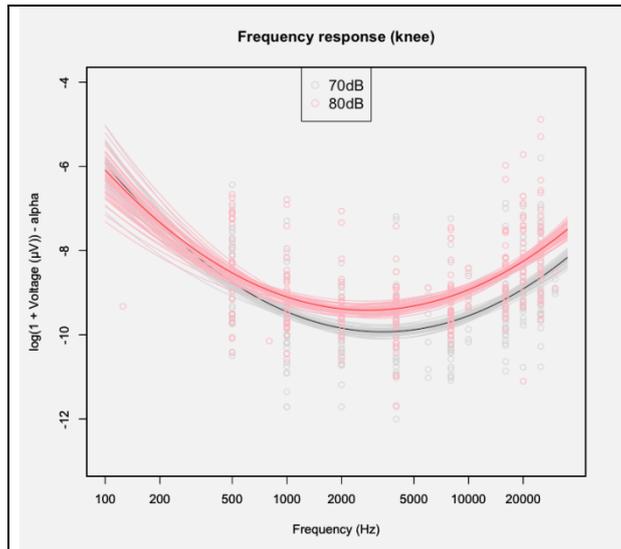

Fig. S31 Effect of the amplitude of the sound stimulus for knee measurements

## Si 18 **Comparison of the Wegel curve of the acoustic pressures, versus the curve of the piezoelectric tensions (pT)**

For all the individuals, and for the two sides, the minimum value of the curve is obtained for a frequency equal to about $f_{pTmin}$ =1632 hz. The frequency $f_{QUIX} \approx$ 3000 hz (fig. S32)would be the frequency which marks the region of maximum audibility, i.e. for which there is the lowest threshold threshold[s61 - s61] (hearing *'yellow spot'* of Quix). These two minima (1632 hz and 3000 hz) are relatively different. Yet we may see that $f_{QUIX}$ (3000 hz) $\approx$ 2 $f_{pTmin}$ (1632 hz); ie $f_{QUIX}$ is approximately the harmonic of $f_{pTmin}$.

$$\mathbf{f}_{QUIX} \approx 2\ \mathbf{f}_{pTmin}$$

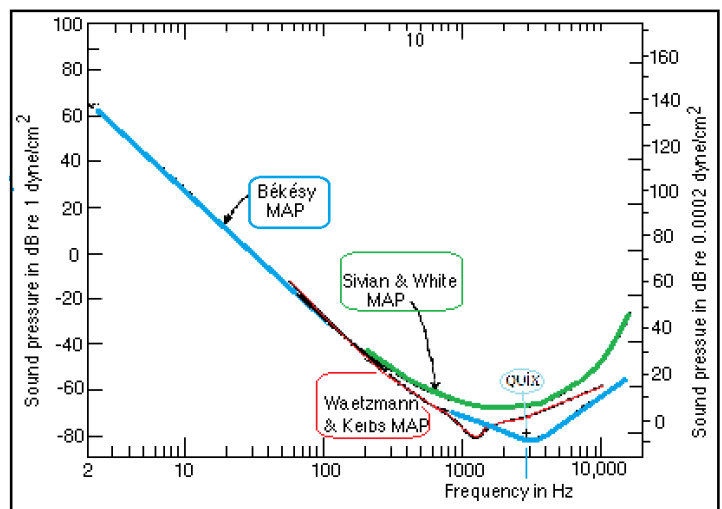

Figure S32
Wegel diagram and other classical schemas excerpted and modified from
Licklider, Handbook of Experimental Psychology, S.S. Stevens, ed., 1951

The frequency of Quix seems to be less fixed than its author thought. The frequency of Quix is that which is audible by the lowest acoustic stimulation compared to the other frequencies. In our experiments, the stimulation of the eardrums used relatively high amplitudes in relation to the threshold of hearing. As a result, the frequency of the minimum of the pT curve is lower than the frequency of Quix, but the two frequencies may be closer, or coincide, if we use liminal acoustic stimuli, in both cases (acoustic and piezo-electric).

The Wegel curve and the many related curves of the physiology of hearing give the minimum pressure required for each frequency to be heard. The Piezo-electric potentials depending on the frequency, have a variation that seems homogeneous to the physical pressures (dB SPL) of the audiometric thresholds that one has in the curve of Wegel. If the minima coincided one could say - depending on the frequency - that the less acoustic energy is necessary to reach the audibility, the less the piezo-electric response is high. The more acoustic energy is needed, the more the piezo-electric response is high!

## Si 19 individual effect…

## Si 20

### Si 20A Judicious redundancy

We think that multiple channels enhance the quality of information received with respect to the information emitted: If the transmission consists of several channels contaminated by radically different noise sources, and if, upon reception, you retain only what is common to these different collected signals (judicious redundancy), the result might be a cleaner signal, partially freed of parasitic noise[S61].

### Si 20B Vicariances[61]

Moreover, the deficiency of one of the pathways might be partially compensated by the vicariance of the other (or other) pathway(s). Hearing simultaneously uses a mechanical transmission (acoustical TW) and electric transmission (pT sent to the trickystor). The noise polluting the overt path (mechanical transmission by ossicles, TW and mobilization of the stereociliae), is different from the noise polluting the covert path (piezoelectric activity of the tympanum, electrical signal transmitted through GJs, functioning of the trickystor). If we select what is common to the outputs of each of these two pathways, we can reduce the specific noise of each of the two.

The multiplicity of channels used to transmit the acoustic information, even if some of these channels are impeded[S61], is beneficial for an optimal hearing. This multiplicity allows some parts of the signal to reach the neurological structures with sufficient practical value.

In addition, the electrical signal itself, is transmitted using a very large number of pathways, isolated from each other, so that their final summation reinforces what is common (signal) and weakens their differences (noise).

## Si 21 Supplementary Information Bibliography